\documentclass[sigconf,authorversion,nonacm]{acmart}
\AtBeginDocument{%
  }

\setcopyright{cc}
\copyrightyear{2025}
\usepackage{esvect}
\usepackage{listings}
\usepackage{dashrule}
\usepackage{tikz}
\usetikzlibrary{shapes.geometric, arrows, positioning}
\usepackage{mathpartir}

\newcommand{\from}{\ensuremath{\colon}}

\newcommand{\La}[0]{\ensuremath{\langle}}
\newcommand{\Ra}[0]{\ensuremath{\rangle}}
\newcommand{\abracket}[1]{\ensuremath{\La #1 \Ra}}

\newcommand{\stx}[1]{\expandafter\ifx\expandafter\relax
    \detokenize{#1}\relax\ensuremath{S_{\mathcal{X}}}\else\ensuremath{S_{\mathcal{X}\text{-#1}}}\fi}
\newcommand{\Mx}[1]{\expandafter\ifx\expandafter\relax
    \detokenize{#1}\relax\ensuremath{\mathcal{M}_{\mathcal{X}}}\else\ensuremath{\mathcal{M}_{\mathcal{X}\text{-#1}}}\fi}

\newcommand{\mcalx}[1]{\expandafter\ifx\expandafter\relax
    \detokenize{#1}\relax\ensuremath{\mathcal{X}}\else\ensuremath{\mathcal{X}\text{-#1}}\fi}

\newcommand{\M}[1]{\ensuremath{\mathcal{M}_{\mathit{#1}}}}
\newcommand{\trs}[1]{\mbox{\ensuremath{\mathcal{M_{\text{#1}}}= \langle S_{\text{#1}}, \xrightarrow[\text{#1}]{}, L_{\text{#1}} \rangle}}}
\newcommand{\atrs}[1]{\mbox{\ensuremath{\mathcal{M_{\text{#1}}}= \langle S_{\text{#1}}, A_{\text{#1}}, \xrightarrow[\text{#1}]{}, L_{\text{#1}}\rangle}}}

\newcommand{\trans}[0]{\ensuremath{ \rightarrow }}
\newcommand{\xtrans}[1]{\ensuremath{ \xrightarrow[#1]{} }}
\newcommand{\atrans}[2]{\ensuremath{ \xrightarrow[#1]{#2} }}
\newcommand\scorr[5]{\ensuremath{\mathit{scorr(#1,#2,#3,#4,#5)}}}
\newcommand\smatch[3]{\ensuremath{\mathit{smatch(#1,#2,#3)}}}
\newcommand\acorr[5]{\ensuremath{\mathit{acorr(#1,#2,#3,#4,#5)}}}
\newcommand\amatch[3]{\ensuremath{\mathit{amatch(#1,#2,#3)}}}
\newcommand{\ie}[0]{\emph{i.e.}, }
\newcommand{\eg}[0]{\emph{e.g.}, }
\newcommand{\ea}[0]{et al.}

\newcommand{\mybigcirc}[0]{\ensuremath{\mathop{\text{\Large$\circ$}}}}

\newcommand{\ic}{\textit{IC}}
\newcommand{\act}{\textit{ACT}}
\def\myiff{\ensuremath{\mathit{iff}}}
\newcommand\Comment[1]{}
\newcommand{\fp}{\textit{fp}}
\newcommand{\fpa}{\textit{fp-a}}
\newcommand{\inc}{\textit{INC}}
\newcommand{\partition}[3]{\ensuremath{\mathbin{^{#2}\hspace{-1.5pt}#1^{#3}}}}
\newcommand{\partitionsub}[4]{\ensuremath{\mathbin{^{#2}\hspace{-1.5pt}#1^{#3}_{#4}}}}
\newcommand{\ra}{\ensuremath{\rightarrow\;}}
\newcommand{\disjtrs}[3]{\ensuremath{\langle S_{#1} \uplus S_{#2},\xrightarrow[#1]{} \uplus \xrightarrow[#2]{}, \labf{}_{#3} \rangle }}
\newcommand{\disjatrs}[4]{\ensuremath{\langle S_{#1} \uplus S_{#2}, A_{#4}, \xrightarrow[#1]{} \uplus \xrightarrow[#2]{}, \labf{}_{#3} \rangle }}
\newcommand{\labf}{\ensuremath{\mathcal{L}}}
\newcommand{\incache}{\ensuremath{\mathit{in\textnormal{-}cache}}}
\newcommand{\True}{\texttt{true}}
\newcommand{\False}{\texttt{false}}
\newcommand{\lmem}{\ensuremath{\mathit{LM}}}
\newcommand{\ISAM}{\texttt{ISA}}
\newcommand{\ISAICM}{\texttt{ISA-IC}}
\newcommand{\ISAICAM}{\texttt{ISA-IC-A}}

\newcommand{\MAM}{\texttt{MA}}

\newcommand{\MAICM}{\texttt{MA-IC}}
\newcommand{\MAICAM}{\texttt{MA-IC-A}}
\newcommand{\MAICHM}{\texttt{MA-IC-H}}

\newcommand{\MAICNM}{\texttt{MA-IC-N}}
\newcommand{\MAGICM}{\texttt{MA-G-IC}}

\newcommand{\tda}[0]{\text{-}}

\newcommand{\mtsxc}[1]{\ensuremath{#1\tda\tsx\tda{}c}}
\newcommand{\mrsfi}[1]{\ensuremath{#1\tda\rsf\tda{}i}}
\newcommand{\mrsfe}[1]{\ensuremath{#1\tda\rsf\tda{}e}}
\newcommand{\mrsfw}[1]{\ensuremath{#1\tda\rsf\tda{}w}}
\newcommand{\mcc}[1]{\ensuremath{#1\tda\textit{cmem}\tda{}c}}
\newcommand{\mhccc}[1]{\ensuremath{#1\tda\textit{ccmem}\tda{}c}}
\newcommand{\mhceffc}[1]{\ensuremath{#1\tda\textit{ceff}\tda{}c}}
\newcommand{\mhceffw}[1]{\ensuremath{#1\tda\textit{ceff}\tda{}w}}
\newcommand{\mhl}[1]{\ensuremath{#1\tda\textit{hl}}}
\newcommand{\mhlrm}[1]{\ensuremath{#1\tda\textit{hl}\tda{}rm}}
\newcommand{\mhhlw}[1]{\ensuremath{#1\tda\textit{hl}\tda{}w}}
\newcommand{\mhlrs}[1]{\ensuremath{#1\tda\textit{hl}\tda{}rs}}
\newcommand{\mhlmc}[1]{\ensuremath{#1\tda\textit{hl}\tda{}mc}}
\newcommand{\mhli}[1]{\ensuremath{#1\tda\textit{hl}\tda{}i}}
\newcommand{\mhscrm}[1]{\ensuremath{#1\tda\textit{sc}\tda{}rm}}

\newcommand{\mhscc}[1]{\ensuremath{#1\tda\textit{sc}\tda{}c}}
\newcommand{\mhsci}[1]{\ensuremath{#1\tda\textit{sc}\tda{}i}}
\newcommand{\mrsi}[1]{\ensuremath{#1\tda\textit{rgs}\tda{}i}}
\newcommand{\mrsc}[1]{\ensuremath{#1\tda\textit{rgs}\tda{}c}}
\newcommand{\mpcc}[1]{\ensuremath{#1\tda\textit{pc}\tda{}c}}
\newcommand{\mrfvc}[1]{\ensuremath{#1\tda\rfv\tda{}c}}
\newcommand{\mrbi}[1]{\ensuremath{#1\tda\rob\tda{}i}}
\newcommand{\mrbw}[1]{\ensuremath{#1\tda\rob\tda{}w}}
\newcommand{\mrbc}[1]{\ensuremath{#1\tda\rob\tda{}c}}

\newcommand{\busyrs}[0]{\textit{rs-busy?}}

\newcommand{\allowcommit}[0]{\textit{comm?}}
\newcommand{\allowstart}[0]{\textit{strt?}}

\newcommand{\skiprel}[1]{\ensuremath{B_{\textit{#1}}}}

\newcommand{\good}{entangled}
\newcommand{\goodshort}{ent}

\renewcommand{\And}{\ensuremath{\,\wedge\,}}

\let\origthelstnumber\thelstnumber
\makeatletter
\newcommand*\Suppressnumber{%
  \lst@AddToHook{OnNewLine}{%
    \let\thelstnumber\relax%
     \advance\c@lstnumber-\@ne\relax%
    }%
}

\newcommand*\Reactivatenumber{%
  \lst@AddToHook{OnNewLine}{%
   \let\thelstnumber\origthelstnumber%
   \advance\c@lstnumber\@ne\relax}%
}

\newcommand{\plusc}{\ensuremath{\oplus}}
\newcommand{\minusc}{\ensuremath{\ominus}}
\newcommand{\timesc}{\ensuremath{\otimes}}
\newcommand{\andc}{\ensuremath{\mathbin{\&}}}
\newcommand{\natc}{\ensuremath{\mathbb{N}_{32}}}
\newcommand{\nats}{\ensuremath{\mathbb{N}}}

\newcommand{\pc}{\textbf{pc}}
\newcommand{\rf}[1]{\textbf{rf}(#1)}
\newcommand{\rfv}{\textbf{rf}}
\newcommand{\imem}{\textbf{imem}}
\newcommand{\dmem}{\textbf{dmem}}
\newcommand{\tsx}{\textbf{tsx}}
\newcommand{\halted}{\textbf{halt}}

\newcommand{\bools}{\ensuremath{\mathbb{B}}}
\newcommand{\cmem}{\textbf{cache}}

\newcommand{\tsxactive}{\textbf{tsx-act}}
\newcommand{\tsxrf}{\textbf{tsx-rf}}
\newcommand{\tsxfallback}{\textbf{tsx-fb}}

\newcommand{\regs}{\ensuremath{\mathcal{R}}}

\newcommand{\rob}{\textbf{rob}}
\newcommand{\rsf}{\textbf{rs-f}}
\newcommand{\regstat}{\textbf{reg-st}}
\newcommand{\cyclec}{\textbf{cyc}}
\newcommand{\fetchpc}{\textbf{fetch-pc}}
\newcommand{\prefetch}{\textbf{prefetch}}

\newcommand{\funcfont}[1]{\textit{#1}}

\newcommand{\fetchfunc}[2]{\ensuremath{\funcfont{fetch}(#1, #2)}}
\newcommand{\fetchicfunc}[2]{\ensuremath{\funcfont{fetch}_\textit{IC}(#1, #2)}}
\newcommand{\partialget}[2]{\ensuremath{\funcfont{get}_{#1}(#2)}}

\newcommand{\emptyrf}{\ensuremath{\mathcal{R}_\emptyset}}
\newcommand{\goodaddr}{\ensuremath{\textbf{ga}}}
\newcommand{\goodaddrfn}[1]{\ensuremath{\goodaddr(#1)}}

\newcommand{\fndom}{\mathrm{dom}}

\newcommand{\instr}[1]{\fbox{\ensuremath{#1}}}

\newcommand{\rsmop}{\textbf{rs-mop}}
\newcommand{\rsid}{\textbf{rs-id}}
\newcommand{\rsqk}{\textbf{qk}}
\newcommand{\rsqj}{\textbf{qj}}
\newcommand{\rsvk}{\textbf{vk}}
\newcommand{\rsvj}{\textbf{vj}}
\newcommand{\rscpc}{\textbf{cpc}}
\newcommand{\rsbusy}{\textbf{busy}}
\newcommand{\rsexec}{\textbf{exec}}
\newcommand{\rsdst}{\textbf{dst}}
\newcommand{\rspc}{\textbf{rb-pc}}

\newcommand{\robid}{\textbf{rob-id}}
\newcommand{\robmop}{\textbf{rob-mop}}
\newcommand{\robdst}{\textbf{rdst}}
\newcommand{\robrdy}{\textbf{rdy}}
\newcommand{\robval}{\textbf{val}}
\newcommand{\robexcp}{\textbf{excep}}

\newcommand{\regstbusy}{\textbf{busy}}
\newcommand{\regstreorder}{\textbf{reorder}}

\newcommand{\mafetchnum}{\textbf{FETCH-NUM}}
\newcommand{\marobsize}{\textbf{MAX-ROB}}
\newcommand{\mamaxdecode}{\textbf{MAX-DECODE}}

\newcommand{\instop}{\funcfont{inst-op}}
\newcommand{\uinstop}{\funcfont{minst-op}}

\newcommand\mapp{\ensuremath{\mathbin{+\mkern-6mu+}}}

\makeatletter
\newcommand{\bigmapp}{%
  \DOTSB\mathop{\mathpalette\mattos@bigmapp\relax}\slimits@
}
\newcommand\mattos@bigmapp[2]{%
  \vcenter{\hbox{%
    \sbox\z@{$#1\sum$}%
    \resizebox{!}{0.9\dimexpr\ht\z@+\dp\z@}{\raisebox{\depth}{$\m@th#1\mapp$}}%
  }}%
  \vphantom{\sum}%
}
\makeatother

\newcommand{\sequence}[1]{\ensuremath{\langle #1 \rangle}}
\newcommand{\emptyseq}{\emptyset}

\newcommand{\boolf}{\texttt{false}}
\newcommand{\boolt}{\texttt{true}}
\newcommand{\ops}[1]{\ensuremath{\mathcal{O}_{\textit{#1}}}}
\newcommand{\insts}[1]{\ensuremath{\mathcal{I}_{\textit{#1}}}}
\newcommand{\uops}[1]{\ensuremath{\mathcal{V}_{\textit{#1}}}}
\newcommand{\uinsts}[1]{\ensuremath{\mathcal{U}_{\textit{#1}}}}
\newcommand{\roblines}[1]{\ensuremath{\mathcal{RBL}_{\textit{#1}}}}
\newcommand{\resstations}[1]{\ensuremath{\mathcal{R\mkern-1mu S}_{\textit{#1}}}}

\newcommand{\rsids}{\mathcal{R\mkern-2mu S\mkern-1mu I}}
\newcommand{\robids}{\ensuremath{\mathcal{R\mkern-1mu B}}}
\newcommand{\regstatentries}{\mathcal{R\mkern-1mu G\mkern-1mu S}}

\newcommand{\mahstatuses}{\mathcal{S}}
\newcommand{\mahstatuslines}{\mathcal{SL}}

\newcommand{\mahcommitcyc}{\textbf{comm-cy}}
\newcommand{\mahstartcyc}{\textbf{start-cy}}
\newcommand{\mahcommitcache}{\textbf{comm-cache}}
\newcommand{\mahcacheeffects}{\textbf{ch-eff}}
\newcommand{\mahlines}{\textbf{hist-lines}}

\newcommand{\mahslpc}{\textbf{sl-pc}}
\newcommand{\mahslrob}{\textbf{sl-rob-id}}
\newcommand{\mahsllines}{\textbf{statuses}}

\newcommand{\powerset}[1]{\ensuremath{\mathcal{P}(#1)}}
\DeclareMathOperator{\Ima}{Im}

\begin{document}

\title{Global Microprocessor Correctness in the Presence of Transient Execution}

\author{Andrew T. Walter}
\email{walter.a@northeastern.edu}
\orcid{0000-0002-7588-263X}
\affiliation{%
  \institution{Northeastern University}
  \city{Boston}
  \state{Massachusetts}
  \country{USA}
}

\author{Konstantinos Athanasiou}
\email{kathanas@mathworks.com}
\orcid{0000-0001-8745-2189}
\affiliation{
  \institution{The MathWorks}
  \city{Natick}
  \state{Massachusetts}
  \country{USA}
}
\authornote{Work done while at Northeastern University}
\author{Panagiotis Manolios}
\email{pete@ccs.neu.edu}
\orcid{0000-0003-0519-9699}
\affiliation{%
  \institution{Northeastern University}
  \city{Boston}
  \state{Massachusetts}
  \country{USA}
}



\begin{abstract}
  Correctness for microprocessors is generally understood to be
  conformance with the associated instruction set architecture (ISA).
  This is the basis for one of the most important abstractions in
  computer science, allowing hardware designers to develop
  highly-optimized processors that are functionally ``equivalent'' to
  an ideal processor that executes instructions atomically. This
  specification is almost always informal, \eg commercial
  microprocessors generally do not come with conformance
  specifications.  In this paper, we advocate for the use of formal
  specifications, using the theory of refinement. We introduce notions
  of correctness that can be used to deal with transient execution
  attacks, including Meltdown and Spectre. Such attacks have shown
  that ubiquitous microprocessor optimizations, appearing in numerous
  processors for decades, are inherently buggy. Unlike alternative
  approaches that use non-interference properties, our notion of
  correctness is global, meaning it is single specification that:
  formalizes conformance, includes functional correctness and is
  parameterized by an microarchitecture. We introduce action skipping
  refinement, a new type of refinement and we describe how our notions
  of refinement can be decomposed into properties that are more
  amenable to automated verification using the the concept of
  shared-resource commitment refinement maps. We do this in the
  context of formal, fully executable bit- and cycle-accurate models
  of an ISA and a microprocessor. Finally, we show how light-weight
  formal methods based on property-based testing can be used to
  identify transient execution bugs.
\end{abstract}





\maketitle

\section{Introduction}
\label{sec:intro}
Modern microprocessors are highly optimized systems that employ a
variety of techniques designed to efficiently execute code.  As with
any optimized system, correctness is a fundamental concern, but it is
especially important for microprocessors since they form the base of a
stack of systems that provide powerful abstractions used by all of the
software running on the microprocessors.

The specification of correctness for microprocessors is generally
taken to be conformance to the corresponding instruction set
architecture (ISA).  From Computer Organization and
Design~\cite{comp-org-design}: ``The instruction set architecture
includes anything programmers need to know to make a binary machine
language program work correctly, including instructions, I/O devices,
and so on.''  Conformance is a \emph{global} notion, meaning that it
is a single specification that captures functional correctness. The
ISA defines the hardware-software interface and is widely considered
to be one of the most important abstractions in computer
science. Ideally, it allows hardware designers to develop novel,
powerful techniques that lead to optimized processors which are
functionally ``equivalent'' to the much simpler ISAs which are the
programming models used by software engineers.

Unfortunately, for many modern processors, the hardware-software
abstraction is leaky.
%It has been rendered porous and insecure for
%various reasons, including the pursuit of ever more optimized and
%complicated microprocessor designs.
% This places a significant burden
% on programmers, as transient execution attacks, such as Meltdown and
% Spectre, have made abundantly clear.  In short, modern microprocessors
% can speculatively execute instructions by making assumptions such as
% whether a read instruction has access to the memory it is trying to
% read.  If such assumptions are mostly correct, that makes it possible
% for the microprocessor to efficiently execute programs and when the
% assumptions turn out to be wrong, the microprocessor reverts and
% proceeds as per the ISA semantics.  Unfortunately, speculative
% execution can modify the cache in a way that allows clever programs to
% use performance counters to read all of kernel memory.
To understand why, consider a user space process executing the x86
instructions of Listing~\ref{listing:core-melt}.
%%First we explain the code using x86 ISA semantics.
Instruction~1 loads a byte from the memory address stored in register
\lstinline{ecx} into register \lstinline{eax}. Suppose that
\lstinline{ecx} points to process memory and \lstinline{ebx} points to
an array of bytes. Then instruction~2 moves the \lstinline{eax}-th
element of the array into \lstinline{ebx}.  What if \lstinline{ecx}
points to kernel memory?  Since user processes do not have access to
kernel memory, instruction~1 leads to an exception.

\begin{center}
\begin{lstlisting}[%language={Assembler},
basicstyle=\fontsize{7}{7}\ttfamily,
numbers=left,
firstnumber=1,
xleftmargin=5.0ex,
label={listing:core-melt},
caption={Core Meltdown.}, captionpos=b,]
movsx eax, byte [ecx] ;; ecx: kernel address. |\Suppressnumber|
	              ;; The line below is never executed. |\Reactivatenumber|
mov ebx, [ebx+eax]    ;; Load contents of ebx+eax, an address |\Suppressnumber|
		      ;; that depends on the contents of [ecx]. |\Reactivatenumber|
\end{lstlisting} 
\end{center}

%A modern x86 microprocessor uses a wide range of optimizations to
%achieve high performance.
One optimization present in any modern x86 microprocessor is
\emph{pipelining}, where instruction execution is broken down into
stages to allow processors to fetch and dispatch multiple instructions
at the same time.  To maximize instruction-level parallelism, the goal
is to keep the pipelines as full as possible. However, data and
control flow dependencies will stall pipelines and result in wasted
CPU cycles, hindering performance.  Therefore, modern CPUs execute
instructions ``optimistically'' (i) by making predictions on control
flow information and data dependencies, resulting in \emph{speculative
  execution} and (ii) by executing instructions \emph{out-of-order}
(OoO). The results of instructions executed speculatively or OoO are
not committed until the CPU can determine their validity. In
speculative execution, if the predictions are correct, the
speculatively executed instructions are committed and processor
execution continues.
%% performance due to high instruction parallelism.
Otherwise, the instructions are squashed and execution resumes from
the point where the prediction was made, this time adhering to the ISA
semantics. In OoO execution, instructions are executed as
soon as their data dependencies are resolved ensuring no CPU cycles go
to waste, but are usually committed \emph{in-order} to ensure
correctness of the output.

So, let us now consider how a microprocessor with speculative
execution might execute Listing~\ref{listing:core-melt}.
Instruction~1 is executed in stages.  First, we fetch a byte from the
memory address stored in register \lstinline{ecx}. We do not yet
update \lstinline{eax} because we need to check that the process has
permission to access the memory, but since exceptions are the
exception, the processor speculatively executes instruction~2, while
in parallel checking permissions. So, the processor will compute
\lstinline{ebx+eax} and will fetch the contents of this memory
location.  All of these intermediate results are stored internally and
are not committed until permission checking succeeds.  If
\lstinline{eax} points to process memory, once these checks complete,
the instructions commit, the registers are updated and the processor
continues executing subsequent instructions. In the case where
\lstinline{ecx} points to kernel data, the checks fail, none of the
instructions commit and no secret data is moved into \lstinline{eax}
or \lstinline{ebx}; instead an exception occurs.  Notice that
speculation allows the processor to execute instructions
optimistically. In the common case, where no exceptions occur, this
leads to significant performance improvements and in the exceptional
case, no instructions commit, so the processor conforms to the ISA
semantics.

A large class of security
vulnerabilities~\cite{Canella2018ASE,Lipp2018meltdown,Kocher2018spectre,CVE-2018-3639,schwarz2019zombieload,van2020lvi,canella2019fallout,foreshadow}
has shown that the side-effects of instructions executed
optimistically can be exploited by means of \emph{covert
channels}. While executing speculatively, instruction~2
performs a memory load which alters the microarchitectural state of
the processor by bringing the secret-dependent address
\texttt{ebx+eax} in the cache.  A different user process can now
launch a cache attack and deduce the secret byte stored in kernel
address \texttt{[ecx]} by determining the amount of time the load
instructions \texttt{mov ebx, [ebx + $v$]} take, for all possible
values $v \in[0,255]$.  If none of these locations were in the cache
before the attack was launched, then only one of the locations will be
in the cache after the attack, so the value $v_{min}$ whose load
required the least amount of time corresponds to the secret kernel
data.

The described attack, named Meltdown \cite{Lipp2018meltdown}, was
viable, when it was discovered, on most operating systems running on a
CPU that implements OoO execution. This attack can easily
read kernel data at rates of about 500 KB/s. As operating systems
at the time commonly mapped physical memory, kernel processes, and
other running user space processes into the address space of every
process, Meltdown effectively broke any form of process isolation.

Meltdown exploits a delay in handling unprivileged memory reads and
occurs during instructions that execute \emph{transiently} after the
unprivileged read. Meltdown is an example of a \emph{transient
  execution attack} (TEA) which exploits instructions that are
executed optimistically by the microprocessor, based on some
prediction, and are eventually discarded.

Spectre~\cite{Kocher2018spectre} is another TEA which exploits
instructions executing transiently after a branch prediction. %Broadly,
Using Spectre, an attacker can trick the microprocessor's branch
predictor in a way that forces a victim process to reveal information
it did not intend to.

\begin{center}
\begin{lstlisting}[%language={Assembler},
basicstyle=\fontsize{7}{7}\ttfamily,
numbers=left,
firstnumber=1,
xleftmargin=5.0ex,
label={listing:spectrev1},
caption={Spectre C code.}, captionpos=b,]
if (x < array1_size)
  y = array2[array1[x]];
\end{lstlisting} 
\end{center}

Listing \ref{listing:spectrev1} shows the core C code of a victim
process that is vulnerable to the Spectre attack. Assuming that
\lstinline{x} is a program input, the victim process checks in Line 1
that the input is appropriate for indexing \lstinline|array1|, \ie
that it is within the array's bounds. If so, on Line 2 the victim
process uses \lstinline{x} to index \lstinline{array1} and then uses
the contents of location \lstinline{array1[x]} to index
\lstinline{array2}. Otherwise, if \lstinline{x} is outside the bounds
of \lstinline{array1}, Line 2 is not executed.

In Spectre, an attacker exploits a microprocessor's branch predictor and its
speculative execution capabilities, in order to trick the process into
performing out-of-bounds array reads. Initially, the attacker provides multiple
inputs \lstinline{x} which are less than \lstinline{array1_size} in order to
train the branch predictor of the microprocessor to take the true branch of
the conditional statement. After this training phase, the attacker provides
an out of bounds value \lstinline{x'} to the victim process. Due to
the training, the microprocessor will take the branch, reading
transiently from the out-of-bounds locations \lstinline{array1[x']}, and using
its contents to index \lstinline{array2}. The microprocessor will eventually
identify the misprediction and discard the results of instructions executed
transiently. However, as in Meltdown, the transient execution of Line 2 will
bring the contents of the memory location \lstinline{array2[array1[x']]} into the
cache. An attacker that knows the memory address of \lstinline{array2} can
then utilize a cache attack to reverse engineer the contents of the victim
process at memory location \lstinline{array1[x']}.

The simple variant of Spectre we described above demonstrates how
standard microprocessor optimizations open up possibilities for victim
processes to inadvertently leak information. Implementations of such
attacks that deal with their practicalities are accounted for in
detail in the original work on Spectre \cite{Kocher2018spectre},
where, \eg a website can read private data of the browser process it
executes its JavaScript code in. Followup research
\cite{mambretti2020bypassing} has also demonstrated how transient
execution attacks can be used to bypass standard memory protection
mechanisms like control flow integrity or language-based memory safety
mechanisms.

In this paper, we advocate for a research program whose goals are to
provide \textbf{global} formal specifications using refinement.  A
global notion of correctness is a single specification that includes
functional correctness and is independent of the details of
microarchitectural implementation, enabling the development of
software independently of those details.  We take a step in the
direction of such a program by considering the question of global
formal specifications in the context of TEAs that exploit the cache as
a side channel.  As we expand upon in our discussion of related work,
we believe that a global notion of correctness is ideal, in constrast
to non-interference-based approaches that factor out functional
correctness. We cannot imagine a situation where a user desires TEA
security but not functional correctness and as our discussion will
highlight, approaches capable of verifying that hardware satisfies
non-interference properties also depend on the functional correctness
of the hardware!%, so it will need to be proven in any case.

Formal specifications are required if we are to provide a viable
hardware-software interface.  There is no way to really do this unless
we agree on the specification, \ie we agree on exactly what it means
for a processor to correctly implement an ISA. Our approach is the
first notion, to our knowledge, which can be used to show conformance
between ISAs and microprocessor models that accounts for optimizations
such as pipelining, OoO execution, prefetching, superscalar execution
and caching.  A microprocessor that allows TEAs that exploit caches as
covert channels to exfiltrate secret information will not satisfy our
notion of correctness.
%Our notion of refinement is presented in the context of
%reactive systems, which is general and appropriate, given that
%microprocessors are used for networking and operating systems.
To fully carry out this research program will require significant
effort to establish consensus and acceptance of the observer models,
abstractions and techniques needed to handle the capabilities of
modern processors.

We introduce two notions of correctness in the paper. The first,
described in Section~\ref{sec:meltdown-correctness}, is capable of
disallowing Meltdown-type vulnerabilities that leverage cache side
channels. This notion of correctness is based on \emph{witness
  refinement}, a novel variant of skipping refinement, as well as the
\emph{in-cache} abstraction, a novel approach that we use to model
cache covert channels and simplify reasoning about performance
counters. Like skipping refinement, witness refinement (and therefore
our notion of correctness) covers both safety and liveness, enabling
us to show that running any program, terminating or not, leads to a
conforming run on the ISA side. In Section~\ref{sec:meltdown-decomp}
we describe how our notion of correctness for Meltdown can be soundly
decomposed into simpler properties that are more amenable to automated
verification. We also introduce the novel ideas of \good\ states and
\emph{shared-resource commitment refinement maps}, which we combine to
eliminate unreachable counterexamples in an
automated-verification-friendly
way. Section~\ref{sec:spectre-correctness} contains a presentation of
our notion of correctness for Spectre, introducing \emph{intent
  models} that use \emph{virtual instructions} to define highly
non-deterministic ISA semantics that allow multiple microprocessor
implementations with arbitrary prefetching and eviction policies. This
section also covers a novel kind of refinement, \emph{action witness
  skipping refinement}, that our notion of correctness for Spectre is
based on. We subsequently describe how we decompose our notion of
correctness for Spectre in Section~\ref{sec:spectre-decomp}, using
similar methods as were used for Meltdown. To demonstrate our notions
of correctness and their effectiveness, we define minimal formal
models of an ISA and a microprocessor that are fully executable as
well as bit- and cycle-accurate using the ACL2s theorem prover.  These
are available as artifacts~\cite{artifact-repo} and are discussed in
Section~\ref{sec:formal-models}. We describe how we evaluated our
notions of correctness with lightweight verification techniques in
Section~\ref{sec:lightweight-verification}, discuss related work in
Section~\ref{sec:related-attacks} and conclude in
Section~\ref{sec:conclusions}.

\section{Meltdown Correctness}
\label{sec:meltdown-correctness}
Our notion of correctness for Meltdown is based on \emph{witness
refinement}, a novel variant of skipping refinement. In this section,
we begin with a description of transition systems, which are used
to formalize ISAs %(Instruction Set Architectures)
and MAs (Micro
Architectures). We then define skipping refinement, introduce the
\incache\ abstraction, formalize correctness and present witness
refinement, which facilitates automated verification.

\begin{definition}[Labeled Transition System]
  A labeled transition system (TS) is a structure \abracket{S, \trans,
    L}, where $S$ is a non-empty (possibly infinite) set of states,
  $\trans \subseteq S \times S$, is a left-total transition relation
  (every state has a successor), and $L$ is a labeling function with
  domain $S$.
\end{definition}

Function application is sometimes denoted by an infix dot ``$.$'' and
is left-associative.  The composition of relation $R$ with itself $i$
times (for $0 < i \leq \omega$) is denoted $R^i$ ($\omega = \nats$ and is
the first infinite ordinal).  Given a relation $R$ and $1<k\leq
\omega$, $R^{<k}$ denotes $\bigcup_{1 \leq i < k} R^i$ and $R^{\geq
  k}$ denotes $\bigcup_{\omega > i \geq k} R^i$ .  Instead of $R^{<
  \omega}$ we often write the more common $R^+$.  $\uplus$ denotes the
disjoint union operator.  Quantified expressions are written as
$\abracket{\emph{Q}x \from r \from t }$, where \emph{Q} is the
quantifier (\eg $\exists, \forall, \mathit{min}, \bigcup$), $x$ is a
bound variable, $r$ is an expression that denotes the range of
variable $x$ (\emph{true}, if omitted), and $t$ is a term.

Let \trs{} be a TS. An \M{}-path is a sequence of
states such that for adjacent states, $s$ and $u$, $s \trans u$. The
$j^{th}$ state in an \M{}-path $\sigma$ is denoted by $\sigma.j$. An
\M{}-path $\sigma$ starting at state $s$ is a \emph{fullpath}, denoted
by $\fp.\sigma.s$, if it is infinite. An \M{}-segment,
$\abracket{v_1, \ldots, v_k}$, where $k \geq 1$ is a finite \M{}-path
and is also denoted by $\vv{v}$. The length of an \M{}-segment
$\vv{v}$ is denoted by $|\vv{v}|$. Let $\inc$ be the set of strictly
increasing sequences of natural numbers starting at 0. The $i^{th}$
partition of a fullpath $\sigma$ with respect to $\pi \in \inc$,
denoted by $\partition{\sigma}{\pi}{i}$, is given by an \M{}-segment
$\abracket{\sigma(\pi.i), \ldots, \sigma(\pi(i + 1) - 1)}$.

% Here we describe a notion of correctness that is capable of
% disallowing Meltdown behavior while allowing flexibility in the
% behavior of the MA's cache. We do this by considering ISA and MA
% implementations with an additional instruction, the \incache\ abstract
% instruction. We will first discuss the \incache\ abstract instruction
% in more detail before discussing relevant behavior of our ISA and MA
% models and concluding with our statement of correctness.

\subsection{Skipping Refinement}
\label{sec:skipping-refinement}
We now define skipping simulation refinement, the
weakest notion of refinement we use in this paper.
This definition was introduced by Jain \ea~\cite{jain2015skipping}
% , the
% weakest notion of refinement we use in this paper. This definition was introduced by Jain \ea~\cite{jain2015skipping}.
and uses the notion of
\emph{matching}. Informally, a fullpath $\sigma$ matches a fullpath
$\delta$ under the relation $B$ iff the fullpaths can be partitioned
in to non-empty, finite segments such that all elements in a segment
of $\sigma$ are related to the first element in the corresponding
segment of $\delta$.
\begin{definition}[smatch]
  \label{def:smatch}
  Let \trs{} be a TS, $\sigma,\delta$ be fullpaths in
  \M{}. For $\pi, \xi \in \inc$ and binary relation
  $B \subseteq S \times S$, we define
  \begin{flalign*}
    &\scorr{B}{\sigma}{\pi}{\delta}{\xi} \equiv \La \forall i \in
    \omega:: \La \forall s \in \partition{\sigma}{\pi}{i} ::
    s B \delta(\xi.i) \Ra \Ra \textit{ and } &\\  
    &\smatch{B}{\sigma}{\delta} \equiv \La \exists \pi, \xi \in
    \inc :: \scorr{B}{\sigma}{\pi}{\delta}{\xi}\Ra.
  \end{flalign*}
\end{definition}

%% Using the notion of matching, skipping simulation is defined as
%% follows. Notice that skipping simulation is defined using a single
%% transition system. We will lift this notion to one that relates two
%% transition systems shortly.

\begin{definition}[Skipping Simulation (SKS)]
  \label{def:sks}
  $B \subseteq S \times S$ is a skipping simulation on a TS \trs{}
  \myiff{} for all $s,w$ such that $sBw$, both of the following hold.
  % \begin{enumerate}[(SKS1)]
  \begin{enumerate}
  \item $L.s = L.w$
  \item\abracket{\forall \sigma\from \fp.\sigma.s \from \abracket{
        \exists \delta \from \fp.\delta.w \from
        \smatch{B}{\sigma}{\delta}}}
  \end{enumerate}
\end{definition}

We use skipping simulation, a notion defined in terms of a single
TS, to define skipping refinement, a notion that
relates \emph{two} TSes: an \emph{abstract} transition
system (\eg an ISA) and a \emph{concrete} TS (\eg an
MA). Informally, if a concrete system is a skipping refinement of an
abstract system, then its observable behaviors are also behaviors of
the abstract system, modulo skipping. Skipping allows an MA to
stutter, \ie to take steps that do not change ISA-visible components,
as happens when loading the pipeline. Skipping also allows the MA to
commit multiple ISA instructions at once, which is possible due to
superscaling.
% (which includes stuttering).
The notion is parameterized by a \emph{refinement map}, a function
that maps concrete states to their corresponding abstract states. A
refinement map along with a labeling function determines what is
observable at a concrete state.

\begin{definition}[Skipping Refinement]
  \label{def:skipref} %\\
  Consider TSs \trs{A} and \trs{C} and let
  $\mathit {r : S_C \ra S_A}$ be a refinement map.  We say $\M{C}$ is
  a \textit{skipping refinement} of $\M{A}$ with respect to $r$, written $\M{C} \lesssim_r \M{A}$,
  if there exists a binary relation
  $B$ such that all of the following hold.
  \begin{enumerate}
  \item $\La \forall s \in S_C :: sBr.s\Ra $ \emph{and}
  \item $B$ is an SKS on \disjtrs{C}{A}{} where $\labf.s = L_A(s)$ for
    $s \in S_A$, and $\labf.s = L_A(r.s)$ for $s \in S_C$.
  \end{enumerate}
\end{definition}

% In our models, we consider a variant of Local Well-founded Skipping
% refinement that does not include rank functions, as these functions
% are used to show that stuttering and skipping steps are bounded, and
% this is not needed to derive counterexamples to transient execution
% attacks.
% We also use stronger notions such as Bisimulation refinement
% and Stuttering refinement. These notions are composable, \eg the
% composition of Skipping and Stuttering refinement results in a
% Skipping refinement (the general rule is we wind up with the weaker
% notion).

\subsection{In-Cache Abstract Instruction}
\label{sec:in-cache}
Caches are used to improve processor performance by providing
\emph{fast} access to data that is frequently used. Instruction set
architectures leave caches partially unspecified to allow the
implementer of processors to choose cache configurations most suitable
to their requirements. As illustrated in Section~\ref{sec:intro}, the
memory caches are instrumental in TEAs where
they are used to extract secret data from MA states obtained after
incorrect speculation, using performance counters to determine what
addresses are cached.

In order to reason about such attacks we introduce the \emph{in-cache
abstraction}: we include an \incache\ abstract instruction which given
an address, $a$, will return \True\ if $a$ is in the cache and
\False\ otherwise. We want the ISA to allow all reasonable behaviors
so that we have a single specification that can be used to reason
about any MA machine. Therefore, our \incache{} instruction is
nondeterministic.  Let $A$ be the set of addresses of the ISA and let
\lmem{} be the set of addresses to which an ISA program under
consideration has read and write access, \ie all addresses for which
reads and writes do not generate errors.  At the MA, this instruction
just checks to see if some address is in the MA's cache. But, at the
ISA, this instruction non-deterministically returns a Boolean, subject
only to the following constraint, where $a$ is an address and $s$ is
an ISA state.
\begin{equation}
 a \not \in \lmem \Rightarrow \incache(a,s) = \False \label{eq:in-cache}
 \end{equation}
With the \incache{} instruction, the ISA includes behaviors in which
any subset of addresses that the ISA program can access are in the
cache at any program point.  Thus, the ISA allows all possible correct
cache implementations and prefetching strategies in the MA.  The
abstraction imposes no restrictions on how the memory cache component
will be defined, \eg it does not constrain the size of the cache, its
replacement policy, etc.
%% Correct cache implementations should not
%% include addresses that the ISA program cannot access, thus the
%% \incache{} instruction returns \False\ in such cases.  Put in another
%% way, the \incache\ abstract instruction provides an overapproximation
%% of the addresses in the cache.  %As discussed above, With the
%% \incache\ abstraction, we can reason about the correctness of an MA in
%% the context of Meltdown-type attacks.

%% On the MA, the \incache{} instruction acts as though it is preceded by
%% and followed by memory barriers. This is to say, the \incache{}
%% instruction cannot start execution while there are memory instructions
%% in-flight, and no memory instructions can start execution while an
%% \incache{} instruction is in-flight. While this limits the kinds of
%% cache membership queries that can be modeled by the \incache{}
%% instruction, it covers the common approaches for performing cache
%% membership queries in the context of Meltdown attacks.

\subsection{Statement of Correctness}
\label{sec:meltdown-correctness-statement}

Let \trs{\textit{ISA}} and \trs{\textit{MA}} be TSes
modeling an ISA and an MA that both support the
\incache\ instruction. Let
$r : S_{\textit{MA}} \rightarrow S_{\textit{ISA}}$ be a refinement
map. At a high level, we expect $r$ to map an MA state to an ISA state
that agrees in the programmer-visible values of its components---\eg\
the register file of the mapped ISA state should correspond to the
committed register file for the MA. Then, we say that \M{\textit{MA}}
is a correct implementation of \M{\textit{ISA}} with respect to
Meltdown iff \M{\textit{MA}} is a skipping refinement of
\M{\textit{ISA}} with respect to $r$.

\subsection{Witness Refinement}
\label{sec:witness-refinement}

The definition of skipping refinement (Definition~\ref{def:skipref})
is not amenable to mechanized verification, as it requires reasoning
about infinite traces. We can drastically simplify the proofs by
specializing to certain kinds of
TSes. %using characteristics of MAs and ISAs.
We do this by computing the number of stuttering and skipping steps
needed to apply to one of the systems to match a single step of the
other. MAs have bounds for both of these: any MA has a finite
collection of resources that it can use, thereby bounding the number
of instructions that may be committed in a single step. Similarly, any
MA has a pipeline with a finite number of stages, and each instruction
takes a finite number of cycles to execute. This means that the number
of steps that an MA must take before it commits at least one
instruction is also bounded.  Functions that compute these values
essentially act as Skolem functions, eliminating the need to solve
existential quantifiers when proving that a relation is a witness
skipping relation.  In the context of hardware verification, this is
important for automating proofs, since the search needed to resolve an
existential will be dramatically harder for hardware verification
techniques.
%We also assume the existence of a \funcfont{run}
%function that 

% Instead, for the purposes of formal
% verification, we use a sound characterization that only requires
% reasoning about states and their predecessors within a bounded number
% of steps. This characterization is derived from the notion of reduced
% well-founded skipping introduced by Jain and
% Manolios~\cite{jain2015skipping}.

\begin{definition} [Witness Skipping]
  \label{def:witness-sk}
  $B \subseteq S \times S $ is a witness skipping relation on TS
  \trs{} with respect to functions
  $\funcfont{stutter-wit} : S \times S \rightarrow \nats $,
  $\funcfont{skip-wit} : S \times S \rightarrow \nats \setminus \{ 0
  \}$ and $\funcfont{run} : S \times S \times S \rightarrow S$ iff:
  \vspace{-\topsep} {\setlength{\abovedisplayskip}{1pt}
    \begin{enumerate}
    \item [(WSK1)]  $\abracket{\forall w, s, u \from sBw \wedge s \trans u \from w \trans^{\funcfont{skip-wit}(s, u)} \funcfont{run}(w, s, u)}$
    \item [(WSK2)] $\La \forall s,w \in S : s B w : L.s =
      L.w\Ra$
    \item [(WSK3)] 
      \begin{flalign*}
        &\forall s,u,w \in S: sBw \And s \trans u: &\\
        &\text{\quad (1) } (uBw \And \funcfont{stutter-wit}(u, w) < \funcfont{stutter-wit}(s, w))\ \vee&\\
        &\text{\quad (2) } uB(\funcfont{run}(w, s, u))&
      \end{flalign*}
    \end{enumerate}
  }
\end{definition}

In the above definition,
$w \trans^{\funcfont{skip-wit}(s, u)} \funcfont{run}(w, s, u)$
indicates that there is a path of length
$\funcfont{skip-wit}(s, u)$ from $w$ to $\funcfont{run}(w, s, u)$
in $\rightarrow$, the transition relation of TS.
This means that the function $\funcfont{run}$ runs TS for 
$\funcfont{skip-wit}(s, u)$ steps. If TS is deterministic, then
the path is uniquely defined.
\begin{theorem}
  \label{thm:witness-skipping-sound}
  (Soundness) If $B$ is a witness skipping relation on TS
  \trs{}, then it is an SKS on \M{}.
\end{theorem}
\noindent \emph{Proof Sketch}.  Let $B$ be a witness skipping relation
on TS \trs{} with respect to the $\funcfont{stutter-wit}$,
$\funcfont{skip-wit}$ and $\funcfont{run}$ functions. We show that $B$
is a skipping simulation on \M{}. Condition 1 of
Definition~\ref{def:sks} holds due to WSK2.  To satisfy condition 2
of Definition~\ref{def:sks}, we construct the matching partitions
using WSK1 and WSK3 as follows. Let $s, w$ correspond to the initial
states of a partition. If $\smatch{B}{\sigma}{\delta}$ holds, then
$\pi, \xi$ can be chosen so for every corresponding partition, at
least one of the partitions consists of exactly one state. There are
now three cases to consider. If both partitions include a single
state, then $\funcfont{run}(w, s, u)$ runs $w$ for one step and
WSK3(2) holds.  If the partition starting at $s$ consists of multiple
states but the partition from $w$ consists of one state, then $u$ has
to be related to $w$ and $\funcfont{stutter-wit}(u, w) <
\funcfont{stutter-wit}(s, w)$, which means we can only make this move
a finite number of times. Finally, if the partition starting from $s$
has one state but the partition from $w$ has multiple states, this
case is covered by WSK3(2), which requires that $u$ is related to a
successor of $w$. $\square$

% It is
% straightforward to use these functions to show that
%  there exist appropriate rank functions f\cite{jain2015skipping}

% Jain and Manolios showed that if $B$ is a reduced well-founded
% skipping simulation on \M{} then $B$ is also an SKS on \M{}. So, we
% only need to show that if $B$ is a witness skipping relation on \M{}
% then it also is a reduced well-founded skipping simulation on \M{}.

Using the above soundness result, we can prove Skipping Refinement
using witness skipping instead of Skipping Simulation.  The advantage
is that designers can provide definitions for
$\funcfont{stutter-wit}$, $\funcfont{skip-wit}$ and $\funcfont{run}$,
which leads to verification obligations that are over finite steps of
the TSes and are therefore amenable to automated verification.

Notice that our notion of correctness is global: it is a single
specification that formalizes conformance, includes functional
correctness, is essentially independent of the MA, and can be used to
analyze any MA. This means that it can be used by architects to show
conformance of a MA design while also providing an abstraction based
on the ISA that allows programmers to reason about code that runs on
any conforming MA, without needing to reason about the MA.

\section{Formal Models}
\label{sec:formal-models}
To evaluate our notions of correctness, we developed models of an ISA
and MA that are vulnerable to both Meltdown and Spectre.
% We applied our notion of correctness to models of an ISA and MA.
The models were designed to be complex enough to exhibit TEAs and
express interesting programs. The models are
both executable and formal and are defined using the ACL2s theorem
prover. The ISA is x86-like, except that it features general purpose
registers and uses the Harvard memory model by having disjoint data
and instruction memories. Memory addresses are 32-bits in length, and
there is a basic form of memory protection: a subset of the address
space is taken to be ``kernel memory'' that is not accessible to the
running program (attempting to access it will result in an
exception). The MA model follows a textbook definition of a four-stage
pipeline, multi-issue, OoO microprocessor with exception
handling \cite{hennessy2011computer}, branch prediction, microcode,
and memory prefetching. OoO execution is implemented using
Tomasulo's algorithm \cite{tomasulo1967efficient} with a reorder
buffer (ROB). The MA contains a set of reservation stations (RSes)
that handle execution of most
instructions. Equation~\ref{eqn:ic-insts} provides a listing of the
instructions supported by our models, where $r_d$, $r_1$ and $r_2$
represent register operands and $c$ represents an immediate operand.
\begin{flalign}
  \label{eqn:ic-insts}
  \insts{IC} ::=\ &\texttt{halt}\ \vert\ \texttt{noop}\ \vert\ \texttt{loadi}\ r_d\ c\ \vert\ \texttt{addi}\ r_d\ r_1\ c\ \vert\
                                   \texttt{add}\ r_d\ r_1\ r_2\ \vert \notag\\
                                 & \texttt{mul}\ r_d\ r_1\ r_2\ \vert\ \texttt{and}\ r_d\ r_1\ r_2\ \vert\ \texttt{cmp}\ r_d\ r_1\ r_2\ \vert\ \texttt{jg}\ r_1\ c\ \vert\ \texttt{jge}\ r_1\ c\ \vert \notag\\
                  &\texttt{ldri}\ r_d\ r_1\ c\ \vert\ \texttt{ldr}\ r_d\ r_1\ r_2\ \vert\
                                   \texttt{tsx-start}\ c\ \vert\ \texttt{tsx-end}\ \vert \notag\\
                  &\texttt{in-cache}\ r_d \ r_1 \ c
\end{flalign}
A full description of the models accounting for complexities like ROBs
and RSes is quite involved, taking over 15 pages and is provided in
Appendix~\ref{sec:semantics}. Here, we provide an overview of how the
two models work by presenting a \emph{transition rule} for each that
describes part of its behavior. We begin with the ISA, which is a
TS \trs{\ISAICM}. Members of the set of states
$S_{\ISAICM}$ are structures containing several fields. The fields
relevant to the transition rule we will show are $\pc$ the program
counter, $\rfv$ a mapping from registers to data (the register file),
$\halted$ a bit indicating whether or not the ISA is halted, $\imem$ a
mapping from addresses to instructions and $\tsx$ a TSX structure. A
TSX structure consists of three fields, $\tsxactive$ which indicates
whether the ISA is inside a TSX region, $\tsxrf$ which is the register
file to restore upon a TSX error and $\tsxfallback$ which is the
address to jump to upon a TSX error. %TSX regions will be explained in
more detail shortly.

The behavior of \M{\ISAICM} is described using two auxiliary
TSes. This is done since the deterministic part of the behavior of
\M{\ISAICM} is shared with another TS, \M{\ISAICAM}, which is used in
the notion of correctness for Spectre.
The TS handling the deterministic behavior is
\trs{\ISAICM\texttt{-ISA}} where
$S_{\ISAICM\texttt{-ISA}} =
S_{\ISAICM}$. Equation~\ref{eqn:tsx-start-trans} shows one of the
transition rules for \M{\ISAICM\texttt{-ISA}} relating to TSX
instructions. A transition rule consists of a set of premises (written
above a horizontal line) and a conclusion (written below).  If the
conjunction of the premises hold, the conclusion must also hold. That
is, Equation~\ref{eqn:tsx-start-trans} indicates that if
$\fetchfunc{\imem}{\pc} = \texttt{tsx-start}\ c$ and $\neg\halted$
hold with respect to some $S \in S_{\ISAICM\texttt{-ISA}}$, it must be
the case that $S$ transitions to the state indicated by the right-hand
side of the conclusion. In transition rules, we freely use the names
of fields to refer to the value of a field in the state corresponding
to the left-hand side of the transition rule. For example, $\halted$
in the second premise refers to the value of the $\halted$ field of
$S$.
$[\pc \mapsto \pc \plusc 1, \tsx \mapsto \abracket{\boolt, \rfv, c}]S$
represents a structure where the $\pc$ field is equal to the $\pc$
field of $S$ plus one, the $\tsx$ field is equal to
$\abracket{\boolt, \rfv, c}$ where $c$ is constrained by the first
premise of the transition rule and all other fields have the same
value as in $S$. Equation~\ref{eqn:tsx-start-trans} uses the function
$\fetchfunc{\imem}{\pc}$, which gets the instruction that $\pc$ maps
to in $\imem$, or \texttt{noop} if $\pc$ is not mapped.
\begin{equation}
  \label{eqn:tsx-start-trans}
  \inferrule[tsx-start]{\fetchfunc{\imem}{\pc} = \texttt{tsx-start}\ c \\ \neg \halted}{S \xtrans{\ISAICM\texttt{-ISA}} [\pc \mapsto \pc \plusc 1, \tsx \mapsto \abracket{\boolt, \rfv, c}]S}
\end{equation}
Equation~\ref{eqn:tsx-start-trans} describes how the
\texttt{tsx-start} instruction modifies the TSX state. These
instructions are intended to model the behavior of the transactional
region instructions provided by the TSX x86 ISA
extension~\cite{inteloptimize}. These instructions are used in the
original Meltdown exploit~\cite{Lipp2018meltdown} as an optimization
to suppress exceptions caused by attempted reads of kernel memory. In
short, these instructions allow one to specify a temporary exception
handler over a region of code. A region starts when a
\texttt{tsx-start} instruction is executed and ends when a
\texttt{tsx-end} instruction is executed. If an exception is raised
when executing an instruction inside of a TSX region, the
ISA will undo any modifications to memory and the register file that
were made inside the region and jump to the TSX fallback address
provided in the \texttt{tsx-start} instruction that started the
region.  The problem that Meltdown exploits is that modifications to
the cache are not undone and therefore one can use the cache as a side
channel to exfiltrate data from speculatively executed instructions
inside of a TSX region.

We now discuss the TS for the MA, \trs{\MAICM}. Like
with \M{\ISAICM}, members of the set of states $S_{\MAICM}$ are
structures containing several fields. These structures contain all of
the fields that the members of $S_{\ISAICM}$ have, plus the following
that are relevant for the transition rule we will discuss: $\cyclec$ a
cycle counter, $\rob$ a sequence of ROB lines and $\rsf$ a sequence of
RSes. ROB lines are structures that track the progress of
\emph{microinstructions} as the are issued and executed. Each
instruction in $\insts{IC}$ is turned into one or two
microinstructions when it is issued by \M{\MAICM}, corresponding to
the atomic actions that must be taken to complete execution of the
instruction. ROB lines must keep track of several pieces of
information, the most critical being $\robrdy$ which indicates whether
the ROB line is ready to be committed, $\robid$ which is an identifier
for each ROB line, $\robval$ which stores the result of the execution
of the microinstruction in the ROB line and $\robexcp$ which is a bit
indicating whether executing the microinstruction resulted in an
exception. A RS is also a structure, with fields including $\rscpc$
which denotes the cycle on which the result of the microinstruction
execution will be ready, $\rsbusy$ and $\rsexec$ indicating whether
the RS has an instruction loaded and is currently executing a
microinstruction respectively and $\rsdst$ storing the identifier
for the ROB line that should hold the result of execution.

The behavior of \M{\MAICM} is described using a number of auxiliary
TSes. There is a TS for each component of the \M{\MAICM} state, each
of which is made up of between two and four TSes that roughly speaking
each handle one stage of the pipeline. This organization helps reduce
the complexity of the transition rules. We show a transition rule from
\trs{\mrbw{\MAICM}},\\ which handles part of the behavior of the reorder
buffer, in particular the part that handles RSes that complete the
execution of their associated microinstruction. $S_{\mrbw{\MAICM}}$ is
the product of $S_{\MAICM}$ with sequences of RSes. In essence,
\M{\mrbw{\MAICM}} starts off with the the sequence of RSes in $S$ and
iterates through them, updating the ROB as needed. The premise
$Q = \textit{rs} \bullet Q^\prime$ indicates that the sequence of RSes
is not empty and that we refer to the first element in the sequence as
$\textit{rs}$ and the remainder of the sequence as $Q^\prime$. We
refer to the value of a field of a structure stored in a variable
using a subscript on the field name. For example,
$ \rscpc_\textit{rs}$ indicates the value of the $\rscpc$ field of the
RS structure referred to by $\textit{rs}$.

\begin{equation}
  \label{eqn:rob-wrb-rdy}
  \inferrule[rob-wrb-rdy]{Q = \textit{rs} \bullet Q^\prime \\ \cyclec = \rscpc_\textit{rs} \\ \rsbusy_\textit{rs} \\ \rsexec_\textit{rs} \\ \abracket{\exists i \from i \in \mathbb{N} \from \robid_{\rob(i)} = \rsdst_\textit{rs}} \\ \text{Let } i = \min_{j \in \mathbb{N} \wedge \robid_{\rob(j)} = \rsdst_\textit{rs}} j \\ \neg\halted}{\abracket{S, Q} \xtrans{\mrbw{\MAICM}} \La[\rob \mapsto [i \mapsto [\robval \mapsto \funcfont{comp-val}(rs, S),\\ \robexcp \mapsto \funcfont{comp-exc}(\textit{rs})]\rob(i)]\rob]S, Q^\prime\Ra}
\end{equation}

Equation~\ref{eqn:rob-wrb-rdy} describes how the ROB is updated when a
RS becomes ready. Its definition hinges on two functions:
$\funcfont{comp-val}$, which uses the source operand values in the
RS to compute the result of the RS's microoperation, and
$\funcfont{comp-exc}$ which determines if the microoperation
should result in an exception instead. The appropriate ROB entry is
updated with the result of these two functions. We expect that
$\robid$ values will be unique among all ROB lines and that whenever
an RS is executing, its $\rsdst$ corresponds to a ROB line
present in the ROB. These assumptions are relevant to the idea of
\good\ states in Section~\ref{sec:entangled-states}.
%that will be discussed next.

% In addition to providing executable ACL2s versions of \M{\ISAICM} and
% \M{\MAICM} as well as their variants, we describe their behavior
% mathematically using inference rules.
% %We will provide a short overview
% %of some of the relevant behaviors of the aforementioned models, but
% A
% full listing of the definitions of all of the model variants and their
% inference rules is provided in Appendix~\ref{sec:semantics}.
%A summary
%of the notation used follows.

\section{Meltdown Decomposition Proof}
\label{sec:meltdown-decomp}

% We defined both an ISA model and an MA model implementing the
% \incache\ instruction, \M{\ISAICM} and \M{\MAICM}
% respectively.
% %\M{\ISAICM} is described in Section~\ref{sec:isa-ic-m}
% %and \M{\MAICM} is described in Section~\ref{sec:ma-ic-m}.
% The full set of transition rules for both systems can be found in
% Appendix~\ref{sec:semantics}.
% We now will discuss how we decompose our notion of correctness for
% Meltdown.  Recall that one of the enablers for our notion of
% correctness is the idea of shared-resource commitment refinement
% maps. We first will explain how this works.

To decompose the refinement property corresponding to our notion of
correctness for Meltdown into several simpler properties, we will
introduce several variants of the \M{\MAICM} and \M{\ISAICM}
TSes and take advantage of the following important
algebraic property of refinement.
\begin{theorem}
  \label{thm:compositionality}
  Consider TSs \trs{A}, \trs{B} and \trs{C} and refinement maps
  $\mathit {p : S_A \ra S_B}$ and \\ $\mathit {q : S_B \ra S_C}$ such
  that $\M{A}$ is an $\alpha$-\textit{refinement} of $\M{B}$ with
  respect to $p$ and $\M{B}$ is a $\beta$-\textit{refinement} of
  $\M{C}$ with respect to $q$.  We allow $\alpha$ and $\beta$ to be
  one of: bisimulation, simulation or skipping. Skipping is the
  weakest notion, followed by simulation, followed by bisimulation,
  \eg a bisimulation refinement is both a simulation refinement and a
  skipping refinement. Now $\gamma$ is the weakest of $\alpha, \beta$.
  Then $\M{A}$ is a $\gamma$-\textit{refinement} of $\M{C}$ with
  respect to the composition of $p$ and $q$.
\end{theorem}
\noindent
\emph{Proof sketch}. Without loss of generality, assume that $\beta$
is weaker than $\alpha$. Then $\M{A}$ is also a
$\beta$-\textit{refinement} of $\M{B}$ with respect to $p$. Since we
already have that $\M{B}$ is a $\beta$-\textit{refinement} of $\M{C}$
with respect to $q$, we can appeal to the compositionality of
bisimulation/simulation/skipping refinement to conclude that $\M{A}$
is a $\gamma$-\textit{refinement} of $\M{C}$ with respect to the
composition of $p$ and $q$. $\square$

The above theorem allows us to decompose refinement proofs into a
sequence of simpler refinement proofs and this turns out to be quite
useful, as it allows us to reason about conceptually distinct aspects
of our models in a way that is amenable to automated verification. For
example, each refinement proof contains only a finite unwinding of the
MA or ISA transition relations. One core idea that we leverage to
decompose the refinement proofs is the shared-resource commitment
refinement map. We will first describe how the model variants relate
to each other before discussing what the shared-resource commitment
refinement map is, how we are able to implement it using our models
and conclude with the proof obligations that our approach gives rise
to.

\subsection{Model Variants}
\begin{figure}
  \includegraphics[width=\columnwidth]{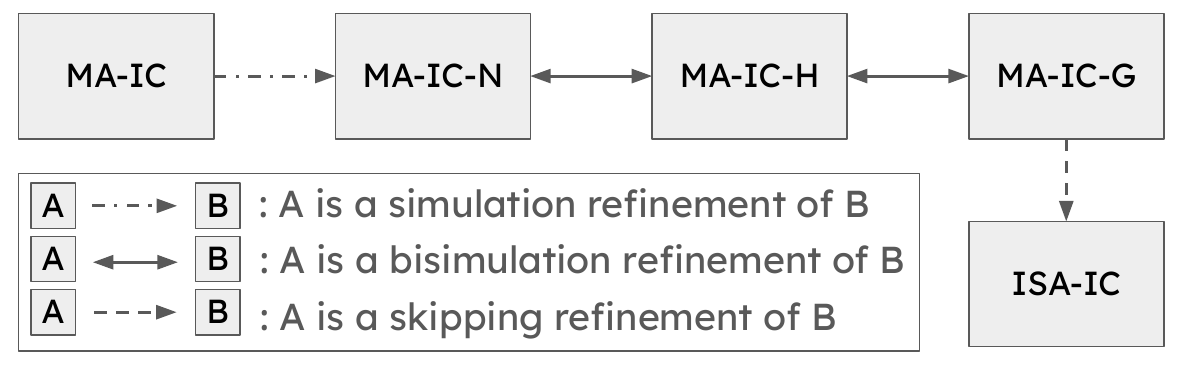}
  \Description{The behaviors of the MA-IC model are a subset of the behaviors of the MA-IC-N model. The MA-IC-N model is bisimilar to the MA-IC-H model, which is itself bisimilar to the MA-IC-G model. The MA-IC-G model is a witness skipping refinement of the ISA-IC model.}
  \caption{The model variants that we use to decompose the proof of
    the Meltdown refinement property, and how they relate to each
    other.}
  \label{fig:meltdown-refinement-models}
\end{figure}

Figure~\ref{fig:meltdown-refinement-models} shows the different model
variants that we use when decomposing the proof of our notion of
correctness for Meltdown, as discussed in
Section~\ref{sec:meltdown-correctness}. We start with \M{\ISAICM} and
\M{\MAICM}, which are models of the ISA and MA respectively that
support the \incache\ instruction as described in
Section~\ref{sec:in-cache}. We will use the \incache\ instruction to
identify differences in the cache behavior of \M{\ISAICM} and
\M{\MAICM} that indicate a Meltdown attack is possible. Next, we have
\M{\MAICNM}, which is a variant of \M{\MAICM} that is nondeterministic
in resource allocation decisions. The behavior of \M{\MAICM} should be
a subset of the behavior of \M{\MAICNM}. \M{\MAICNM} is bisimilar to
\M{\MAICHM}, a variant of \M{\MAICM} that maintains history
information in its state, in addition to the components that
$S_{\MAICM}$ contains.
%The fact that a bisimulation exists between
%\M{\MAICNM} and \M{\MAICHM} implies that the history information is
%sufficient to determine what resource allocation decisions are made
%(\eg\ enough to resolve the nondeterminism in \M{\MAICNM}).
The combination of the history generated by \M{\MAICHM} and the
nondeterminism of \M{\MAICNM} provides the core of the shared-resource
commitment refinement map, as together they can determine whether or
not a particular state needs to be considered when performing a
refinement proof against \M{\ISAICM}. Finally, \M{\MAGICM} is a
variant of \M{\MAICHM} that is defined only over those states that
need to be considered.

\subsection{Entangled States and the Shared-Resource Commitment Refinement Map}
\label{sec:entangled-states}

% To understand why we need \M{\MAICHM} and \M{\MAICNM},
To understand why the shared-resource commitment refinement map is
useful, we must first discuss reachability and how it is relevant to
refinement. Given a TS
$\trs{\ensuremath{\mathcal{X}}}$ and a set of
initial states $S^\textit{init}_\mathcal{X} \subseteq \stx{}$,
the set of reachable states is:% $S^\textit{reach}_\mathcal{X}$:
\begin{definition}[Reachable States]
  \label{def:reachability}
  \begin{equation*}
S^\textit{reach}_\mathcal{X} = \{ s \in \stx{} \from
\abracket{\exists s_i \from s_i \in S^\textit{init}_\mathcal{X} \from
  s_i \xtrans{\mathcal{X}}^\ast s } \}
\end{equation*}
\end{definition}
Say that $\Mx{}$ is a model of an MA. When implementing an
MA, one will often define the system's state as a structure with
fields that vary over bounded domains (\eg\ a program counter is a
64-bit unsigned integer). However, the MA will often only behave
correctly if additional constraints between fields are
satisfied. These constraints are called \emph{invariants}. For
example, \M{\MAICM} expects that the lines in the ROB have unique
$\robid$ fields. Such a system can still be shown to be correct if the
initial states $\stx{}^{\textit{init}}$ satisfy the
invariants and the invariants are shown to be preserved by the
system's transition relation (they are \emph{inductive invariants},
\eg\ if they hold for a state, they hold for all successors of
the state).

The refinement we show when using our notion of correctness requires
us to provide a refinement map that describes how to map \M{\MAICM}
states to \M{\ISAICM} states. We use the \emph{commitment approach}
described by Manolios~\cite{Manolios00} to do so: we map a \M{\MAICM}
state to a \M{\ISAICM} state by retaining only the programmer visible
components of the state. This can be thought of as invalidating the
pipeline: ``throwing out'' any in-flight instructions and only
considering the effects of committed instructions on the state.

A problem with allowing unreachable MA states in general is that, at
least for our models, some unreachable states will cause the MA to
behave incorrectly. This behavior includes getting stuck, retiring
instructions that do not exist in the instruction memory and loading
incorrect values from memory or the register file. When showing that
the MA is a refinement of an ISA model, it is unclear how we would map
these unreachable states to ISA states such that
refinement also enforces that the MA behavior on reachable states is
correct.  We do not care what the MA does when started from an invalid
state, so this behavior is undesirable.

One approach to resolving this issue is to create a version of
\M{\ISAICM} that only operates over reachable states.  However,
expressing and evaluating the reachability predicate is challenging,
since it requires resolving multiple existentials, which is likely to
be problematic for the kinds of solvers that are often used in
hardware verification. An alternative to expressing the reachability
predicate is to explicitly encode the relationships between different
components of the MA model's state that should hold for all reachable
states. Devising and expressing these relationships formally requires
substantial effort from someone with a deep understanding of the
microarchitecture. We instead choose to define a model over a superset
of the reachable states that is easier to express and reason about. We
introduce a recipe that can be used to define particular kinds of
supersets, allowing one to select the superset with the appropriate
trade-offs for their specific MA model and verification tooling. We
call these supersets of the reachable states sets of \emph{\good\
  states}.

Our approach is derived from the following insight about checking if a
state is reachable: if a state happened to be generated by starting
from an initial state and running it forward some number of steps,
then it is possible to resolve a similar existential to that seen in
the reachability predicate by maintaining some additional
\emph{history information} that indicates what the initial state
was. %As previously mentioned, the number of steps since the initial
%state is potentially very large (or even unbounded), so the history
%information needed for such an approach would also be potentially very
%large.
A key insight of our approach is that for our MAs, and we suspect for
many MAs, we can still identify many unreachable states by only
maintaining history information for instructions that have not yet
been retired. To check a state, we can \emph{invalidate} the MA state
and then use the history information to run it forward the appropriate
number of steps. Invalidation is required in any pipelined MA, and
refers to the process of discarding instructions that are in the
pipeline and restarting execution from the appropriate address that is
required when an instruction's execution results in an exception, or
when it is determined that a speculatively executed instruction should
not be committed. A complicating factor for this approach is the fact
that the decisions that an MA may make about resource allocation and
how to schedule actions (\eg\ how many instructions to fetch on a
particular cycle, whether to start executing a microinstruction loaded
in a RS, which RS to issue a microinstruction to) are based on the
resources that are available during that cycle, which may be affected
by instructions that had committed by the time the MA state is being
inspected. This is where the second key insight of our approach comes
in: if we have a version of the MA that is nondeterministic in
resource allocation and scheduling decisions, we
can use the history information alongside this new MA variant to
determine whether it is possible to reach a particular state $s$ when
starting at the state corresponding to an invalidated version of $s$,
when making the same resource allocation decisions as were made in the
execution of $s$ for the microinstructions that are in-flight in
$s$.
%This is the intuition for
%\good\ states.

Given:
\begin{itemize}
\item A deterministic MA TS $\trs{\ensuremath{\mathcal{X}}}$
\item \trs{\ensuremath{\mathcal{X}}-H}, a version of $\Mx{}$
  that gathers history such that $\Mx{H}$ is
  deterministic, $\stx{H} = \stx{} \times H$ and
  satisfying the below conditions
\item A version of $\Mx{}$ that is nondeterministic in
  resource allocation decisions, $\trs{\ensuremath{\mathcal{X}}-N}$,
  such that $\stx{N} = \stx{}$ and satisfying the
  below conditions
\item A transition function
  $\funcfont{step-using-h}_{\mcalx{N}} : \stx{} \times H \rightarrow
  \stx{} \times H$ that returns a successor with respect to
  \Mx{N} of the given state, using the given history
  information to resolve nondeterminism
\item An invalidation function
  $\funcfont{invl}_\mathcal{X} : \stx{} \times H \rightarrow \stx{}$
\item A function
  $\funcfont{init-h}_{\mathcal{X}} : \stx{} \rightarrow H$ that
  produces an ``empty'' history for the given state
\item A set of initial states
  $\stx{}^{\textit{init}} \subseteq \stx{}$
\end{itemize}

The set of \good\ states for \Mx{} is
\begin{flalign*}
  \stx{H}^{\textit{\goodshort}} = \{& \abracket{s, h} \in
\stx{H} \from \La\exists n \in \nats, h^\prime \from h^\prime \in
                             H \from\\
  &\funcfont{step-using-h}_{\mcalx{N}}^n(\funcfont{invl}(s, h), h) = \abracket{s, h^\prime}\Ra \}
\end{flalign*}

The additional conditions are:
\begin{itemize}
\item $\Mx{H}\sim_{\funcfont{hist}}\Mx{N}$ where
  $\funcfont{hist}$ is a function such that\\
  $\abracket{\forall s, h \from \abracket{s, h} \in \stx{H}
    \from \funcfont{hist}(\abracket{s, h}) = s}$
\item The behavior of \Mx{} is a subset of the behaviors of
  \Mx{N}. \eg
  $\abracket{\forall s, u \from s, u \in S_{\mathcal{X}} \wedge s
    \xtrans{\mathcal{X}} u \from s \xtrans{\mcalx{N}}
    u}$.
\item $\abracket{\forall s \from s \in \stx{}^{\textit{init}} \from \abracket{s, \funcfont{init-h}_{\mathcal{X}}(s)} \in \stx{H}^{\textit{\goodshort}}}$
\end{itemize}

At first glance the definition of
$\stx{H}^{\textit{\goodshort}}$ may not seem better than the
definition of reachable states. However, there is one more fact that
helps here: for a reachable $\Mx{H}$ state, the maximum
number of $\Mx{N}$ steps from the invalidated version of
that state back to itself is bounded. This means that the unwinding of
the transition function in $\stx{H}^{\textit{\goodshort}}$ is
also bounded. Even better, for our models the number of steps can be
computed from the state and its history information. Assuming that we
have a function
$\funcfont{steps-to-take}_{\mcalx{H}} : \stx{H} \rightarrow \nats$ that
determines the number of steps to take to get from the invalidated
version of a state back to itself, we can provide a simplified
definition for $\stx{H}^{\textit{\goodshort}}$:

\begin{flalign*}
  &\stx{H}^{\textit{\goodshort}} = \{ \abracket{s, h} \in \stx{H}, i = \funcfont{steps-to-take}_{\mcalx{H}}(\abracket{s, h}) \from\\
                                             &\abracket{\exists h \from h^\prime \in H \from \funcfont{step-using-h}^i_{\mcalx{N}}(\funcfont{invl}_{\mathcal{X}}(s, h), h) = \abracket{s, h^\prime}}\}
\end{flalign*}

% Finally, it must be the case that the conjunction of
% $\Mx{H}$, $\Mx{N}$, $\funcfont{invl}$,
% $\funcfont{step-using-h}$ and $\funcfont{steps-to-take}$ are
% \emph{replay-consistent}, meaning that

Finally, we must show that $\stx{H}^{\textit{\goodshort}}$ is
closed under $\xtrans{\mcalx{H}}$. This, in conjunction with the
fact that
$\abracket{\forall s \from s \in \stx{}^{\textit{init}} \from
  \abracket{s, \funcfont{init-h}_{\mathcal{X}}(s)} \in
  \stx{H}^{\textit{\goodshort}}}$, implies that
$\stx{H}^{\textit{\goodshort}}$ is a superset of the reachable
states of $\Mx{H}$, if we specify that the initial states
of $\Mx{H}$ are elements of
$\stx{}^{\textit{init}}$ paired with the history produced by
running $\funcfont{init-h}_{\mathcal{X}}$ on that element.

Now we can
define a TS that is $\Mx{H}$ but limited to
$\stx{H}^{\textit{\goodshort}}$:
$\trs{\ensuremath{\mathcal{X}}-G}$ where
$\stx{G} = \stx{H}^\textit{\goodshort}$,
$\xtrans{\mcalx{G}} = \xtrans{\mcalx{H}} \cap
\stx{H}^{\textit{\goodshort}} \times
\stx{H}^{\textit{\goodshort}}$ and
$\abracket{\forall s \from s \in \stx{H}^{\textit{\goodshort}}
  \from L_{\mcalx{G}} = L_{\mcalx{H}}}$. If we only care about
the behavior of $\Mx{}$ on reachable states, we can instead
reason about that of $\Mx{G}$. This is the core
addition of the shared-resource commitment refinement map to the
commitment refinement map.

\subsection{History Information}
One of the key observations regarding the idea of \good\ states is
that an MA will make resource allocation and scheduling decisions
based on instructions that have already been committed, and we
maintain history information to allow us to make the same scheduling
decisions that the MA did in the transitions leading up to a
particular state.
% We will give an overview of what kinds of resource
% allocation and scheduling decisions \M{\MAICM} makes, how \M{\MAICNM}
% is defined in such a way that those decisions can be made
% nondeterministically and how \M{\MAICHM} gathers enough information to
% be able to resolve nondeterminism in \M{\MAICNM} to exactly replicate
% the decisions made for the in-flight instructions in a \M{\MAICHM}
% state.
The resource allocation and scheduling decisions that
\M{\MAICM} makes are as follows: (1) the number of instructions to
fetch, (2) which RSes to issue fetched instructions to, (3) whether a
busy RS can begin execution and (4) whether a ready ROB entry should
be retired.

\trs{\MAICNM} is a nondeterministic TS.
$S_{\MAICNM} = S_{\MAICM}$. For each transition of \M{\MAICNM}, a
nondeterministic selection is made for the number of instructions to
fetch and issue,
%a mapping from ROB IDs that
%will be assigned to instructions that need RSes this cycle to the RS
%ID that they will be assigned to,
the set of RSes which are unavailable, the set of ROB lines which are
allowed to commit and the set of RSes which are allowed to begin
execution.  Notice that the nondeterministic choices only allow
\M{\MAICNM} to behave as though fewer resources are
available. \M{\MAICM} can be thought of as a version of \M{\MAICNM}
where the ``maximal'' choices are always selected.
%: $n = \funcfont{max-fetch-n}(S)$,
%$\allowcommit = \robids$, $\allowstart = \rsids$,
%$\busyrs = \emptyset$.

\trs{\MAICHM} can be thought of as \\\M{\MAICM} augmented with history
information: $S_\MAICHM = S_\MAICM \times H_\MAICM$. The behavior of
\M{\MAICHM} on the $S_\MAICM$ part of the state is identical to
\M{\MAICM}:
$\abracket{\forall s, h, s^\prime, h^\prime \from \abracket{s, h}
  \xtrans{\MAICHM} \abracket{s^\prime, h^\prime} \from s
  \xtrans{\MAICM} s^\prime}$. The history information is such that
given a reachable state $\abracket{s, h}$, if $s$ is invalidated and
then run forward using \M{\MAICNM} such that any resource allocation
or scheduling decisions are made in the same way they were in $s$ for
any in-flight instructions, the resulting state should be identical to
$s$. In other words, the history information must be sufficient to
allow us to reconstruct the nondeterministic choices that will
reproduce the behavior of \M{\MAICM}. We give a brief overview of the
gathered history information here, but
Appendix~\ref{sec:maichm-semantics} contains a full description.

$H_\MAICM$ is a structure consisting of several components. Here we
focus on $\mahlines$ and $\mahstartcyc$. % since these are the most
%relevant for reconstructing resource allocation
%decisions.
$\mahstartcyc$ is the first cycle for which this history
state has data and $\mahlines$ is a sequence of status lines,
representing information about the progress of all in-flight
microinstructions.
% \begin{itemize}
%   \item $\mahcommitcyc : \natc$ is the cycle during which the most recent commit occurred
%   \item $\mahstartcyc : \natc$ is the first cycle for which this history state has data
%   \item $\mahcommitcache : \natc \rightharpoonup \natc$ is the cache state, without any updates that may have occurred since the last instruction commit
%   \item $\mahcacheeffects : \robids \rightharpoonup (\natc \rightharpoonup \natc)$ maps a ROB identifier to the cache entries that should be added to the cache after committing that ROB line's microinstruction
%   \item $\mahlines : \mahstatuslines^\ast$ contains information about the progress of all in-flight microinstructions
%   \end{itemize}
%
% Information about the MA's resource allocation decisions is maintained
% in $\mahlines$. 
For any state $w$ such that $s\xtrans{\MAICHM}^\ast w$ from an initial
state $s \in S^{\textit{init}}_{\MAGICM}$, $\mahlines_w$ will contain
for each in-flight microinstruction a sequence of statuses indicating
what operation was performed %on that microinstruction
for each cycle starting at and including the cycle when the
microinstruction was issued. This information allows us to determine
when a microinstruction was issued and what resources were allocated
for it, when the microinstruction started execution and when it was
committed.

\subsection{Decomposition}
We will now describe the proof obligations that arise from using our
notion of correctness for Meltdown on \M{\ISAICM} and
\M{\MAICM}. First, we will instantiate the set of \good\ states with
$\mathcal{X} = \MAICM$.
The definition of \good\ states
requires that we provide $\M{\MAICNM}$, $\M{\MAICHM}$,
$\funcfont{step-using-h}_{\MAICNM}$, $\funcfont{invl}_{\MAICM}$,
$\funcfont{init-h}_{\MAICM}$ and $S^{\textit{init}}_{\MAICM}$. We
briefly discussed $\M{\MAICNM}$ and $\M{\MAICHM}$ above and full
definitions can be found in Appendices~\ref{sec:maicnm-semantics} and
\ref{sec:maichm-semantics} respectively. Similarly we provide
definitions for all of the required functions in
Section~\ref{sec:appendix-meltdown-obligations} and touch only on
$\funcfont{step-using-h}_{\MAICNM}$ here. That function operates by
using the history to calculate the appropriate values for the
nondeterministic choices made previously, and then transitions
\M{\MAICNM} using those choices. Our notion of \good\ states imposes four proof obligations:
% \begin{equation}
%   \label{eqn:maicn-superset-behavior-maic}
%   \abracket{\forall s, u \from s, u \in S_{\MAICM} \wedge s
%     \xtrans{\MAICM} u \from s \xtrans{\MAICNM} u}
% \end{equation}
% \begin{equation}
%   \label{eqn:maicn-bisim-maich}
%   \begin{aligned}
%   &\M{\MAICHM}\sim_{\funcfont{hist}}\M{\MAICNM} \text{ where }
%   \funcfont{hist} \text{ is a function such that }\\
%   &\abracket{\forall s, h \from \abracket{s, h} \in S_{\MAICHM}
%     \from \funcfont{hist}(\abracket{s, h}) = s}
%   \end{aligned}
% \end{equation}
\begin{align}
  &\abracket{\forall s, u \from s, u \in S_{\MAICM} \wedge s
    \xtrans{\MAICM} u \from s \xtrans{\MAICNM} u}\label{eqn:maicn-superset-behavior-maic}\\
  &\begin{aligned}
  &\M{\MAICHM}\sim_{\funcfont{hist}}\M{\MAICNM} \text{ where }
  \funcfont{hist} \text{ is a function such that }\\
  &\abracket{\forall s, h \from \abracket{s, h} \in S_{\MAICHM}
    \from \funcfont{hist}(\abracket{s, h}) = s}
  \end{aligned}\label{eqn:maicn-bisim-maich}\\
  &\abracket{\forall s \from s \in S^{\textit{init}}_{\MAICM} \from \abracket{s, \funcfont{init-h}_{\MAICM}(s)} \in S^\textit{\goodshort}_{\MAICHM}}\label{eqn:maic-init-states-entangled}\\
  &\abracket{\forall s \from s \in S^{\textit{\goodshort}}_{\MAICHM} \from \abracket{\forall w \from s \xtrans{\MAICHM} w \from w \in S^\textit{\goodshort}_{\MAICHM}}}\label{eqn:maich-closed-under-entangled}
\end{align}
In addition, our notion of correctness for Meltdown requires that
\M{\MAGICM} is a witness skipping refinement of \M{\ISAICM} with
respect to our refinement map $\funcfont{r-ic}$, defined below. This
is proved by showing the existence of a witness skipping relation on
the TS produced by taking the disjoint union of
\M{\MAGICM} and \M{\ISAICM}. Let
$\M{\textit{ic}} = \disjtrs{\MAGICM}{\ISAICM}{}$ be this system. Let
$S_{\textit{ic}} = S_{\MAGICM} \uplus S_{\ISAICM}$ and
$\xtrans{\textit{ic}} = \xtrans{\MAGICM} \uplus \xtrans{\ISAICM}$. We
instantiate Definition~\ref{def:witness-sk}, providing
$\funcfont{skip-wit-ic} : S_{\textit{ic}} \times S_{\textit{ic}}
\rightarrow \nats \setminus \{ 0 \}$ for $\funcfont{skip-wit}$,
$\funcfont{stutter-wit-ic} : S_{\textit{ic}} \times S_{\textit{ic}}
\rightarrow \nats$ for $\funcfont{stutter-wit}$,
$\funcfont{run-ic} : S_{\textit{ic}} \times S_{\textit{ic}} \times
S_{\textit{ic}} \rightarrow S_{\textit{ic}} $ for $\funcfont{run}$,
and $\skiprel{ic} \subseteq S_{\textit{ic}} \times S_{\textit{ic}} $
for $B$. The obligations generated are as follows:
% \begin{equation}
%   \label{eqn:wsk1-meltdown}
%   \La \forall s \in S_{\MAGICM} :: s \skiprel{ic} \funcfont{r-ic}.s\Ra
% \end{equation}
% \begin{equation}
%   \label{eqn:wsk2-meltdown}
%   \abracket{\forall w, s, u \from s \skiprel{ic} w \wedge s \xtrans{\textit{ic}} u \from w \xtrans{\textit{ic}}^{\funcfont{skip-wit-ic}(s, u)} \funcfont{run-ic}(w, s, u)}
% \end{equation}
% \begin{equation}
%   \label{eqn:wsk3-meltdown}
%   \begin{aligned}
%   &\forall s,u,w \in S_{\textit{ic}}: s \skiprel{ic} w \And s \xtrans{\textit{ic}} u: &\\
%   &\text{\quad } (u \skiprel{ic} w \And \funcfont{stutter-wit-ic}(u, w) < \funcfont{stutter-wit-ic}(s, w))\ \vee&\\
%   &\text{\quad } u \skiprel{ic} (\funcfont{run-ic}(w, s, u))&
%   \end{aligned}
% \end{equation}
\begin{align}
  \label{eqn:wsk1-meltdown}
  &\La \forall s \in S_{\MAGICM} :: s \skiprel{ic} \funcfont{r-ic}.s\Ra\\
  \label{eqn:wsk2-meltdown}
  &\abracket{\forall w, s, u \from s \skiprel{ic} w \wedge s \xtrans{\textit{ic}} u \from w \xtrans{\textit{ic}}^{\funcfont{skip-wit-ic}(s, u)} \funcfont{run-ic}(w, s, u)}\\
  &\begin{aligned}
  &\forall s,u,w \in S_{\textit{ic}}: s \skiprel{ic} w \And s \xtrans{\textit{ic}} u: &\\
  &\text{\quad } (u \skiprel{ic} w \And \funcfont{stutter-wit-ic}(u, w) < \funcfont{stutter-wit-ic}(s, w))\ \vee&\\
  &\text{\quad } u \skiprel{ic} (\funcfont{run-ic}(w, s, u))&
  \end{aligned}\label{eqn:wsk3-meltdown}
\end{align}

\section{Correctness for Spectre}
\label{sec:spectre-correctness}

Using witness skipping refinement with the \incache\ abstraction
allows one to identify susceptibility to Meltdown attacks that rely on
cache side channels, but such a notion of correctness is not violated
by a system which is vulnerable to Spectre.
%Broadly, the reason why
%Spectre attacks do not violate refinement in the presence of the
%\incache\ abstract instructions is because they do not access
%unprivileged memory, as Meltdown attacks do, but instead access legal
%memory locations that are however not accessed by the ISA
%semantics.
In short, this is because Spectre attacks do not access unprivileged
memory, as Meltdown attacks do, but instead access legal memory
locations that are however not accessed by the ISA semantics.  We note
here the connection between \emph{prefetching} and Spectre
attacks. Prefetching is a hardware mechanism that allows MAs to bring
data (that they have access to) into the cache in advance, \ie before
the ISA explicitly accesses them. Spectre attacks exploit transient
execution to have the MA bring memory locations in the cache that
otherwise wouldn't have been accessed according the ISA semantics. The
key challenge here is, how can we detect the difference between a
desirable hardware mechanism such as prefetching and a hardware
behavior that can result in TEAs? We propose a solution using
\emph{intent models} alongside a novel notion of refinement.
%We propose the use of
%\emph{intent models} alongside a novel notion of refinement to resolve
%this challenge.
%We will first discuss intent
%models and prefetching in more detail before describing the MA and ISA
%implementations that we developed and concluding with our statement of
%correctness.

\subsection{Intent Models}
\label{sec:intent}
Modern microprocessors include hardware prefetch units which extend
cache units by monitoring memory accesses and fetching data before it
is needed \cite{intel-prefetch}. The goal of prefetching is to reduce
memory latencies by eliminating cache misses and can be viewed as
predicting which data will be required in the future. Consider a
simple form of hardware prefetching, \emph{next-line}
prefetching. After an access to memory address $a$, a next-line
prefetcher will request that the next $N$ cache lines $a+1,\ldots a+N$
be cached.  A slightly more complicated approach is to perform
\emph{stride} prefetching, wherein an access to memory location $a$
results in the prefetching of addresses $a+N, a+2N,\ldots$ for a
selected stride $N$.  For a thorough examination of cache prefetching,
we refer the reader to Mittal's survey \cite{mittal2016survey}.

Similar to caches, prefetch units are intentionally left
underspecified at the ISA level to allow implementer flexibility in
defining hardware prefetchers based on the MA's requirements.
% Our
% proposed modeling of cache behaviors, the \incache\ abstract
% instruction described in the previous section, freely allows the
% implementation of arbitrary prefetching strategies, so long as the
% prefetch units are only fetching addresses that the process has access
% to.
% We show that while this maximally permissive approach of modeling
% the cache behavior is sufficient to catch hardware security bugs that
% result in some Meltdown attacks, \ie by having their refinement based
% notion of correctness fail, it cannot successfully detect some
% transient execution attacks, such as Spectre.
% Broadly, the reason why Spectre attacks do not violate refinement in
% the presence of the \incache\ abstract instructions, is because they
% do not access unprivileged memory, as Meltdown attacks do, but instead
% access legal memory locations that are however not accessed by the ISA
% semantics. We note here the connection between prefetching and Spectre
% attacks. Prefetching is a hardware mechanism that allows MAs to bring
% data (that they have access to) into the cache in advance, \ie before
% the ISA explicitly accesses them. Spectre attacks exploit transient
% execution to have the MA bring memory locations in the cache that
% otherwise wouldn't have been accessed according the ISA semantics. The
% key challenge here is, how can we detect the difference between a
% desirable hardware mechanism such as prefetching and a hardware
% behavior that can result in transient execution attacks?
To provide a notion of correctness that is able to catch Spectre
attacks while allowing MAs to freely implement hardware
prefetchers, we introduce the idea of \emph{intent} models. The key
idea of intent models is that during each step, the MA emits
information regarding the set of addresses it intended to cache due to
each instruction, which we call \emph{intent virtual
  instructions}. The ISA can be stepped in a way that conforms to
this information, \eg\ by prefetching the same set of addresses the MA
did upon committing the same instruction.
% The key idea of intent
% models is that during each step
% ISA is allowed to cache or prefetch
% any memory address as long as i) it is part of the legal memory \lmem,
% and ii) it explicitly mentions its \emph{intent}, \ie that this
% address was brought in the cache as a result of prefetching, and not
% because it was explicitly accessed by an instruction. Eviction is
% similarly handled.
Intent virtual instructions do not appear in instruction memory---they
are emitted at runtime by the MA. The expectation is that the designer
of an MA will implement an intent version of the MA by providing a
function that describes the intended cache modifications corresponding
to instructions committed in a given step.

% In particular, whenever the MA loads
% or evicts addresses from the cache due to prefetching, it emits
% information that maps the instruction that triggered the the
% prefetching to the set of addresses that were prefetched. This
% information can then be threaded into the ISA, which after executing
% each instruction, will execute any intent virtual instructions that
% were emitted by the MA due to that instruction. This causes the ISA to
% prefetch the same addresses.

% Intent models are variants of ISA/MA models that allow \emph{load} and
% \emph{evict virtual instructions}. We call these instructions virtual
% as they are not part of ISA programs that programmers write. Virtual
% instructions are emitted at runtime by both the ISA and MA machines in
% the prefetch intent model and they have a single operand, namely the
% address being prefetched. To do so, the ISA extends its semantics by
% allowing all non-virtual ISA instructions to non-deterministically
% issue a finite number of virtual instructions. The virtual
% instructions are executed atomically directly after the non-virtual
% ISA instruction. The semantics of the MA for non-virtual instructions
% remain the same. At every step, the MA consults its prefetch unit and
% commits load virtual instructions every time the latter decides to
% prefetch an address and similarly for evict instructions.

\subsection{Action Labeled Transition System}
\label{sec:action-labeled-tranistion-system}
Mathematically, we represent an intent model using an \emph{action
  labeled transition system} (ALT). An ALT $\atrs{}$ is a structure
consisting of a set of states $S$, a set of actions $A$ where each
action is a sequence over elements $\mathcal{A}$, a transition
relation $\trans{} \subseteq S \times A \times S$ such that
$\abracket{\forall s \in S \from \abracket{\exists u \in S, a \in A
    \from (s, a, u) \in\ \trans{}}}$ and a state label function
$\funcfont{L}$ with domain $S$. Given $s, u \in S, a \in A$, we write
$s \atrans{}{a} u$ as a shorthand for
$\abracket{s, a, u} \in\ \trans{}$. For our notion of correctness for
Spectre, the set of actions will consist of sequences of intent
virtual instructions.

\subsection{Action Skipping Refinement}

We generalize the notion of skipping refinement, which is defined
on TSes that do not have labels on transitions (actions), to ALTs. To
do this, we need to update the definition of matching used in
skipping refinement to account for actions.
%We will extend the concept of matching used in the definition of
%skipping refinement to action labeled transition systems.
Intuitively, we do this by specifying that two paths in an ALT match
iff they can each be partitioned in such a way that both the states
and actions in two corresponding partitions match, rather than just
the states.

Let \atrs{} be an ALT. An A-path for \M{} is a tuple
$\abracket{\sigma, \delta}$ where $\sigma$ is a sequence of states
from $S$ and $\delta$ is a sequence of states from $A$ such that for
every pair of adjacent states in $\sigma$ $s = \sigma.j$ and
$u = \sigma.(j+1)$, it is the case that $s \atrans{}{\delta.j} u$. As
a convention, given an A-path $\tau$ we use $\tau_S$ to refer to the
first element of the tuple (the sequence of states) and $\tau_A$ to
refer to the second element of the tuple (the sequence of actions). An
A-path $\sigma$ starting at state $s$ is an \emph{A-fullpath}, denoted
by $\fpa.\sigma.s$, if both $\sigma_S$ and $\sigma_A$ are infinite. An
\emph{A-segment},
$\abracket{\abracket{v_1, \ldots, v_k}, \abracket{a_1, \ldots,
    a_{k-1}}}$, where $k \geq 1$ is a finite A-path and is denoted by
$\vv{v}$. The length of an A-segment $\vv{v}$ is denoted by
$|\vv{v}|$.
%Let $\inc$ be the set of strictly increasing sequences of
%natural numbers starting at 0.
The $i^{th}$ partition of an A-fullpath $\sigma$ with respect to
$\pi \in \inc$, denoted by $\partition{\sigma}{\pi}{i}$, is given by
an A-segment
$\abracket{\abracket{\sigma_{S}(\pi.i), \ldots, \sigma_{S}(\pi(i + 1)
    - 1)}, \abracket{\sigma_{A}(\pi.i), \ldots, \sigma_{A}(\pi(i + 1)
    - 1)}}$.
%$\partition{\sigma}{\pi}{i}_S$ and
%$\partition{\sigma}{\pi}{i}_A$ denote the first and second component
%of $\partition{\sigma}{\pi}{i}$ respectively.
Given a sequence of sequences $\sigma$, $\mybigcirc(\sigma)$ denotes
the concatenation of all of the sequences in $\sigma$, in order.  Note
that concatenation is sufficient for our case, but in general, given
an set of actions where an action can ``undo'' another action, it
might be necessary to introduce a more complicated notion of combining
actions.
% We require that the set of actions $A$ has an operator
% $\circ : A \times A \rightarrow A$ that in some sense combines the
% effects of two actions into a single action. Given an A-segment
% $\sigma$, $\mybigcirc(\sigma)$ denotes the result of using $\circ$ to
% combine all of the elements of $\sigma_A$ in order, \eg\
% $\sigma_{A}(1) \circ \sigma_{A}(2) \circ \ldots \circ
% \sigma_{A}(|\sigma|)$.
%We also require that there is an ``empty''
%action $\epsilon_A \in A$ that does not modify the 
\begin{definition}[amatch]
  \label{def:amatch}
  Let \atrs{} be an ALT and $\sigma,\delta$
  be A-fullpaths in \M{}. For $\pi, \xi \in \inc$ and binary relation
  $B \subseteq S \times S$ we define two functions:
  \begin{flalign*}
    &\acorr{B}{\sigma}{\pi}{\delta}{\xi} \equiv &\\
    &\quad\La \forall i \in
    \omega:: \La \forall s \in \partitionsub{\sigma}{\pi}{i}{S} ::
    s B \delta_{S}(\xi.i) \Ra \wedge \mybigcirc(\partitionsub{\sigma}{\pi}{i}{A}) = \mybigcirc(\partitionsub{\delta}{\xi}{i}{A}) \Ra \textit{ and } &\\
    &\amatch{B}{\sigma}{\delta} \equiv \La \exists \pi, \xi \in
      \inc :: \acorr{B}{\sigma}{\pi}{\delta}{\xi}\Ra.
  \end{flalign*}
\end{definition}
We can now define the notion of an action skipping simulation using
$\funcfont{amatch}$.
\begin{definition}[Action Skipping Simulation (ASKS)]
  \label{def:asks}
  $B \subseteq S \times S$ is an action skipping simulation on an action TS
  \atrs{} \myiff{} for all $s,w$ such that $sBw$, both of the following hold:
  \begin{enumerate}
  \item $L.s = L.w$
  \item\abracket{\forall \sigma\from \fpa.\sigma.s \from \abracket{
        \exists \delta \from \fpa.\delta.w \from
        \amatch{B}{\sigma}{\delta}}}
  \end{enumerate}
\end{definition}

\subsection{Statement of Correctness}
\label{sec:spectre-correctness-statement}

Let \atrs{\textit{ISA}} and \atrs{\textit{MA}} be ALTs modeling an ISA
and an MA such that the two systems both support the \incache\
instruction. The actions $A_{\textit{ISA}} = A_{\textit{MA}}$ used by
these systems are used to indicate what changes to the cache are
authorized, a notion decided by the designer of the system. Let
$r : S_{\textit{MA}} \rightarrow S_{\textit{ISA}}$ be a refinement
map. At a high level, we expect $r$ to map an MA state to an ISA state
that agrees in the programmer-visible values of its components---\eg
the register file of the mapped ISA state should correspond to the
committed register file for the MA. We require that $r$ projects the
cache component of the \M{\textit{MA}} state. Then, we say that
\M{\textit{MA}} is a correct implementation of \M{\textit{ISA}} with
respect to Spectre iff \M{\textit{MA}} is an action skipping
refinement of \M{\textit{ISA}} with respect to $r$.

%In practice, we expect that the designer 

% Therefore, we use a different notion of correctness. Let
% \atrs{\textit{ISA}} and \atrs{\textit{MA}} be action labeled
% transition systems modeling an ISA and an MA, where both support the
% \incache\ instruction. Let
% $r : S_{\textit{MA}} \rightarrow S_{\textit{ISA}}$ be a refinement
% map. At a high level, we expect $r$ to map an MA state to an ISA state
% that agrees in the programmer-visible values of its components---\eg\
% the register file of the mapped ISA state should correspond to the
% committed register file for the MA. Then, we say that \M{\textit{MA}}
% is a correct implementation of \M{\textit{ISA}} with respect to
% Spectre iff \M{\textit{MA}} is an action witness skipping refinement
% of \M{\textit{ISA}} with respect to $r$.

\section{Spectre Decomposition Proof}
\label{sec:spectre-decomp}

At a high level, our strategy involves breaking
down correctness into our notion of correctness for Meltdown on an MA
and an ISA, plus a property that expresses that the MA only performs
updates to its cache that are ``authorized'' according to the system
designer.

We start with two ALTs, \M{\MAICAM} and \M{\ISAICAM}, denoting the MA
and the ISA respectively. \M{\MAICAM} and \M{\ISAICAM} have the same
set of actions $A$, \eg\ $A_{\MAICAM} = A_{\ISAICAM} = A$, where A
consists of sequences of \emph{authorized cache actions}:
\[
  A = (\{ \texttt{prefetch}\ a\ \vert\ a \in \natc \} \cup \{ \texttt{cache}\ a\ \vert\ a \in \natc \} )^\ast
\]
Let $\funcfont{r-a} : S_{\MAICAM} \rightarrow S_{\ISAICAM}$ be a
refinement map that maps all of the corresponding components of an
\M{\MAICAM} state to an \M{\ISAICAM} state, including the cache.

\atrs{\MAICAM} is an ALT that can be thought of as \M{\MAICM} but
restricted so that for any transition from state $s$ to state $u$ in
\M{\MAICM}, \M{\MAICAM} allows that transition only under the action
consisting of the sequence of authorized cache actions that were
performed during the transition.

From an operational perspective, we have a function
$\funcfont{auth-actions} : S_{\MAICM} \times S_{\MAICM} \rightarrow A$
that, when given states $s, u \in S_{\MAICM}$ such that
$s \xtrans{\MAICM} u$, produces a sequence of authorized cache actions
corresponding to the behavior of \M{\MAICM} during that
transition. This can be thought of as a specification that needs to be
provided by a system designer, describing what changes to the cache
should be visible due to a \M{\MAICM} transition. We define
$\xtrans{\MAICAM}$ as follows:
\[
  s \atrans{\MAICAM}{a} u \iff s \xtrans{\MAICM} u \wedge a = \funcfont{auth-actions}(s, u)
\]
% Notice that here we require the sequence of authorized actions for a
% particular transition to be unique. This is an assumption we will use
% here, but as we will see this can be generalized such that
% $\funcfont{auth-actions}$ is a relation rather than a function.

\atrs{\ISAICAM} is an ALT that can be thought of as \M{\ISAICM} but
with restricted nondeterminism. $S_{\ISAICAM} = S_{\ISAICM}$. In
particular, the top-level transition rule of \M{\ISAICAM} composes the
\M{\ISAICM\texttt{-ISA}} auxiliary TS with a new
\M{\ISAICAM\texttt{-C}} ALT that applies the authorized cache actions
for this transition. 

\paragraph{Action Skipping}

The notion of correctness with respect to Spectre and \funcfont{r-a} for \M{\ISAICAM} and
\M{\MAICAM} is:
\begin{equation}
  \label{eqn:spectre-correctness}
  \M{\MAICAM} \sqsubseteq_{\funcfont{r-a}} \M{\ISAICAM}
\end{equation}

We decompose this into the conjunction of the following two
statements, where $\cmem_s$ refers to the cache memory component of
the state $s$.

\begin{align}
  &\M{\MAICM} \lesssim_{\funcfont{r-a}} \M{\ISAICM} \label{eqn:spectre-meltdown}\\
&\begin{aligned}
  &\La\forall s, u \in S_{\MAICAM}, a \in A, w \in S_{\ISAICAM} \from s \atrans{\MAICAM}{a} u \wedge\\
  &\quad\quad\quad \cmem_s = \cmem_w \wedge L_{\ISAICAM}(\funcfont{r-a}(s)) = L_{\ISAICAM}(w)\from\\
  &\quad \La\exists v \in S_{\ISAICAM}, \sigma \in A^\ast \from \mybigcirc(\sigma) = a \wedge w \atrans{\ISAICAM}{\sigma}^\ast v \from\\
  &\quad\quad\quad\cmem_u = \cmem_v \wedge L_{\ISAICAM}(\funcfont{r-a}(u)) = L_{\ISAICAM}(v)\Ra\Ra
\end{aligned} \label{eqn:spectre-cache-same}
\end{align}

% \begin{equation}
%   \label{eqn:spectre-meltdown}
%   \M{\MAICM} \lesssim_{\funcfont{r-a}} \M{\ISAICM}
% \end{equation}

% \begin{equation}
%   \label{eqn:spectre-cache-same}
% \begin{aligned}
%   &\La\forall s, u \in S_{\MAICAM}, a \in A, w \in S_{\ISAICAM} \from s \atrans{\MAICAM}{a} u \wedge\\
%   &\quad\quad\quad \cmem_s = \cmem_w \wedge L_{\ISAICAM}(\funcfont{r-a}(s)) = L_{\ISAICAM}(w)\from\\
%   &\quad \La\exists v \in S_{\ISAICAM}, \sigma \in A^\ast \from \mybigcirc(\sigma) = a \wedge w \atrans{\ISAICAM}{\sigma}^\ast v \from\\
%   &\quad\quad\quad\cmem_u = \cmem_v \wedge L_{\ISAICAM}(\funcfont{r-a}(u)) = L_{\ISAICAM}(v)\Ra\Ra
% \end{aligned}
% \end{equation}

We now argue that proofs of Equations~\ref{eqn:spectre-meltdown} and
\ref{eqn:spectre-cache-same} imply a proof of
Equation~\ref{eqn:spectre-correctness}, given our machines.
%We will do
%this by proving the contrapositive.
Say that we have proofs of Equations~\ref{eqn:spectre-meltdown} and
\ref{eqn:spectre-cache-same}. This means that there exists a $B$ over
$S_{\ic} \times S_{\ic}$ such that
$\La \forall s \in S_{\MAICM} :: sBr.s\Ra$ and $B$ is an SKS on
$\M{\ic} =$ \disjtrs{\MAICM}{\ISAICM}{\ic} where
$\labf_{\ic}.s = L_A(s)$ for $s \in S_{\ISAICM}$, and
$\labf_{\ic}.s = L_A(r.s)$ for $s \in S_{\MAICM}$. This means that for
all $s, w \in S_{\ic}$ such that $s B w$, the following two statements
hold: $L_{\ic}.s = L_{\ic}.w$,
\[
  \abracket{\forall \sigma\from
  \fp.\sigma.s \from \abracket{ \exists \delta \from \fp.\delta.w
    \from \smatch{B}{\sigma}{\delta}}}
\]

To prove Equation~\ref{eqn:spectre-correctness}, we must show the
existence of a $B^\prime$ over $S_{\act} \times S_{\act}$ such that
$\La \forall s \in S_{\MAICAM} :: sB\funcfont{r-a}.s\Ra$ and $B^\prime$ is an ASKS
on $\M{\act} =$ \disjatrs{\MAICM}{\ISAICM}{\act}{\act} where
$\labf_{\act}.s = L_{\ISAICAM}(s)$ for $s \in S_{\ISAICAM}$, and
$\labf_{\act}.s = L_{\ISAICAM}(\funcfont{r-a}.s)$ for $s \in S_{\MAICAM}$. We will argue
that $B^\prime = B$ satisfies these conditions.

% Since $L_{\ISAICAM} = L_{\ISAICM}$

Since $B$ is an SKS on \M{\ic}, we can assume that
condition (2) from the definition for SKS holds. Expanding gives:

\begin{equation}
  \label{eqn:sks-cond-2-assume}
  \begin{aligned}
  \La \forall s, w & \from s B w \from
                     \La\forall \sigma\from \fp.\sigma.s \from \La
    \exists \delta \from \fp.\delta.w \from
                     \La\exists \pi, \xi \in \inc \from\from \\
                   &\La\forall i \in \omega \from\from\abracket{\forall s \in \partition{\sigma}{\pi}{i} \from \from s B \delta(\xi.i)}\Ra\Ra
                     \Ra\Ra\Ra
  \end{aligned}
\end{equation}

We will now show that $B$ is also an action skipping simulation on the
corresponding $\M{\act} =
\disjatrs{\MAICAM}{\ISAICAM}{\act}{\act}$. Since $S_{\act} = S_{\ic}$
and $L_{\ISAICAM} = L_{\ISAICM}$, we can discharge the obligation that
the labels of states related by $B$ are equal, as this must hold for
$B$ to be a SKS on \M{\ic}. We need to show that condition (2) of
the definition for ASKS holds. Expansion gives:

\begin{equation}
  \label{eqn:asks-must-discharge}
  \begin{aligned}
    \La \forall s, w & \from s B^\prime w \from \La \forall \rho \from \fpa.\rho.s \from \La\exists \tau \from \fpa.\tau.w \from \La\exists \pi, \xi \in \inc \from\from\\
    &\La\forall i \in \omega \from\\
  &\textbf{\quad\quad(a)} \abracket{\forall s \in \partitionsub{\sigma}{\pi}{i}{S} \from \from s B^\prime \delta_{S}(\xi.i)} \wedge\\
  &\textbf{\quad\quad(b)}\mybigcirc(\partitionsub{\rho}{\pi}{i}{A}) =
    \mybigcirc(\partitionsub{\tau}{\xi}{i}{A})\Ra\Ra\Ra
  \end{aligned}
\end{equation}

Pick an arbitrary $s$, $w$ and $\sigma$, and then pick the $\delta$,
$\pi$, and $\xi$ that Equation~\ref{eqn:sks-cond-2-assume} asserts
exist. Let $\rho$ be an A-fullpath such that $\rho_S = \sigma$ and
$\tau$ be an A-fullpath such that $\tau_S = \delta$. Since
$B^\prime = B$, $\rho_S = \sigma$ and $\tau_S = \delta$, condition (a)
of Equation~\ref{eqn:asks-must-discharge} follows from
Equation~\ref{eqn:sks-cond-2-assume}. For condition (b), we will
consider several cases. Let $x$ be the first state in a
$\partitionsub{\rho}{\pi}{i}{S}$ and $y$ be the first state in
$\partitionsub{\tau}{\xi}{i}{S}$.  First, if $x$ is halted, then $y$
must also be halted, and no actions will be emitted in this partition
or a future partition, so (b) trivially holds. We now consider the
case where $s \in S_{\MAICAM}$ and $w \in S_{\ISAICAM}$. Notice that
$x$ must have the same cache as
$y$. %, since $B^\prime$ requires that $L_{\ISAICAM}$ agrees on the
All of the states in $\partitionsub{\rho}{\pi}{i}{S}$ must be related
via $B^\prime$ to $y$, so if
$\partitionsub{\rho}{\pi}{i}{A}(j) \neq \emptyseq$ for any
$j < \vert\partitionsub{\rho}{\pi}{i}{A}\vert$, that action did not
have an effect on the cache. The last action in
$\partitionsub{\rho}{\pi}{i}{A}$ corresponds to the transition from
the last state in $\partitionsub{\rho}{\pi}{i}{S}$ to the first state
in $\partitionsub{\rho}{\pi}{i+1}{S}$. The actions in
$\partitionsub{\tau}{\xi}{i}{A}$ correspond to the transitions
comprising the A-segment starting at $y$ and ending at the first state
of $\partitionsub{\tau}{\xi}{i+1}{S}$. We can instantiate
Equation~\ref{eqn:spectre-cache-same} here, which gives us that for
the transition from the last state in $\partitionsub{\rho}{\pi}{i}{S}$
to the first state in $\partitionsub{\rho}{\pi}{i+1}{S}$, given that
$y$ matches with $x$, there exists some $v \in S_{\ISAICAM}$ and some
$\chi \in A^\ast$ such that $\mybigcirc(\chi)$ is equal to the last
action in $\partitionsub{\rho}{\pi}{i}{A}$,
$y \atrans{\ISAICAM}{\chi}^\ast v$, and $v$ has the same cache and
label as the first state in $\partitionsub{\rho}{\pi}{i+1}{S}$. We now
argue that $v$ must be the first state of
$\partitionsub{\tau}{\xi}{i+1}{S}$, and $\chi$ must be
$\partitionsub{\tau}{\xi}{i}{A}$.

Note that the definitions of $L_{\ISAICAM}$ and $\funcfont{r-a}$ are
such that any \M{\MAICAM} state has a unique related \M{\ISAICAM}
state. Therefore, $v$ must be the first state of
$\partitionsub{\tau}{\xi}{i+1}{S}$, since both must be related to
$\partitionsub{\rho}{\pi}{i+1}{S}$ via $B^\prime$. We can assume that
$v$ only appears once in the path from $y$ to $v$, since \M{\MAICAM}
is not capable of committing an instruction twice in a single skipping
step. Therefore, the actions in $\chi$ and
$\partitionsub{\tau}{\xi}{i}{A}$ start at the same cache and produce
the same cache, meaning they must be equivalent up to ``noop''
actions.

Notice that the purpose of \M{\ISAICAM} in
Equation~\ref{eqn:spectre-cache-same} is in some sense to give us a
specification for how actions should affect \M{\MAICAM} state. We can
simplify that equation by directly defining the effects that actions
should have on the cache and checking that \M{\MAICAM} is behaving
appropriately. That is, we would like to check that \M{\MAICAM} is not
modifying the cache through some unauthorized actions that it is not
emitting. This gives Equation~\ref{eqn:spectre-cache-same-simple},
where $\funcfont{apply-action}$ is a function that takes in a
\M{\MAICAM} state $s$, a cache $c$ and an action $a$, and produces $c$
after applying $a$ to it.
% Let \funcfont{c-addr} be a function that updates the given cache so
% that it contains the given address.
%
% \[
%   \funcfont{c-addr}(s, c, a) = [a \mapsto \partialget{\dmem_s}{a, 0}]c
% \]
%
% \begin{flalign*}
%   &\funcfont{apply-action}(s, c, a) =\\
%   &\begin{cases}
%     c &\text{if } a = \emptyseq\\
%     \funcfont{apply-action}(s, \funcfont{c-addr}(s, c, d), a^\prime) &\text{if } a = \texttt{prefetch } d \bullet a^\prime\\
%     \funcfont{apply-action}(s, \funcfont{c-addr}(s, c, d), a^\prime) &\text{if } a = \texttt{cache } d \bullet a^\prime
%   \end{cases}
% \end{flalign*}
\begin{equation}
  \label{eqn:spectre-cache-same-simple}
  \abracket{\forall s, u \in S_{\MAICAM}, a \in A \from s \atrans{\MAICAM}{a} u\from\cmem_u = \funcfont{apply-action}(s, a)}
\end{equation}

\subsection{Proof Obligations for Spectre}

We now describe the proof obligations that arise from using our
notion of correctness for Spectre on \M{\ISAICAM} and
\M{\MAICAM}. First, we will decompose
$\M{\MAICM} \lesssim_{\funcfont{r-a}} \M{\ISAICM}$ using the same
approach that we used for
$\M{\MAICM} \lesssim_{\funcfont{r-ic}} \M{\ISAICM}$ in
Section~\ref{sec:meltdown-decomp}. The only new obligation is
Equation~\ref{eqn:spectre-cache-same-simple}.

Several of the proof obligations for
$\M{\MAICM} \lesssim_{\funcfont{r-a}} \M{\ISAICM}$ are identical to
those that are needed here. The main differences appear in the context
of witness skipping refinement, as our refinement map and
$\funcfont{run}$ witness function both differ.

$\skiprel{ic-a} \subseteq S_{\ic} \times S_{\ic}$ is the witness
skipping relation over \M{\ic} that we must show exists and satisfies
the below properties,
Equations~\ref{eqn:wsk1-spectre}-\ref{eqn:wsk3-spectre}.

$\funcfont{run-ic-c}(w, s, u)$ is a function that steps $w$
$\funcfont{skip-wit-ic}(s, u)$ times, using $s$ and $u$ to resolve
nondeterminism when there are multiple successors to the \M{\ISAICM}
state. Unlike \funcfont{run-ic}, it does not update the \M{\ISAICM}'s
cache prior to execution, only afterwards. This is because the
refinement map includes the cache, meaning that the caches of $w$ and
$s$ are already known to be identical.

Notice that we apply \funcfont{r-a} to states in $S_{\MAGICM}$, despite the
fact that \funcfont{r-a} is defined over members of $S_{\MAICM}$. This is
shorthand for applying \funcfont{r-a} to the $S_{\MAICM}$ component of the
\M{\MAGICM} state.
\begin{align}
  &\La \forall s \in S_{\MAGICM} :: s \skiprel{ic-a} \funcfont{r-a}.s\Ra\label{eqn:wsk1-spectre}\\
  &\abracket{\forall w, s, u \from s \skiprel{ic-a} w \wedge s \xtrans{\textit{ic}} u \from w \xtrans{\textit{ic}}^{\funcfont{skip-wit-ic}(s, u)} \funcfont{run-ic-c}(w, s, u)} \label{eqn:wsk2-spectre}\\
  &\begin{aligned}
  &\forall s,u,w \in S_{\textit{ic}}: s \skiprel{ic} w \And s \xtrans{\textit{ic}} u: &\\
  &\text{\quad } (u \skiprel{ic-a} w \And \funcfont{stutter-wit-ic}(u, w) < \funcfont{stutter-wit-ic}(s, w))\ \vee&\\
  &\text{\quad } u \skiprel{ic-a} (\funcfont{run-ic-c}(w, s, u))&
  \end{aligned}\label{eqn:wsk3-spectre}
\end{align}

\section{Evaluation and Lightweight Verification}
\label{sec:lightweight-verification}

We evaluated our notions of correctness on \M{\MAICM} and \M{\ISAICM}%
% , our bit- and cycle-accurate formal models of a MA and an ISA.
. \M{\MAICM} is vulnerable to both Meltdown and Spectre attacks, so
both of our notions of correctness should be falsified when applied to
\M{\MAICM} and \M{\ISAICM}. We also developed \M{\MAM} and \M{\ISAM},
versions of \M{\MAICM} and \M{\ISAICM} that do not have
\incache. %abstract instruction.
Our notion of correctness for Meltdown is not violated by \M{\MAM},
since it does not allow one to query cache membership in an
architecturally-visible way.

We developed our models inside of the ACL2 Sedan
(ACL2s)~\cite{dillinger-acl2-sedan, acl2s11}, an extension of the ACL2
theorem prover\cite{acl2-car, acl2-acs, acl2-web}. On top of the
capabilities of ACL2, ACL2s provides a powerful type system via the
defdata data definition framework~\cite{defdata} and the
\texttt{definec} and \texttt{property} forms, which support typed
definitions and properties, and counterexample generation capability
via the cgen framework, which is based on the synergistic
integration of theorem proving, type reasoning and testing~\cite{cgen,
  harsh-fmcad, harsh-dissertation}. 
We used cgen to perform property-based testing of refinement proof
obligations.  This enabled us to find and repair several functional
correctness bugs, including:
\begin{enumerate}
% \item an RS becoming deadlocked when an instruction depending on
%   register $X$ was issued to it during the same cycle that another RS
%   corresponding to an instruction writing to register $X$ finished
%   executing,
\item RSes becoming deadlocked due to a race condition when an
  instruction's dependencies are forwarded from other RSes,
\item failures to invalidate the ROB and register status file in
  certain situations where they should have been,
\item branch instructions using the wrong PC value to compute the
  relative jump target,
%\item the MA committing ROB lines that should not have been committed
%  after committing a taken branch instruction in that same cycle and
\item differences in how \ISAICM\ and \MAICM\ handled \lstinline{halt} and
  \lstinline{jge} instructions.
\end{enumerate}

Table~\ref{tab:found-bugs} shows for each configuration the number of
functional correctness bugs and TEAs that were found. %We updated the
% models to eliminate the functional correctness bugs and
We exhibited TEA bugs in both buggy configurations.
% After doing so, only TEA bugs
% remained. 
The discovery of TEA bugs in the buggy systems but not in the safe one
suggests that our notions of correctness are useful in distinguishing
TEA-vulnerable MAs.
%Appendix~\ref{sec:obligations-proved} provides
%additional information on which obligations counterexamples were found
%to.

\begin{table}[]
  \caption{The number of discovered functional correctness bugs and TEA bugs across three configurations of machine and notion of correctness.}
  \label{tab:found-bugs}
    \begin{tabular}{lll}
    \hline
    Config. & Func. bugs & TEA bugs        \\
    \hline
      Safe (\M{\MAM} \& \M{\ISAM}) + Melt.    & 18               & 0     \\
      Buggy (\M{\MAICM} \& \M{\ISAICM}) + Melt.   & 5                & 1     \\
      Buggy (\M{\MAICM} \& \M{\ISAICM}) + Spect.  & 0                & 2     \\
    \hline
  \end{tabular}
  % \begin{tabular}{llll}
  %   \hline
  %   Config. & Func. corr. bugs & TEA bugs & Obls. proved \\
  %   \hline
  %   Safe + Melt.    & 18               & 0        &              \\
  %   Buggy + Melt.   & 5                & 1        &              \\
  %   Buggy + Spect.  & 0                & 2        &              \\
  %   \hline
  % \end{tabular}
\end{table}

To perform property-based testing in ACL2s, one must describe what
kinds of values each free variable may take (a ``type'' for each free
variable). We encoded both the \ISAICM\ and \MAICM\ states as types in
ACL2s' defdata data definition framework. Defining a type in defdata's
DSL results in the generation of an \emph{enumerator}---a function
that takes in a natural number and produces an element of that data
type. Enumerators are then used by cgen to generate data for
testing. Something as complicated as the set of \good\ states is not
possible to encode in defdata's DSL, so we used defdata's custom type
facilities instead. We defined a predicate that holds only on MA
states that are \good, as well as an enumerator that generates such
states. Generating \good\ states is fairly straightforward: we
generate an arbitrary MA state, invalidate it, and then run it forward
for a randomly selected (but bounded) number of steps. We also
modified the way in which MA states are generated---instead of the
default approach that generated sparse instruction memories where
instructions were scattered throughout the address space and choosing
an arbitrary program counter, we generate contiguous sequences of
instructions and choose program counter values that are ``close'' to
those sequences.

We benchmarked the execution speed of \M{\ISAICM} and \M{\MAICM} on an
assembly program that performs na{\"i}ve primality testing. On an M4
Apple Silicon processor, \M{\MAICM} executed an average of 46,000
steps per second, whereas \M{\ISAICM} executed an average of 2.3
million steps per second.
%The code for all of the models plus the
%proof obligations and testing code consists of about 8000 lines of
%code, excluding blank lines and comments.
Running all of the proofs and tests for the three configurations shown
in Table~\ref{tab:found-bugs} takes around 30 minutes.
%, and involves
%over 15,000 MA steps.

%\drew{TODO add high-level data on performance here}

\section{Related Work}
\label{sec:related-attacks}

% \paragraph{Related Work}
A number of high-quality survey papers have been published regarding
transient execution attacks. For an overview of Meltdown and Spectre
attacks, we recommend the surveys of Canella \ea~\cite{Canella2018ASE}
and Fiolhais and
Sousa~\cite{transient-execution-computer-architect-perspective}.

Our work uses an approach based on refinement, which has a long
history of being used to specify and reason about implementations of
complex systems. See Abadi and Lamport~\cite{abadi-91} for an early
and influential paper in this area. In 2000, Manolios introduced the
use of refinement for specifying the correctness of pipelined
processors in a way that implies equivalidity of both safety and
liveness properties between the ISA and the MA~\cite{Manolios00}. In
that work, Manolios argues that previous notions of correctness were
insufficient since they were satisfied by machines that were clearly
incorrect, like an MA implementation that never commits an
instruction. The notion of correctness that Manolios proposes, using
Well-founded Equivalence Bisimulation (WEB) and commitment refinement
maps, has many strengths: it is capable of reasoning about pipelined
in-order MAs running arbitrary programs and that feature interrupts
and exceptions is amenable to mechanical verification as it only
involves local reasoning (reasoning about states and their immediate
successors) and is compositional, enabling large and complicated
refinement proofs to be soundly decomposed into an independent
sequence of smaller and simpler ones. Further work by Jain and
Manolios extended the theory of refinement to support skipping while
retaining local and compositional reasoning~\cite{jain2015skipping,
  manolios2019sks}. Note that support for skipping is necessary to
reason about any of the commercial MA designs that were vulnerable to
the original Meltdown attack, as they are multi-issue and allow
multiple instructions to commit in a cycle. These strengths indicate
to us that a refinement-based approach has promise in enabling both
the specification of a global notion of correctness capable of
identifying TEAs and its verification.

Another approach that has been proposed for validating that hardware
is not vulnerable to TEAs is the use of confidentiality
properties. These properties state that if a program behaves
equivalently with respect to ISA semantics when run starting from two
ISA states that differ only in places that are specified to be
confidential, the program should also behave equivalently when run
using MA semantics from corresponding MA states. Notice that a
confidentiality property is not a global notion of correctness, as it
does not require that the observable behavior of an ISA and MA are
equivalent in any way, and it does not apply to all possible
programs. The framework of hardware-software
contracts~\cite{hardware-software-contracts-secure-speculation}
(referred to simply as \emph{contracts} here) provides one way to
express these confidentiality properties as conditional
non-interference properties.

We discuss three works that describe methods for verifying that MA
designs modeled at the RTL level satisfy confidentiality contracts:
Unique Program Execution Checking (UPEC)~\cite{fadiheh-upec-date,
  fadiheh-upec}, LEAVE~\cite{wang-side-channel-verification} and
shadow contracts~\cite{tan-asplos-25}. These works were all evaluated
on several open-source RISC-V MA implementations, and UPEC was able to
identify novel bugs. All three ultimately formulate proof obligations
that do not refer to the ISA, using a mapping technique similar to the
commitment refinement map to calculate the ISA observation from an MA
state. This is sound if the MA is functionally correct with respect to
the ISA, which is an assumption made in each case. Nonetheless, this
highlights that the fact that compliance of hardware to a
confidentiality contract is not a global notion of correctness, as it
can be proven independently of the ISA's semantics.  LEAVE and shadow
contracts both prove properties that hold only on programs satisfying
particular conditions, unlike UPEC and our work. UPEC requires that
the user provide a substantial number of invariants to eliminate
unreachable counterexamples and while LEAVE automatically generates
certain invariants, it is necessary to provide additional invariants
to support even simple OoO designs. Generating invariants is a
challenging task on its own, and both approaches assume the
inductivity of invariants that are considered ``functional'' which is
critical to ensuring that no reachable states are omitted from
consideration during reasoning. This is precisely why we introduced
the notion of \good\ states, which enable us to eliminate many
unreachable counterexamples without the need to develop and verify
design-specific invariants. Like the shadow logic approach, the \good\
state approach requires microarchitects to implement additional
machinery (history information), but unlike the shadow logic paper
(which assumes that the shadow logic is correct) our work explains how
one can test and verify that this additional machinery is correct.
% Similarly, the shadow logic paper
% omits a description of the work needed to validate that the shadow
% logic is implemented correctly. 
Finally, as Tan \ea\ showed, LEAVE would require manually provided
invariants to verify an OoO design and the shadow logic approach was
only shown to scale to small OoO designs for proofs with extremely
limited structure sizes: up to an 8-entry ROB and 16-entry register
file and data memory~\cite{tan-asplos-25}. Compare this to our
approach, which has a 19-entry ROB, 12 registers and a memory of size
up to $2^{32}$ bytes.

Mathure \ea\ use an approach based on stuttering refinement to state a
Spectre invulnerability property and verify that MA models with and
without certain mitigations are vulnerable or invulnerable to Spectre
attacks~\cite{refinement-spectre-invulnerability-verification}. The
property described in their work does not imply functional
correctness, as they use abstraction to factor out parts of the MA and
ISA model under consideration that should behave
identically. Additionally, their approach requires that the designer
provide inductive invariants to eliminate some unreachable states in a
way that is dependent on the property being proven, as opposed to our
notion of \good\ states.
Other interesting works include that of Cabodi \ea\
\cite{model-checking-spec-dependent-security-properties} and the
Pensieve framework~\cite{pensieve}, both of which reason about
abstract models of MAs, enabling the identification of abstract
information leakage pathways and design-phase evaluation of TEA
defenses respectively.

\section{Conclusions and Future Work}
\label{sec:conclusions}
% Mention that we are only considering Meltdown and related attacks, but
% there are many other microprocessor attacks that one should consider
% in a related way to answer questions such as: what is the observer
% model? what abstractions do we need? what's the right specification?
% etc. until we reach a fixpoint that covers all known attacks and
% communities on both sides of the abstraction are in agreement.
% This is a disciplined, formal-methods-based approach to tackling the
% general issue of microprocessor attacks.

We proposed formal notions of microprocessor conformance based on
refinement that can be used to show the absence of transient execution
attacks such as Meltdown and Spectre. We described how we decomposed
each notion of correctness into properties that are more amenable to
automated verification, making use of the novel shared-resource
commitment refinement map and \good\ states. We demonstrated the
effectiveness of our approach by using the ACL2s theorem prover to
define executable MA and ISA models and to construct counterexamples
to the refinement conjecture of correctness, showing how our work can
identify Meltdown and Spectre vulnerabilities in a simple pipelined,
out-of-order MA that supports speculative execution.
%We showed how our
%work can be used to find other Meltdown/Spectre vulnerabilities and
%can find problems with previous work that introduced local mitigations
%for transient execution attacks, which, were later found to introduce
%other exploits.
As far as we know, these are the first global notions of correctness
that address transient execution attacks. For future work we plan to
extend our work to handle other side
channels~\cite{transient-execution-computer-architect-perspective,
  lose-transient-execution-war} and richer observer models that
prevent unwanted leakage of information, while providing a simple
hardware/software interface that gives architects the flexibility to
design performant processors and maintaining a simple programming
model for programmers.

\begin{acks}
  The authors thank John Matthews and Brian Huffman for their support
  of this work. This work was partially funded by the Intel
  Corporation.
\end{acks}

\bibliographystyle{ACM-Reference-Format}
\bibliography{references}


\begin{thebibliography}{39}

%%% the Web address.

\ifx \showCODEN    \undefined \def \showCODEN     #1{\unskip}     \fi
\ifx \showISBNx    \undefined \def \showISBNx     #1{\unskip}     \fi
\ifx \showISBNxiii \undefined \def \showISBNxiii  #1{\unskip}     \fi
\ifx \showISSN     \undefined \def \showISSN      #1{\unskip}     \fi
\ifx \showLCCN     \undefined \def \showLCCN      #1{\unskip}     \fi
\ifx \shownote     \undefined \def \shownote      #1{#1}          \fi
\ifx \showarticletitle \undefined \def \showarticletitle #1{#1}   \fi
\ifx \showURL      \undefined \def \showURL       {\relax}        \fi
% invisible to TeX
\providecommand\bibfield[2]{#2}
\providecommand\bibinfo[2]{#2}
\providecommand\natexlab[1]{#1}
\providecommand\showeprint[2][]{arXiv:#2}

\bibitem[int(2014)]%
        {intel-prefetch}
 \bibinfo{year}{2014}\natexlab{}.
\newblock \bibinfo{title}{Hardware Prefetcher on Intel Processors}.
\newblock
\urldef\tempurl%
\url{https://software.intel.com/content/www/us/en/develop/articles/disclosure-of-hw-prefetcher-control-on-some-intel-processors.html}
\showURL{%
\tempurl}


\bibitem[CVE(2018)]%
        {CVE-2018-3639}
 \bibinfo{year}{2018}\natexlab{}.
\newblock \bibinfo{title}{{CVE}-2018-3639: Speculative Store Bypass}.
\newblock \bibinfo{howpublished}{Available from MITRE}.
\newblock
\urldef\tempurl%
\url{https://cve.mitre.org/cgi-bin/cvename.cgi?name=CVE-2018-3639}
\showURL{%
\tempurl}


\bibitem[Abadi and Lamport(1991)]%
        {abadi-91}
\bibfield{author}{\bibinfo{person}{Mart{\'{\i}}n Abadi} {and}
  \bibinfo{person}{Leslie Lamport}.} \bibinfo{year}{1991}\natexlab{}.
\newblock \showarticletitle{The Existence of Refinement Mappings}.
\newblock \bibinfo{journal}{\emph{Theor. Comput. Sci.}} \bibinfo{volume}{82},
  \bibinfo{number}{2} (\bibinfo{year}{1991}), \bibinfo{pages}{253--284}.
\newblock
\href{https://doi.org/10.1016/0304-3975(91)90224-P}{doi:\nolinkurl{10.1016/0304-3975(91)90224-P}}


\bibitem[Bulck et~al\mbox{.}(2018)]%
        {foreshadow}
\bibfield{author}{\bibinfo{person}{Jo~Van Bulck}, \bibinfo{person}{Marina
  Minkin}, \bibinfo{person}{Ofir Weisse}, \bibinfo{person}{Daniel Genkin},
  \bibinfo{person}{Baris Kasikci}, \bibinfo{person}{Frank Piessens},
  \bibinfo{person}{Mark Silberstein}, \bibinfo{person}{Thomas~F. Wenisch},
  \bibinfo{person}{Yuval Yarom}, {and} \bibinfo{person}{Raoul Strackx}.}
  \bibinfo{year}{2018}\natexlab{}.
\newblock \showarticletitle{Foreshadow: Extracting the Keys to the Intel {SGX}
  Kingdom with Transient Out-of-Order Execution}. In
  \bibinfo{booktitle}{\emph{27th {USENIX} Security Symposium, {USENIX} Security
  2018}}, \bibfield{editor}{\bibinfo{person}{William Enck} {and}
  \bibinfo{person}{Adrienne~Porter Felt}} (Eds.). \bibinfo{publisher}{{USENIX}
  Association}, \bibinfo{pages}{991--1008}.
\newblock
\urldef\tempurl%
\url{https://www.usenix.org/conference/usenixsecurity18/presentation/bulck}
\showURL{%
\tempurl}


\bibitem[Cabodi et~al\mbox{.}(2019)]%
        {model-checking-spec-dependent-security-properties}
\bibfield{author}{\bibinfo{person}{Gianpiero Cabodi}, \bibinfo{person}{Paolo
  Camurati}, \bibinfo{person}{Fabrizio~F. Finocchiaro}, {and}
  \bibinfo{person}{Danilo Vendraminetto}.} \bibinfo{year}{2019}\natexlab{}.
\newblock \showarticletitle{Model Checking Speculation-Dependent Security
  Properties: Abstracting and Reducing Processor Models for Sound and Complete
  Verification}. In \bibinfo{booktitle}{\emph{Codes, Cryptology and Information
  Security - Third International Conference, {C2SI} 2019, Proceedings - In
  Honor of Said El Hajji}} \emph{(\bibinfo{series}{Lecture Notes in Computer
  Science}, Vol.~\bibinfo{volume}{11445})},
  \bibfield{editor}{\bibinfo{person}{Claude Carlet}, \bibinfo{person}{Sylvain
  Guilley}, \bibinfo{person}{Abderrahmane Nitaj}, {and}
  \bibinfo{person}{El~Mamoun Souidi}} (Eds.). \bibinfo{publisher}{Springer},
  \bibinfo{pages}{462--479}.
\newblock
\href{https://doi.org/10.1007/978-3-030-16458-4\_27}{doi:\nolinkurl{10.1007/978-3-030-16458-4\_27}}


\bibitem[Canella et~al\mbox{.}(2019a)]%
        {canella2019fallout}
\bibfield{author}{\bibinfo{person}{Claudio Canella}, \bibinfo{person}{Daniel
  Genkin}, \bibinfo{person}{Lukas Giner}, \bibinfo{person}{Daniel Gruss},
  \bibinfo{person}{Moritz Lipp}, \bibinfo{person}{Marina Minkin},
  \bibinfo{person}{Daniel Moghimi}, \bibinfo{person}{Frank Piessens},
  \bibinfo{person}{Michael Schwarz}, \bibinfo{person}{Berk Sunar},
  {et~al\mbox{.}}} \bibinfo{year}{2019}\natexlab{a}.
\newblock \showarticletitle{Fallout: Leaking data on meltdown-resistant cpus}.
  In \bibinfo{booktitle}{\emph{Proceedings of the 2019 ACM SIGSAC Conference on
  Computer and Communications Security}}. \bibinfo{pages}{769--784}.
\newblock


\bibitem[Canella et~al\mbox{.}(2019b)]%
        {Canella2018ASE}
\bibfield{author}{\bibinfo{person}{Claudio Canella}, \bibinfo{person}{Jo
  Van~Bulck}, \bibinfo{person}{Michael Schwarz}, \bibinfo{person}{Moritz Lipp},
  \bibinfo{person}{Benjamin Von~Berg}, \bibinfo{person}{Philipp Ortner},
  \bibinfo{person}{Frank Piessens}, \bibinfo{person}{Dmitry Evtyushkin}, {and}
  \bibinfo{person}{Daniel Gruss}.} \bibinfo{year}{2019}\natexlab{b}.
\newblock \showarticletitle{A systematic evaluation of transient execution
  attacks and defenses}. In \bibinfo{booktitle}{\emph{28th $\{$USENIX$\}$
  Security Symposium ($\{$USENIX$\}$ Security 19)}}. \bibinfo{pages}{249--266}.
\newblock


\bibitem[Chamarthi et~al\mbox{.}(2011b)]%
        {acl2s11}
\bibfield{author}{\bibinfo{person}{Harsh Chamarthi}, \bibinfo{person}{Peter~C.
  Dillinger}, \bibinfo{person}{Panagiotis Manolios}, {and}
  \bibinfo{person}{Daron Vroon}.} \bibinfo{year}{2011}\natexlab{b}.
\newblock \showarticletitle{The "{ACL2}" Sedan Theorem Proving System}. In
  \bibinfo{booktitle}{\emph{Tools and Algorithms for the Construction and
  Analysis of Systems (TACAS)}}.
\newblock
\href{https://doi.org/10.1007/978-3-642-19835-9\_27}{doi:\nolinkurl{10.1007/978-3-642-19835-9\_27}}


\bibitem[Chamarthi(2016)]%
        {harsh-dissertation}
\bibfield{author}{\bibinfo{person}{Harsh~Raju Chamarthi}.}
  \bibinfo{year}{2016}\natexlab{}.
\newblock \emph{\bibinfo{title}{Interactive Non-theorem Disproving}}.
\newblock \bibinfo{thesistype}{Ph.\,D. Dissertation}.
  \bibinfo{school}{Northeastern University}.
\newblock
\href{https://doi.org/10.17760/D20467205}{doi:\nolinkurl{10.17760/D20467205}}


\bibitem[Chamarthi et~al\mbox{.}(2011a)]%
        {cgen}
\bibfield{author}{\bibinfo{person}{Harsh~Raju Chamarthi},
  \bibinfo{person}{Peter~C. Dillinger}, \bibinfo{person}{Matt Kaufmann}, {and}
  \bibinfo{person}{Panagiotis Manolios}.} \bibinfo{year}{2011}\natexlab{a}.
\newblock \showarticletitle{Integrating Testing and Interactive Theorem
  Proving}. In \bibinfo{booktitle}{\emph{International Workshop on the {ACL2}
  Theorem Prover and its Applications}} \emph{(\bibinfo{series}{{EPTCS}})}.
\newblock
\href{https://doi.org/10.4204/EPTCS.70.1}{doi:\nolinkurl{10.4204/EPTCS.70.1}}


\bibitem[Chamarthi et~al\mbox{.}(2014)]%
        {defdata}
\bibfield{author}{\bibinfo{person}{Harsh~Raju Chamarthi},
  \bibinfo{person}{Peter~C. Dillinger}, {and} \bibinfo{person}{Panagiotis
  Manolios}.} \bibinfo{year}{2014}\natexlab{}.
\newblock \showarticletitle{Data Definitions in the {ACL2} Sedan}. In
  \bibinfo{booktitle}{\emph{Proceedings Twelfth International Workshop on the
  {ACL2} Theorem Prover and its Applications}}
  \emph{(\bibinfo{series}{{EPTCS}})}.
\newblock
\href{https://doi.org/10.4204/EPTCS.152.3}{doi:\nolinkurl{10.4204/EPTCS.152.3}}


\bibitem[Chamarthi and Manolios(2011)]%
        {harsh-fmcad}
\bibfield{author}{\bibinfo{person}{Harsh~Raju Chamarthi} {and}
  \bibinfo{person}{Panagiotis Manolios}.} \bibinfo{year}{2011}\natexlab{}.
\newblock \showarticletitle{Automated specification analysis using an
  interactive theorem prover}. In \bibinfo{booktitle}{\emph{International
  Conference on Formal Methods in Computer-Aided Design, {FMCAD} '11}},
  \bibfield{editor}{\bibinfo{person}{Per Bjesse} {and} \bibinfo{person}{Anna
  Slobodov{\'{a}}}} (Eds.). \bibinfo{publisher}{{FMCAD} Inc.},
  \bibinfo{pages}{46--53}.
\newblock
\urldef\tempurl%
\url{https://dl.acm.org/doi/10.5555/2157654.2157665}
\showURL{%
\tempurl}


\bibitem[Dillinger et~al\mbox{.}(2007)]%
        {dillinger-acl2-sedan}
\bibfield{author}{\bibinfo{person}{Peter~C. Dillinger},
  \bibinfo{person}{Panagiotis Manolios}, \bibinfo{person}{Daron Vroon}, {and}
  \bibinfo{person}{J.~Strother Moore}.} \bibinfo{year}{2007}\natexlab{}.
\newblock \showarticletitle{{ACL2s}: ``The {ACL2} Sedan''}. In
  \bibinfo{booktitle}{\emph{Proceedings of the 7th Workshop on User Interfaces
  for Theorem Provers (UITP 2006)}} \emph{(\bibinfo{series}{Electronic Notes in
  Theoretical Computer Science})}.
\newblock
\href{https://doi.org/10.1016/j.entcs.2006.09.018}{doi:\nolinkurl{10.1016/j.entcs.2006.09.018}}


\bibitem[Fadiheh(2022)]%
        {fadiheh-upec}
\bibfield{author}{\bibinfo{person}{Mohammad~Rahmani Fadiheh}.}
  \bibinfo{year}{2022}\natexlab{}.
\newblock \emph{\bibinfo{title}{Unique Program Execution Checking: A Novel
  Approach for Formal Security Analysis of Hardware}}.
\newblock \bibinfo{thesistype}{Ph.\,D. Dissertation}.
  \bibinfo{school}{Technische Universit{\"a}t Kaiserslautern}.
\newblock
\href{https://doi.org/10.26204/KLUEDO/6930}{doi:\nolinkurl{10.26204/KLUEDO/6930}}


\bibitem[Fadiheh et~al\mbox{.}(2019)]%
        {fadiheh-upec-date}
\bibfield{author}{\bibinfo{person}{Mohammad~Rahmani Fadiheh},
  \bibinfo{person}{Dominik Stoffel}, \bibinfo{person}{Clark~W. Barrett},
  \bibinfo{person}{Subhasish Mitra}, {and} \bibinfo{person}{Wolfgang Kunz}.}
  \bibinfo{year}{2019}\natexlab{}.
\newblock \showarticletitle{Processor Hardware Security Vulnerabilities and
  their Detection by Unique Program Execution Checking}. In
  \bibinfo{booktitle}{\emph{Design, Automation {\&} Test in Europe Conference
  {\&} Exhibition, {DATE} 2019, Florence, Italy, March 25-29, 2019}},
  \bibfield{editor}{\bibinfo{person}{J{\"{u}}rgen Teich} {and}
  \bibinfo{person}{Franco Fummi}} (Eds.). \bibinfo{publisher}{{IEEE}},
  \bibinfo{pages}{994--999}.
\newblock
\href{https://doi.org/10.23919/DATE.2019.8715004}{doi:\nolinkurl{10.23919/DATE.2019.8715004}}


\bibitem[Fiolhais and Sousa(2024)]%
        {transient-execution-computer-architect-perspective}
\bibfield{author}{\bibinfo{person}{Lu{\'{\i}}s Fiolhais} {and}
  \bibinfo{person}{Leonel Sousa}.} \bibinfo{year}{2024}\natexlab{}.
\newblock \showarticletitle{Transient-Execution Attacks: {A} Computer Architect
  Perspective}.
\newblock \bibinfo{journal}{\emph{{ACM} Comput. Surv.}} \bibinfo{volume}{56},
  \bibinfo{number}{3} (\bibinfo{year}{2024}), \bibinfo{pages}{74:1--74:38}.
\newblock
\href{https://doi.org/10.1145/3603619}{doi:\nolinkurl{10.1145/3603619}}


\bibitem[Guarnieri et~al\mbox{.}(2021)]%
        {hardware-software-contracts-secure-speculation}
\bibfield{author}{\bibinfo{person}{Marco Guarnieri}, \bibinfo{person}{Boris
  K{\"{o}}pf}, \bibinfo{person}{Jan Reineke}, {and} \bibinfo{person}{Pepe
  Vila}.} \bibinfo{year}{2021}\natexlab{}.
\newblock \showarticletitle{Hardware-Software Contracts for Secure
  Speculation}. In \bibinfo{booktitle}{\emph{42nd {IEEE} Symposium on Security
  and Privacy, {SP} 2021}}. \bibinfo{publisher}{{IEEE}},
  \bibinfo{pages}{1868--1883}.
\newblock
\href{https://doi.org/10.1109/SP40001.2021.00036}{doi:\nolinkurl{10.1109/SP40001.2021.00036}}


\bibitem[Hennessy and Patterson(2011)]%
        {hennessy2011computer}
\bibfield{author}{\bibinfo{person}{John~L Hennessy} {and}
  \bibinfo{person}{David~A Patterson}.} \bibinfo{year}{2011}\natexlab{}.
\newblock \bibinfo{booktitle}{\emph{Computer architecture: a quantitative
  approach}}.
\newblock \bibinfo{publisher}{Elsevier}.
\newblock


\bibitem[{Intel Corporation}(2016)]%
        {inteloptimize}
\bibfield{author}{\bibinfo{person}{{Intel Corporation}}.}
  \bibinfo{year}{2016}\natexlab{}.
\newblock \bibinfo{booktitle}{\emph{{Intel\textsuperscript{\textregistered} 64
  and IA-32 Architectures Optimization Reference Manual}}}.
\newblock \bibinfo{publisher}{{Intel Corporation}}.
\newblock


\bibitem[Jain and Manolios(2015)]%
        {jain2015skipping}
\bibfield{author}{\bibinfo{person}{Mitesh Jain} {and}
  \bibinfo{person}{Panagiotis Manolios}.} \bibinfo{year}{2015}\natexlab{}.
\newblock \showarticletitle{Skipping refinement}. In
  \bibinfo{booktitle}{\emph{International Conference on Computer Aided
  Verification}}. Springer, \bibinfo{pages}{103--119}.
\newblock


\bibitem[Jain and Manolios(2019)]%
        {manolios2019sks}
\bibfield{author}{\bibinfo{person}{Mitesh Jain} {and}
  \bibinfo{person}{Panagiotis Manolios}.} \bibinfo{year}{2019}\natexlab{}.
\newblock \showarticletitle{Local and Compositional Reasoning for Optimized
  Reactive Systems}. In \bibinfo{booktitle}{\emph{CAV}}.
\newblock


\bibitem[Kaufmann et~al\mbox{.}(2000a)]%
        {acl2-car}
\bibfield{author}{\bibinfo{person}{Matt Kaufmann}, \bibinfo{person}{Panagiotis
  Manolios}, {and} \bibinfo{person}{J~Strother Moore}.}
  \bibinfo{year}{2000}\natexlab{a}.
\newblock \bibinfo{booktitle}{\emph{Computer-Aided Reasoning: An Approach}}.
\newblock \bibinfo{publisher}{Kluwer Academic Publishers}.
\newblock
\href{https://doi.org/10.1007/978-1-4615-4449-4}{doi:\nolinkurl{10.1007/978-1-4615-4449-4}}


\bibitem[Kaufmann et~al\mbox{.}(2000b)]%
        {acl2-acs}
\bibfield{author}{\bibinfo{person}{Matt Kaufmann}, \bibinfo{person}{Panagiotis
  Manolios}, {and} \bibinfo{person}{J~Strother Moore}.}
  \bibinfo{year}{2000}\natexlab{b}.
\newblock \bibinfo{booktitle}{\emph{Computer-Aided Reasoning: Case Studies}}.
\newblock \bibinfo{publisher}{Kluwer Academic Publishers}.
\newblock
\href{https://doi.org/10.1007/978-1-4757-3188-0}{doi:\nolinkurl{10.1007/978-1-4757-3188-0}}


\bibitem[Kaufmann and Moore(2025)]%
        {acl2-web}
\bibfield{author}{\bibinfo{person}{Matt Kaufmann} {and}
  \bibinfo{person}{J~Strother Moore}.} \bibinfo{year}{2025}\natexlab{}.
\newblock \showarticletitle{{ACL2 homepage}}.
\newblock  (\bibinfo{year}{2025}).
\newblock
\urldef\tempurl%
\url{https://www.cs.utexas.edu/users/moore/acl2/}
\showURL{%
\tempurl}


\bibitem[Kocher et~al\mbox{.}(2019)]%
        {Kocher2018spectre}
\bibfield{author}{\bibinfo{person}{Paul Kocher}, \bibinfo{person}{Jann Horn},
  \bibinfo{person}{Anders Fogh}, \bibinfo{person}{}, \bibinfo{person}{Daniel
  Genkin}, \bibinfo{person}{Daniel Gruss}, \bibinfo{person}{Werner Haas},
  \bibinfo{person}{Mike Hamburg}, \bibinfo{person}{Moritz Lipp},
  \bibinfo{person}{Stefan Mangard}, \bibinfo{person}{Thomas Prescher},
  \bibinfo{person}{Michael Schwarz}, {and} \bibinfo{person}{Yuval Yarom}.}
  \bibinfo{year}{2019}\natexlab{}.
\newblock \showarticletitle{Spectre Attacks: Exploiting Speculative Execution}.
  In \bibinfo{booktitle}{\emph{40th IEEE Symposium on Security and Privacy
  (S\&P'19)}}.
\newblock


\bibitem[Lipp et~al\mbox{.}(2018)]%
        {Lipp2018meltdown}
\bibfield{author}{\bibinfo{person}{Moritz Lipp}, \bibinfo{person}{Michael
  Schwarz}, \bibinfo{person}{Daniel Gruss}, \bibinfo{person}{Thomas Prescher},
  \bibinfo{person}{Werner Haas}, \bibinfo{person}{Anders Fogh},
  \bibinfo{person}{Jann Horn}, \bibinfo{person}{Stefan Mangard},
  \bibinfo{person}{Paul Kocher}, \bibinfo{person}{Daniel Genkin},
  \bibinfo{person}{Yuval Yarom}, {and} \bibinfo{person}{Mike Hamburg}.}
  \bibinfo{year}{2018}\natexlab{}.
\newblock \showarticletitle{Meltdown: Reading Kernel Memory from User Space}.
  In \bibinfo{booktitle}{\emph{27th {USENIX} Security Symposium ({USENIX}
  Security 18)}}.
\newblock


\bibitem[Mambretti et~al\mbox{.}(2020)]%
        {mambretti2020bypassing}
\bibfield{author}{\bibinfo{person}{Andrea Mambretti},
  \bibinfo{person}{Alexandra Sandulescu}, \bibinfo{person}{Alessandro
  Sorniotti}, \bibinfo{person}{Wil Robertson}, \bibinfo{person}{Engin Kirda},
  {and} \bibinfo{person}{Anil Kurmus}.} \bibinfo{year}{2020}\natexlab{}.
\newblock \showarticletitle{Bypassing memory safety mechanisms through
  speculative control flow hijacks}.
\newblock \bibinfo{journal}{\emph{arXiv preprint arXiv:2003.05503}}
  (\bibinfo{year}{2020}).
\newblock


\bibitem[Manolios(2000)]%
        {Manolios00}
\bibfield{author}{\bibinfo{person}{Panagiotis Manolios}.}
  \bibinfo{year}{2000}\natexlab{}.
\newblock \showarticletitle{Correctness of Pipelined Machines}. In
  \bibinfo{booktitle}{\emph{Formal Methods in Computer-Aided Design, Third
  International Conference, {FMCAD} 2000, Proceedings}}.
  \bibinfo{pages}{161--178}.
\newblock
\href{https://doi.org/10.1007/3-540-40922-X\_11}{doi:\nolinkurl{10.1007/3-540-40922-X\_11}}


\bibitem[Mathure et~al\mbox{.}(2022)]%
        {refinement-spectre-invulnerability-verification}
\bibfield{author}{\bibinfo{person}{Nimish Mathure},
  \bibinfo{person}{Sudarshan~K. Srinivasan}, {and} \bibinfo{person}{Kushal~K.
  Ponugoti}.} \bibinfo{year}{2022}\natexlab{}.
\newblock \showarticletitle{A Refinement-Based Approach to Spectre
  Invulnerability Verification}.
\newblock \bibinfo{journal}{\emph{{IEEE} Access}}  \bibinfo{volume}{10}
  (\bibinfo{year}{2022}), \bibinfo{pages}{80949--80957}.
\newblock
\href{https://doi.org/10.1109/ACCESS.2022.3195508}{doi:\nolinkurl{10.1109/ACCESS.2022.3195508}}


\bibitem[Mittal(2016)]%
        {mittal2016survey}
\bibfield{author}{\bibinfo{person}{Sparsh Mittal}.}
  \bibinfo{year}{2016}\natexlab{}.
\newblock \showarticletitle{A survey of recent prefetching techniques for
  processor caches}.
\newblock \bibinfo{journal}{\emph{ACM Computing Surveys (CSUR)}}
  \bibinfo{volume}{49}, \bibinfo{number}{2} (\bibinfo{year}{2016}),
  \bibinfo{pages}{1--35}.
\newblock


\bibitem[Patterson and Hennessy(2013)]%
        {comp-org-design}
\bibfield{author}{\bibinfo{person}{David~A. Patterson} {and}
  \bibinfo{person}{John~L. Hennessy}.} \bibinfo{year}{2013}\natexlab{}.
\newblock \bibinfo{booktitle}{\emph{Computer Organization and Design, Fifth
  Edition: The Hardware/Software Interface} (\bibinfo{edition}{5th} ed.)}.
\newblock \bibinfo{publisher}{Morgan Kaufmann Publishers Inc.},
  \bibinfo{address}{San Francisco, CA, USA}.
\newblock
\showISBNx{0124077269}


\bibitem[Randal(2023)]%
        {lose-transient-execution-war}
\bibfield{author}{\bibinfo{person}{Allison Randal}.}
  \bibinfo{year}{2023}\natexlab{}.
\newblock \showarticletitle{This is How You Lose the Transient Execution War}.
\newblock \bibinfo{journal}{\emph{CoRR}}  \bibinfo{volume}{abs/2309.03376}
  (\bibinfo{year}{2023}).
\newblock
\href{https://doi.org/10.48550/ARXIV.2309.03376}{doi:\nolinkurl{10.48550/ARXIV.2309.03376}}
\showeprint[arXiv]{2309.03376}


\bibitem[Schwarz et~al\mbox{.}(2019)]%
        {schwarz2019zombieload}
\bibfield{author}{\bibinfo{person}{Michael Schwarz}, \bibinfo{person}{Moritz
  Lipp}, \bibinfo{person}{Daniel Moghimi}, \bibinfo{person}{Jo Van~Bulck},
  \bibinfo{person}{Julian Stecklina}, \bibinfo{person}{Thomas Prescher}, {and}
  \bibinfo{person}{Daniel Gruss}.} \bibinfo{year}{2019}\natexlab{}.
\newblock \showarticletitle{ZombieLoad: Cross-privilege-boundary data
  sampling}. In \bibinfo{booktitle}{\emph{Proceedings of the 2019 ACM SIGSAC
  Conference on Computer and Communications Security}}.
  \bibinfo{pages}{753--768}.
\newblock


\bibitem[Tan et~al\mbox{.}(2025)]%
        {tan-asplos-25}
\bibfield{author}{\bibinfo{person}{Qinhan Tan}, \bibinfo{person}{Yuheng Yang},
  \bibinfo{person}{Thomas Bourgeat}, \bibinfo{person}{Sharad Malik}, {and}
  \bibinfo{person}{Mengjia Yan}.} \bibinfo{year}{2025}\natexlab{}.
\newblock \showarticletitle{{RTL} Verification for Secure Speculation Using
  Contract Shadow Logic}. In \bibinfo{booktitle}{\emph{Proceedings of the 30th
  {ACM} International Conference on Architectural Support for Programming
  Languages and Operating Systems, Volume 1, {ASPLOS} 2025}},
  \bibfield{editor}{\bibinfo{person}{Lieven Eeckhout},
  \bibinfo{person}{Georgios Smaragdakis}, \bibinfo{person}{Kaitai Liang},
  \bibinfo{person}{Adrian Sampson}, \bibinfo{person}{Martha~A. Kim}, {and}
  \bibinfo{person}{Christopher~J. Rossbach}} (Eds.).
  \bibinfo{publisher}{{ACM}}, \bibinfo{pages}{970--986}.
\newblock
\href{https://doi.org/10.1145/3669940.3707243}{doi:\nolinkurl{10.1145/3669940.3707243}}


\bibitem[Tomasulo(1967)]%
        {tomasulo1967efficient}
\bibfield{author}{\bibinfo{person}{Robert~M Tomasulo}.}
  \bibinfo{year}{1967}\natexlab{}.
\newblock \showarticletitle{An efficient algorithm for exploiting multiple
  arithmetic units}.
\newblock \bibinfo{journal}{\emph{IBM Journal of research and Development}}
  \bibinfo{volume}{11}, \bibinfo{number}{1} (\bibinfo{year}{1967}),
  \bibinfo{pages}{25--33}.
\newblock


\bibitem[Van~Bulck et~al\mbox{.}(2020)]%
        {van2020lvi}
\bibfield{author}{\bibinfo{person}{Jo Van~Bulck}, \bibinfo{person}{Daniel
  Moghimi}, \bibinfo{person}{Michael Schwarz}, \bibinfo{person}{Moritz Lipp},
  \bibinfo{person}{Marina Minkin}, \bibinfo{person}{Daniel Genkin},
  \bibinfo{person}{Yarom Yuval}, \bibinfo{person}{Berk Sunar},
  \bibinfo{person}{Daniel Gruss}, {and} \bibinfo{person}{Frank Piessens}.}
  \bibinfo{year}{2020}\natexlab{}.
\newblock \showarticletitle{LVI: Hijacking transient execution through
  microarchitectural load value injection}. In \bibinfo{booktitle}{\emph{41th
  IEEE Symposium on Security and Privacy (S\&P’20)}}.
  \bibinfo{pages}{1399--1417}.
\newblock


\bibitem[Walter et~al\mbox{.}(2025)]%
        {artifact-repo}
\bibfield{author}{\bibinfo{person}{Andrew~T. Walter},
  \bibinfo{person}{Konstantinos Athanasiou}, {and} \bibinfo{person}{Panagiotis
  Manolios}.} \bibinfo{year}{2025}\natexlab{}.
\newblock \bibinfo{booktitle}{\emph{Global Microprocessor Correctness in the
  Presence of Transient Execution - Supporting Material}}.
\newblock
\href{https://doi.org/10.5281/zenodo.15706553}{doi:\nolinkurl{10.5281/zenodo.15706553}}


\bibitem[Wang et~al\mbox{.}(2023)]%
        {wang-side-channel-verification}
\bibfield{author}{\bibinfo{person}{Zilong Wang}, \bibinfo{person}{Gideon Mohr},
  \bibinfo{person}{Klaus von Gleissenthall}, \bibinfo{person}{Jan Reineke},
  {and} \bibinfo{person}{Marco Guarnieri}.} \bibinfo{year}{2023}\natexlab{}.
\newblock \showarticletitle{Specification and Verification of Side-channel
  Security for Open-source Processors via Leakage Contracts}. In
  \bibinfo{booktitle}{\emph{Proceedings of the 2023 {ACM} {SIGSAC} Conference
  on Computer and Communications Security, {CCS} 2023}},
  \bibfield{editor}{\bibinfo{person}{Weizhi Meng},
  \bibinfo{person}{Christian~Damsgaard Jensen}, \bibinfo{person}{Cas Cremers},
  {and} \bibinfo{person}{Engin Kirda}} (Eds.). \bibinfo{publisher}{{ACM}},
  \bibinfo{pages}{2128--2142}.
\newblock
\href{https://doi.org/10.1145/3576915.3623192}{doi:\nolinkurl{10.1145/3576915.3623192}}


\bibitem[Yang et~al\mbox{.}(2023)]%
        {pensieve}
\bibfield{author}{\bibinfo{person}{Yuheng Yang}, \bibinfo{person}{Thomas
  Bourgeat}, \bibinfo{person}{Stella Lau}, {and} \bibinfo{person}{Mengjia
  Yan}.} \bibinfo{year}{2023}\natexlab{}.
\newblock \showarticletitle{Pensieve: Microarchitectural Modeling for Security
  Evaluation}. In \bibinfo{booktitle}{\emph{Proceedings of the 50th Annual
  International Symposium on Computer Architecture, {ISCA} 2023}},
  \bibfield{editor}{\bibinfo{person}{Yan Solihin} {and}
  \bibinfo{person}{Mark~A. Heinrich}} (Eds.). \bibinfo{publisher}{{ACM}},
  \bibinfo{pages}{59:1--59:15}.
\newblock
\href{https://doi.org/10.1145/3579371.3589094}{doi:\nolinkurl{10.1145/3579371.3589094}}


\end{thebibliography}

\appendix

\section{Notation}
\label{sec:notation}

Given a nonempty and finite set of integers $S$, $\max(S)$ is the
maximum element in $S$ and $\min(S)$ is the minimum element in
$S$. Given a set $S$, $\powerset{S}$ denotes the power set of $S$ (the
set of all subsets of $S$, including $\emptyset$ and $S$). $\uplus$
denotes disjoint union.

$f : A \rightharpoonup B$ indicates that $f$ is a partial function
from $A$ to $B$ (\eg{} $\fndom(f) \subseteq A$), whereas
$f : A \rightarrow B$ indicates that $f$ is a total function from $A$
to $B$. Given a partial function $f$, $f(x) \uparrow$ indicates that
$a \notin \fndom(f)$ ($a$ is not mapped by $f$) and $f(x) \downarrow$
indicates that $a \in \fndom(f)$. Given any partial function
$f : A \rightharpoonup B$, we define
$\funcfont{get}_f : A \times B \rightarrow B$ such that:
\[
  \partialget{f}{a, b} = \begin{cases}
    f(a) &\text{if } f(a)\downarrow\\
    b &\text{otherwise}
  \end{cases}
\]

A partial function $f : A \rightharpoonup B$ can be treated as a
subset of $A \times B$. This set representation is defined as follows:

\[
  \abracket{\forall a \from a \in A  \from f(a) \downarrow \implies (a, f(a)) \in f}
\]
\[
  \abracket{\forall a \from a \in A \from f(a) \uparrow \implies a \notin f}
\]

A set $s \subseteq \powerset{A \times B}$ can be treated as a partial
function $f : A \rightharpoonup B$ if $s$ satisfies the following
condition:
\[
  \abracket{\forall a \from a \in A \from \vert \{ (x, y) \in s \from x = a \} \vert \leq 1}
\]

If so, the semantics of $f$ are defined as follows:
\[
  \abracket{\forall a, b \from (a, b) \in s \from f(a) \downarrow \wedge f(a) = b}
\]
\[
  \abracket{\forall a \from a \in A \wedge \neg\abracket{\exists b \from b \in B \from (a, b) \in s} \from f(a) \uparrow}
\]

Given a partial function $f : A \rightharpoonup B$ and
$a \in A, b \in B$, $[a \mapsto b]f$ denotes a partial function
$f^\prime$ such that
\[
  f^\prime(x) = \begin{cases}
    b &\text{if } x = a\\
    f(x) &\text{otherwise}
  \end{cases}
\]

Given a partial function $f : A \rightharpoonup B$ and $a \in A$,
$[a \mapsto \uparrow]f$ denotes a partial function $f^\prime$ such that
\[
  f^\prime(x) = \begin{cases}
    f &\text{if } f(a)\uparrow\\
    f \setminus (a, f(a)) &\text{otherwise}
  \end{cases}
\]

Given a function $f : A \rightarrow B$, or a partial function
$f : A \rightharpoonup B$, $\Ima f$ is the set
$\{ y \vert y \in B \wedge \abracket{\exists x \from x \in A \from
  f(x) = y} \}$.

A sequence of elements of a set $A$ is a function from an interval of
the natural numbers to $A$. In this work, any finite sequence we
consider has a domain of the form
$\{ i \from i \in \nats \from 0 < i \leq j \}$ for some $j \in
\nats$. Given a finite sequence $\sigma$, the length of $\sigma$
(denoted by $\vert \sigma \vert$) is the cardinality of the domain of
$\sigma$.

\sequence{x, y} denotes the finite sequence $\sigma$ such that
$\fndom(\sigma) = \{1, 2\}$, $\sigma(1) = x$ and $\sigma(2) = y$.
Given a set $A$, $A^\ast$ denotes the set of all finite sequences over
elements of $A$.  Given a value $e$ and a sequence $a$, $e \bullet a$
denotes the sequence obtained by prepending $e$ onto the
sequence. That is, if $s = e \bullet a$ then:
\[
  s(i) = \begin{cases}
    e &\text{if } i = 1\\
    a(i-1) &\text{otherwise}
  \end{cases}
\]
Given two finite sequences $a$ and $b$, $a \mapp b$ denotes the
sequence obtained by appending the two sequences.
% Given a sequence of
% sequences $\sigma$, $\bigmapp_{\sigma}$ denotes the result of
% appending all of the sequences in $\sigma$ together in order.

Given a sequence $\sigma$, \sequence{f(x) \from x \in \sigma \from
  p(x)} denotes the sequence consisting of $f$ applied to the elements
of $\Ima \sigma$ that satisfy $p$, in the order in which they appeared
in $\sigma$. That is, if $C = \{ (i, x) \in \sigma \vert p(x) \}$ and
$\pi = \sequence{f(x) \from x \in \sigma \from p(x)}$, then
$\abracket{\forall i, x \from (i, x) \in C \from \pi(\vert \{ (j, y)
  \in C \vert j \leq i \} \vert) = f(x)}$.

% Given two sequences $\sigma$ and $\pi$ with the same domain,
% $\sigma \triangledown \pi$ denotes the sequence produced by
% ``zipping'' together corresponding elements of the two sequences. That
% is,
% \[
%   \abracket{\forall i \from i \in \fndom(\sigma) \from (\sigma \triangledown \pi)(i)(1) 
% \]

$\bools = \{\boolt, \boolf\}$ is the set of Boolean
values. $\mathbb{N}_{32} = \{ x \in \mathbb{N} \from x < 2^{32} \}
$. That is, $\mathbb{N}_{32}$ is the set of all unsigned 32-bit
integers. $\plusc$ indicates unsigned 32-bit addition, $\minusc$
indicates unsigned 32-bit subtraction and $\timesc$ indicates unsigned
32-bit multiplication. $\andc$ indicates the bitwise AND operator
applied to two unsigned 32-bit numbers. Let \regs\ be the set of
register specifiers. A register file is a function
$\regs \rightarrow \natc$. The initial register file $\emptyrf$ maps
all registers to 0.

We define transition relations for TSes by providing \emph{inference
  rules}. Each inference rule consists of two parts: a set of
\emph{premises} and a \emph{conclusion}. An inference rule indicates
that when all of its premises hold, the conclusion must also hold. An
inference rule is represented as a whitespace-separated sequence of
premises written above a horizontal line, with the conclusion written
below the horizontal line. An example is given in
Equation~\ref{eqn:inference-rule-example}. In that example, the
premises are $A$, $\neg B$ and $C$ and the conclusion is $D$.

\begin{equation}
\label{eqn:inference-rule-example}
\inferrule{A \\ \neg B \\ C}{D}
\end{equation}

We refer to an inference rule that is used to define a transition
relation as a \emph{transition rule}. The conclusion of a transition
rule will always be an application of a transition
relation. Equation~\ref{eqn:transition-rule-example} provides an
example of a transition rule for a transition system
\trs{\textit{foo}} and a function
$f: S_\textit{foo} \rightarrow
S_\textit{foo}$. Equation~\ref{eqn:transition-rule-example} indicates
that
$\abracket{\forall s \in S_\textit{foo} \from P(s) \wedge \neg Q(s)
  \from (s, f(s)) \in\ \xtrans{\textit{foo}}}$. Note that
Equation~\ref{eqn:transition-rule-example} elides an explicit
definition of the domain of the variable $s$; any variables on the
left-hand side of the transition relation in the conclusion of a
transition rule are inferred to have domains that are appropriate
given the domain of the transition relation. For example, given a
transition system \trs{\textit{qux}} where
$S_{\textit{qux}} = \nats \times \bools$ and
$g: \nats \rightarrow \nats$,
Equation~\ref{eqn:transition-rule-tuple-example} indicates that
$\abracket{\forall x \in \nats, y \in \bools \from U(x) \wedge \neg
  V(y) \from (\abracket{x, y}, \abracket{g(x), y}) \in\
  \xtrans{\textit{qux}}}$. 

\begin{equation}
\label{eqn:transition-rule-example}
  \inferrule{P(s) \\ \neg Q(s)}{s \xtrans{\textit{foo}} f(s)}
\end{equation}

\begin{equation}
\label{eqn:transition-rule-tuple-example}
\inferrule{U(x) \\ \neg V(y)}{\abracket{x, y} \xtrans{\textit{qux}} \abracket{g(x), y}}
\end{equation}

In transition rules, $r$ refers to a register specifier and $c$ refers
to an unsigned 32-bit number. Subscript indices are used when it is
necessary to introduce multiple register specifiers or constants. We
will define the state space of a transition system as consisting of
tuples, each element of which will have a name associated with it. We
will freely use the names of the tuple elements in transition rules to
refer to the value of that tuple element in the starting state of the
rule. For example, say we have a transition system \trs{\textit{foo}}
where $S_\textit{foo} : \abracket{\textbf{bar}, \textbf{baz}}$ where
$\textbf{bar} : \bools$ and $\textbf{baz}: \mathbb{N}$. Say we have
the following transition rule for $\mathcal{M}_\textit{foo}$:

\[
\inferrule{\neg \textbf{bar} \\ \textbf{baz} > 10}{S \xtrans{\textit{foo}} [\textbf{bar} \mapsto \boolf, \textbf{baz} \mapsto \textbf{baz} + 1]S}
\]

This rule should be read the same as:

\[
\inferrule{\neg x \\ y > 10}{\abracket{x, y} \xtrans{\textit{foo}} \abracket{\boolf, \textit{y} + 1}}
\]

Given a variable $x$ over a named tuple, if \textbf{bar} is the name
of a field in that tuple then $\textbf{bar}_x$ refers to the value of
field \textbf{bar} in $x$.

Given a transition system \trs{\textit{foo}} and two states
$s,u \in S_\textit{foo}$, $s \xtrans{\textit{foo}^\ast} u$ indicates
that there exists a path from $s$ to $u$. That is:
\begin{flalign*}
  s \xtrans{\textit{foo}}^\ast u \iff \La&\exists s_1, ..., s_n \from s_1, ..., s_n \in S_\textit{foo} \from s_1 = s \wedge s_n = u \wedge\\
  &\abracket{\forall i \from i \in [1, ..., n-1] \from s_i \xtrans{\textit{foo}} s_{i+1}}\Ra
\end{flalign*}

\section{Formal Semantics}
\label{sec:semantics}

This section contains the formal definitions of several ISA and MA
machine variants.

\subsection{Formal Semantics of \ISAICM}
\label{sec:isa-ic-semantics}

\trs{\ISAICM} is a transition system. Let $\insts{IC}$ be the set of
instructions that the ISA is defined over.
\begin{flalign*}
  \insts{IC} ::=\ &\texttt{halt}\ \vert\ \texttt{noop}\ \vert\ \texttt{loadi}\ r_d\ c\ \vert\ \texttt{addi}\ r_d\ r_1\ c\ \vert\
                                   \texttt{add}\ r_d\ r_1\ r_2\ \vert \\
                                 & \texttt{mul}\ r_d\ r_1\ r_2\ \vert\ \texttt{and}\ r_d\ r_1\ r_2\ \vert\ \texttt{cmp}\ r_d\ r_1\ r_2\ \vert\ \texttt{jg}\ r_1\ c\ \vert\ \texttt{jge}\ r_1\ c\ \vert\\
                  &\texttt{ldri}\ r_d\ r_1\ c\ \vert\ \texttt{ldr}\ r_d\ r_1\ r_2\ \vert\
                                   \texttt{tsx-start}\ c\ \vert\ \texttt{tsx-end}\ \vert\\
  &\texttt{in-cache}\ r_d \ r_1 \ c
\end{flalign*}
We define the set of \M{\ISAICM} states to be a tuple:
\[
  S_{\ISAICM} : \La\pc, \rfv, \tsx, \halted, \imem, \dmem,
  \goodaddr, \cmem\Ra
\]
where each component is as follows:
\begin{itemize}
  \item $\pc : \natc$ is the program counter
  \item $\rfv : \mathcal{R} \rightarrow \natc$ is the register file
  \item $\tsx : \abracket{\tsxactive, \tsxrf, \tsxfallback}$ is the TSX state, described below
  \item $\halted : \mathbb{B}$ is true if the ISA is halted
  \item $\imem : \natc \rightharpoonup \insts{IC}$ is a partial map from addresses to instructions (the instruction memory)
  \item $\dmem : \natc \rightharpoonup \natc$ is a partial map from addresses to data (the data memory)
  \item $\goodaddr : \natc \rightarrow \bools$ is a predicate that is
    true on any data memory address that the running program has
    permission to access.
  \item $\cmem : \natc \rightharpoonup \natc$ is a partial map from addresses to data (the cache)
  \item $\tsxactive : \bools$ is true if the ISA is in an active TSX region
  \item $\tsxrf : \mathcal{R} \rightarrow \natc$ is the register file at the start of the TSX region
  \item $\tsxfallback : \natc$ is the address to resume execution from if an error occurs in an active TSX region
\end{itemize}
 % \begin{flalign*}
%   &\pc : \natc \ \textrm{is the program counter}\\
%   &\rfv : \mathcal{R} \rightarrow \natc \ \textrm{is the register file}\\
%   &\tsx : \abracket{\tsxactive, \tsxrf, \tsxfallback} \ \textrm{is the TSX state, described below}\\
%   &\halted : \mathbb{B} \ \textrm{is true if the ISA is halted}\\
%   &\imem : \natc \rightharpoonup \mathcal{I} \ \textrm{is a partial map from addresses to instructions (the instruction memory)}\\
%   &\dmem : \natc \rightharpoonup \natc \ \textrm{is a partial map from addresses to data (the data memory)}\\
%   &\goodaddr : \natc \rightarrow \bools \ \textrm{is a predicate that is true
%     on any data memory address that the running program}\\
%   &\textrm{has permission to access.}\\
%   &\tsxactive : \bools \ \textrm{is true if the ISA is in an active TSX region}\\
%   &\tsxrf : \mathcal{R} \rightarrow \natc \ \textrm{is the register file at the start of the TSX region}\\
%   &\tsxfallback : \natc \ \textrm{is the address to resume execution from if an error occurs in an active TSX region}
% \end{flalign*}

\M{\ISAICM} is defined as a composition of two auxiliary transition
systems, the deterministic \M{\ISAICM\texttt{-ISA}} and the
nondeterministic \M{\ISAICM\texttt{-C}}. It has a single transition
rule.

\[
  \inferrule[isa-ic]{S \xtrans{\ISAICM\texttt{-C}} S^\prime \\ S^\prime \xtrans{\ISAICM\texttt{-ISA}} S^{\prime\prime} \\ S^{\prime\prime} \xtrans{\ISAICM\texttt{-C}} S^{\prime\prime\prime}}
  {S \xtrans{\ISAICM} S^{\prime\prime\prime}}
\]

Let $\funcfont{fetch} : (\natc \rightharpoonup \insts{IC}) \times \natc \rightarrow \insts{IC}$ be a function such that:
\[
  \fetchfunc{\textit{imem}}{a} = \begin{cases}
    \textit{imem}(a) &\text{if } \textit{imem}(a)\downarrow\\
    \texttt{noop} &\text{otherwise}
  \end{cases}
\]

The deterministic behavior of \M{\ISAICM} is represented using\\
\trs{\ISAICM\texttt{-ISA}}, where\\
$S_{\ISAICM\texttt{-ISA}} = S_{\ISAICM}$. The behavior of this system
is straightforward, and can be summarized as follows: if not halted, it
executes the instruction at address \pc\ in \imem\ (treating it as a
\texttt{noop} otherwise), updates the register file appropriately, and
then either increments the \pc\ or sets it to a different value in a
few special cases (\texttt{jg}/\texttt{jge} when taken, a
\texttt{ldr}/\texttt{ldri} that raises an exception while in a TSX
region). A selection of the transition rules are shown
below. \textsc{ldr-ok-c} describes how a memory load instruction
operates in the case where the computed address is accessible and
\textsc{ic-ga-p}, \textsc{ic-ga-a} and \textsc{isa-not-ga} describe
the behavior of the \incache\ instruction.

\[
  \inferrule[halted]{\halted}{S \xtrans{\ISAICM\texttt{-ISA}} S}
\]

\[
\inferrule[halt]{\fetchfunc{\imem}{\pc} = \instr{\texttt{halt}} \\ \neg \halted}{S \xtrans{\ISAICM\texttt{-ISA}} [\pc \mapsto \pc \plusc 1, \halted \mapsto \boolt]S}
\]

\[
\inferrule[noop]{\fetchfunc{\imem}{\pc} = \instr{\texttt{noop}} \\ \neg \halted}{S \xtrans{\ISAICM\texttt{-ISA}} [\pc \mapsto \pc \plusc 1]S}
\]

\paragraph*{ALU Operations}
\texttt{loadi} loads a constant value into a register. The rest of the
ALU operations are straightforward: the destination register is set to
the result of some operation performed on the two source operands, the first
of which is always a register and the second of which is either a
register or a constant.

\[
  \inferrule[loadi]{\fetchfunc{\imem}{\pc} = \instr{\texttt{loadi}\ r_d\ c}\\ \neg \halted}{S \xtrans{\ISAICM\texttt{-ISA}} [\pc \mapsto \pc \plusc 1, \rfv \mapsto [r_d \mapsto c]\rfv]S}
\]

\[
\inferrule[addi]{\fetchfunc{\imem}{\pc} = \instr{\texttt{addi}\ r_d\ r_1\ c} \\ \neg \halted}{S \xtrans{\ISAICM\texttt{-ISA}} [\pc \mapsto \pc \plusc 1, \rfv \mapsto [r_d \mapsto \rf{r_1} \plusc c]\rfv]S}
\]

\[
\inferrule[add]{\fetchfunc{\imem}{\pc} = \instr{\texttt{add}\ r_d\ r_1\ r_2} \\ \neg \halted}{S \xtrans{\ISAICM\texttt{-ISA}} [\pc \mapsto \pc \plusc 1, \rfv \mapsto [r_d \mapsto \rf{r_1} \plusc \rf{r_2}]\rfv]S}
\]

\[
\inferrule[mul]{\fetchfunc{\imem}{\pc} = \instr{\texttt{mul}\ r_d\ r_1\ r_2} \\ \neg \halted}{S \xtrans{\ISAICM\texttt{-ISA}} [\pc \mapsto \pc \plusc 1, \rfv \mapsto [r_d \mapsto \rf{r_1} \timesc \rf{r_2}]\rfv]S}
\]

\[
\inferrule[and]{\fetchfunc{\imem}{\pc} = \instr{\texttt{and}\ r_d\ r_1\ r_2} \\ \neg \halted}{S \xtrans{\ISAICM\texttt{-ISA}} [\pc \mapsto \pc \plusc 1, \rfv \mapsto [r_d \mapsto \rf{r_1} \andc \rf{r_2}]\rfv]S}
\]

\paragraph*{Comparison and Branch Instructions}
The \texttt{cmp} instruction compares the values referred to by the
two source operands and sets the destination register to a value
based on the result of the comparison. This is used to support the two
conditional jump instructions, \texttt{jg} (``jump if greater than'')
and \texttt{jge} (``jump if greater than or equal to''). The
conditional jump instructions will check the given source operand to
determine if the jump condition holds, and then will jump to a
relative offset (provided as a constant operand) from the current \pc\
if so, or will behave as a \texttt{noop} otherwise.

\[
  \text{Let } \funcfont{compare}(a, b) = \begin{cases}
    1 &\text{if } a = b\\
    2 &\text{if } a > b\\
    0 &\text{otherwise}
    \end{cases}
\]

\[
\inferrule[cmp]{\fetchfunc{\imem}{\pc} = \instr{\texttt{cmp}\ r_d\ r_1\ r_2} \\ \neg \halted}{S \xtrans{\ISAICM\texttt{-ISA}} [\pc \mapsto \pc \plusc 1, \rfv \mapsto [r_d \mapsto \funcfont{compare}(\rf{r_1}, \rf{r_2})]\rfv]S}
\]

\[
\inferrule[jg-taken]{\fetchfunc{\imem}{\pc} = \instr{\texttt{jg}\ r_1\ c} \\ \rf{r_1} = 2 \\ \neg \halted}{S \xtrans{\ISAICM\texttt{-ISA}} [\pc \mapsto \pc \plusc c]S}
\]

\[
\inferrule[jg-not-taken]{\fetchfunc{\imem}{\pc} = \instr{\texttt{jg}\ r_1\ c} \\ \rf{r_1} \neq 2 \\ \neg \halted}{S \xtrans{\ISAICM\texttt{-ISA}} [\pc \mapsto \pc \plusc 1]S}
\]

\[
\inferrule[jge-taken]{\fetchfunc{\imem}{\pc} = \instr{\texttt{jge}\ r_1\ c} \\ \rf{r_1} = 1 \vee \rf{r_1} = 2 \\ \neg \halted}{S \xtrans{\ISAICM\texttt{-ISA}} [\pc \mapsto \pc \plusc c]S}
\]

\[
\inferrule[jge-not-taken]{\fetchfunc{\imem}{\pc} = \instr{\texttt{jge}\ r_1\ c} \\ \neg (\rf{r_1} = 1 \vee \rf{r_1} = 2) \\ \neg \halted}{S \xtrans{\ISAICM\texttt{-ISA}} [\pc \mapsto \pc \plusc 1]S}
\]

\paragraph*{TSX Instructions}
The TSX instructions either begin or end a TSX
region. \texttt{tsx-start} begins a TSX region, setting the TSX active
flag to \boolt, setting the fallback register file to the current
\rfv\ and setting the fallback PC to the instruction's source
operand. \texttt{tsx-end} sets the TSX active flag to \boolf{} and
leaves the other TSX state unchanged. The values of the fallback
register file and fallback PC do not matter when the TSX active flag is \boolf{}.

Note that unlike the similar TSX instructions from Intel's x86 TSX
extension (\texttt{XBEGIN} and \texttt{XEND}), our ISA does not
support nested TSX regions. If a \texttt{tsx-start} instruction
executes when a TSX region is already active, the existing TSX status
information is overwritten. Unlike \texttt{XEND}, \texttt{tsx-end}
does not cause an exception if executed outside of a TSX region (\eg\
when $\neg \tsxactive$). In this case, \texttt{tsx-end} behaves like
\texttt{noop}.

\[
\inferrule[tsx-start]{\fetchfunc{\imem}{\pc} = \instr{\texttt{tsx-start}\ c} \\ \neg \halted}{S \xtrans{\ISAICM\texttt{-ISA}} [\pc \mapsto \pc \plusc 1, \tsx \mapsto \abracket{\boolt, \rfv, c}]S}
\]

\[
\inferrule[tsx-end]{\fetchfunc{\imem}{\pc} = \instr{\texttt{tsx-end}} \\ \neg \halted}{S \xtrans{\ISAICM\texttt{-ISA}} [\pc \mapsto \pc \plusc 1, \tsxactive \mapsto \boolf]S}
\]

\paragraph*{Load Instructions}
The memory load instructions have the most complicated semantics of
any of the instructions in $\mathcal{I}$. The
\emph{effective address} for the load is computed by adding together
the two source operands. If the effective address is valid in the
ISA's address space (as determined by $\goodaddr$), the load proceeds
and the value stored at the effective address in \dmem\ is loaded into
the destination register. If the effective address is not valid and
the ISA is inside of a TSX region, \rfv\ is reset to \tsxrf, \pc\ is
set to \tsxfallback, and the TSX region is marked as inactive. This
effectively restarts execution from the fallback PC specified in the
\texttt{tsx-start} instruction associated with this region. If the
effective address is not valid and the ISA is not inside of a TSX
region, the ISA is halted. This is because our ISA does not support
exception handling (which is normally what would occur in an x86
process when accessing unmapped memory or memory that requires a
higher privilege level than the current one).

Note that the signature of \goodaddr\ and the way that it is is
manipulated by $\M{\ISAICM}$ has implications for the kinds of abstract
behavior that $\M{\ISAICM}$ will allow for later on. In particular,
notice that \goodaddr\ is never modified by any of the $\M{\ISAICM}$
transition rules, implying that the set of addresses that the ISA may
access is known ahead-of-time and is constant across an ISA
execution. This is intentional, and reflects the goal of presenting an
$\M{\ISAICM}$ that is as simple as possible while still having enough
complexity to highlight Meltdown and Spectre.

% \[
% \inferrule[ldri-ok]{\fetchfunc{\imem}{\pc} = \instr{\texttt{ldri}\ r_d\ r_1\ c} \\ \goodaddrfn{\rf{r_1} \plusc c} \\ \neg \halted}{S \xtrans{\ISAICM} [\pc \mapsto \pc \plusc 1, \rfv \mapsto [r_d \mapsto \partialget{\dmem}{\rf{r_1} \plusc c, 0}]\rfv]S}
% \]

\[
  \inferrule[ldri-ok-c]{\fetchicfunc{\imem}{\pc} = \instr{\texttt{ldri}\ r_d\ r_1\ c} \\ \text{Let } a = \rf{r_1} \plusc c \\ \goodaddrfn{a} \\ \text{Let } v = \partialget{\dmem}{a, 0} \\ \neg \halted}
  {S \xtrans{\ISAICM\texttt{-ISA}} [\pc \mapsto \pc \plusc 1, \rfv \mapsto [r_d \mapsto v]\rfv, \\\cmem \mapsto [a \mapsto v]\cmem]S}
\]

\[
\inferrule[ldri-err-tsx]{\fetchfunc{\imem}{\pc} = \instr{\texttt{ldri}\ r_d\ r_1\ c} \\ \neg \goodaddrfn{\rf{r_1} \plusc c} \\ \tsxactive \\ \neg \halted}{S \xtrans{\ISAICM\texttt{-ISA}} [\pc \mapsto \tsxfallback, \rfv \mapsto \tsxrf, \tsxactive \mapsto \boolf]S}
\]

\[
\inferrule[ldri-err-notsx]{\fetchfunc{\imem}{\pc} = \instr{\texttt{ldri}\ r_d\ r_1\ c} \\ \neg \goodaddrfn{\rf{r_1} \plusc c} \\ \neg \tsxactive \\ \neg \halted}{S \xtrans{\ISAICM\texttt{-ISA}} [\halted \mapsto \boolt]S}
\]

% \[
% \inferrule[ldr-ok]{\fetchfunc{\imem}{\pc} = \instr{\texttt{ldr}\ r_d\ r_1\ r_2} \\ \goodaddrfn{\rf{r_1} \plusc \rf{r_2}} \\ \neg \halted}{S \xtrans{\ISAICM} [\pc \mapsto \pc \plusc 1, \rfv \mapsto [r_d \mapsto \partialget{\dmem}{\rf{r_1} \plusc \rf{r_2}, 0}]\rfv]S}
% \]

\[
  \inferrule[ldr-ok-c]{\fetchicfunc{\imem}{\pc} = \instr{\texttt{ldr}\ r_d\ r_1\ r_2} \\ \text{Let } a = \rf{r_1} \plusc \rf{r_2} \\ \goodaddrfn{a} \\ \text{Let } v = \partialget{\dmem}{a, 0} \\ \neg \halted}
  {S \xtrans{\ISAICM\texttt{-ISA}} [\pc \mapsto \pc \plusc 1, \rfv \mapsto [r_d \mapsto v]\rfv, \\\cmem \mapsto [a \mapsto v]\cmem]S}
\]

\[
\inferrule[ldr-err-tsx]{\fetchfunc{\imem}{\pc} = \instr{\texttt{ldr}\ r_d\ r_1\ r_2} \\ \neg \goodaddrfn{\rf{r_1} \plusc \rf{r_2}} \\ \tsxactive \\ \neg \halted}{S \xtrans{\ISAICM\texttt{-ISA}} [\pc \mapsto \tsxfallback, \rfv \mapsto \tsxrf, \tsxactive \mapsto \boolf]S}
\]

\[
\inferrule[ldr-err-notsx]{\fetchfunc{\imem}{\pc} = \instr{\texttt{ldr}\ r_d\ r_1\ r_2} \\ \neg \goodaddrfn{\rf{r_1} \plusc \rf{r_2}} \\ \neg \tsxactive \\ \neg \halted}{S \xtrans{\ISAICM\texttt{-ISA}} [\halted \mapsto \boolt]S}
\]

\paragraph*{\incache}
\[
  \inferrule[ic-ga-p]{\fetchicfunc{\imem}{\pc} = \instr{\texttt{in-cache}\ r_d\ r_1\ r_2} \\ \goodaddrfn{\rf{r_1} \plusc \rf{r_2}} \\ \cmem(\rf{r_1} \plusc \rf{r_2}) \downarrow \\ \neg \halted}
  {S \xtrans{\ISAICM\texttt{-ISA}} [\pc \mapsto \pc \plusc 1, \rfv \mapsto [r_d \mapsto 1]\rfv]S}
\]

\[
  \inferrule[ic-ga-a]{\fetchicfunc{\imem}{\pc} = \instr{\texttt{in-cache}\ r_d\ r_1\ r_2} \\ \goodaddrfn{\rf{r_1} \plusc \rf{r_2}}  \\ \cmem(\rf{r_1} \plusc \rf{r_2}) \uparrow \\  \neg \halted}
  {S \xtrans{\ISAICM\texttt{-ISA}} [\pc \mapsto \pc \plusc 1, \rfv \mapsto [r_d \mapsto 0]\rfv]S}
\]

\[
  \inferrule[ic-not-ga]{\fetchicfunc{\imem}{\pc} = \instr{\texttt{in-cache}\ r_d\ r_1\ r_2} \\ \neg \goodaddrfn{\rf{r_1} \plusc \rf{r_2}} \\ \neg \halted}
  {S \xtrans{\ISAICM\texttt{-ISA}} [\pc \mapsto \pc \plusc 1, \rfv \mapsto [r_d \mapsto 0]\rfv]S}
\]

The nondeterministic behavior of \M{\ISAICM} is represented using
\trs{\ISAICM\texttt{-C}}, where
$S_{\ISAICM\texttt{-C}} = S_{\ISAICM}$. This transition system has a
single transition rule.

\[
  \inferrule[isa-ic-c]{\funcfont{add} \subseteq \powerset{\natc \times \natc} \\  \funcfont{rem} \subseteq \powerset{\natc \times \natc} \\ \abracket{\forall a,d \from (a,d) \in \funcfont{add} \from \goodaddrfn{a} \wedge d = \partialget{\dmem}{a, 0}} \\ \abracket{\forall a,d \from (a,d) \in \funcfont{rem} \from \goodaddrfn{a} \wedge d = \partialget{\dmem}{a, 0}}}
  {S \xtrans{\ISAICM\texttt{-C}} [\cmem \mapsto (\cmem \cup \funcfont{add}) \setminus \funcfont{rem}]S^{\prime}}
\]

\subsection{Formal Semantics of \ISAICAM}
\label{sec:isa-ic-a-semantics}

$\atrs{\ISAICAM}$ is an ALT.

Let
$\mathcal{I} = \insts{IC} \setminus \{
\texttt{in-cache}\ r_d \ r_1 \ c \}$ be the set of instructions that
the ISA is defined over.

We define the set of \M{\ISAICAM} states to be a tuple:
\[
  S_{\ISAICAM} : \La\pc, \rfv, \tsx, \halted, \imem, \dmem,
  \goodaddr, \cmem\Ra
\]
where all of the components except for $\imem$ are identical to the
same components in \M{\ISAICM}.
$\imem : \natc \rightharpoonup \mathcal{I}$ is a partial map from
addresses to instructions.

The set of \M{\ISAICAM} actions consists of sequences of authorized
cache actions:
\[
   A_{\ISAICAM} = (\{ \texttt{prefetch}\ a\ \vert\ a \in \natc \} \cup \{ \texttt{cache}\ a\ \vert\ a \in \natc \} )^\ast
\]

Like \M{\ISAICM}, \M{\ISAICAM} has one transition rule that uses an
auxiliary transition system.

Let
$\funcfont{apply-prefetches} : A_{\ISAICAM} \times (\natc
\rightharpoonup \natc) \times (\natc \rightharpoonup \natc)
\rightarrow (\natc \rightharpoonup \natc)$ be a function that takes in
a sequence of prefetch virtual instructions, a data memory and a cache
memory and returns the cache after applying the given virtual
instructions in order to the starting cache.

\[
  \inferrule[isa-p]{S \xtrans{\ISAICAM\texttt{-ISA}} S^{\prime}}
  {S \atrans{\ISAICAM}{a} [\cmem \mapsto \funcfont{apply-prefetches}(a, \dmem_{S^\prime}, \cmem_{S^\prime})]S^{\prime}}
\]

\trs{\ISAICAM-\textit{ISA}} is a transition system representing the
deterministic behavior of \ISAICAM.
$S_{\ISAICAM-\textit{ISA}} = S_{\ISAICAM}$. The behavior of
\M{\ISAICAM-\textit{ISA}} can be described using the transition rules
for \M{\ISAICM-\textit{ISA}}, except for \textsc{ic-ga-p},
\textsc{ic-ga-a} and \textsc{ic-not-ga}. Notice that none of the
transition rules for \M{\ISAICM-\textit{ISA}} modify the \imem\
component of the state, so \M{\ISAICAM-\textit{ISA}} defined in this way
is indeed closed under its transition relation.

\subsection{Formal Semantics of \MAICM}
\label{sec:ma-ic-semantics}

\subsubsection{Parameters}

$\rsids$ is the set of all reservation station identifiers. $\robids$
is the set of all ROB tags (ROB line identifiers). Both $\rsids$ and
$\robids$ have finite cardinality, and both must be isomorphic to the
standard cyclic group with order equal to their cardinality. The
implication of this isomorphism that is used here is the existence of
functions $\funcfont{next}_{\robids} : \robids \rightarrow \robids$
and $\funcfont{next}_{\rsids} : \rsids \rightarrow \rsids$ that get
the successor for a ROB tag or reservation station identifier
respectively and functions
$\funcfont{prev}_{\robids} : \robids \rightarrow \robids$ and
$\funcfont{prev}_{\rsids} : \rsids \rightarrow \rsids$ that get the
predecessor for a ROB tag or reservation station identifier
respectively. It is the case that $\funcfont{next}_\ast$ and
$\funcfont{prev}_\ast$ are inverses of each other.

\paragraph{\mafetchnum}
is the maximum number of instructions that can be fetched in a
single cycle. The value of this parameter must be a non-zero natural
number.

\paragraph{\mamaxdecode}
is the maximum number of microinstructions that an
instruction can decode into.

\paragraph{\marobsize}
is the maximum number of reorder buffer (ROB) lines supported. The
value of this parameter must be a natural number greater than or equal
to \mamaxdecode. Without this restriction, it is possible to generate
a non-halted machine that is unable to commit any instructions, since
it doesn't have enough resources to issue all of the microinstructions
that the first instruction decodes to.

\subsubsection{Transition System}
{
  \setlength{\parindent}{0pt}
\trs{\MAICM} is a deterministic transition system. We define the
set of \M{\MAICM} states to be a tuple
\begin{flalign*}
  S_{\MAICM} : \La&\pc, \rfv, \tsx, \halted, \imem, \dmem,
    \goodaddr, \cmem, \rob, \rsf, \regstat,\\ &\cyclec, \fetchpc, \prefetch\Ra
\end{flalign*}
where \pc, \rfv, \tsx, \halted, \imem, \dmem, \goodaddr, and \cmem\
are as in \M{\ISAICM}. The other components of $S_{\MAICM}$ are described as follows:

$\insts{IC}$ is the set of instructions that \M{\MAICM} is defined
over, and is the same as in Appendix~\ref{sec:isa-ic-semantics}.

$\uops{IC}$ denotes the set of microoperations corresponding to the
instructions in $\insts{IC}$, and $\uinsts{IC}$ denotes the set of
microinstructions corresponding to the instructions in $\insts{IC}$.
}
\begin{itemize}
  \item $\rob : \roblines{IC}^\ast $ is a sequence of reorder buffer (ROB) lines. ROB line IDs must be unique in this sequence.
  \item $\rsf : \resstations{IC}^\ast $ is a sequence of reservation stations
  \item $\regstat : \mathcal{R} \rightharpoonup \natc$ is the register status file
  \item $\cyclec : \natc $ is a counter that increments on each \M{\MAICM} step
  \item $\fetchpc : \natc $ is the PC of the next instruction to fetch
  \item $\prefetch : \natc \rightarrow \powerset{\natc}$ computes the
    set of addresses that should be prefetched when the given address
    is loaded
\end{itemize}

$\roblines{IC} : \abracket{\robid, \robmop, \robdst, \robrdy, \robval, \robexcp}$
\begin{itemize}
  \item $\robid : \robids $ is the identifier for this ROB line
  \item $\robmop : \uops{IC} $ is the microoperation for this ROB line
  \item $\robdst : \regs \cup \{ \texttt{nil} \} $ is the register that the result of this microinstruction should be written to, or \texttt{nil} if not needed
  \item $\robrdy : \bools $ is \boolt{} iff this ROB line is ready to be committed
  \item $\robval : \natc $ is the result of the microinstruction
  \item $\robexcp : \bools $ is \boolt{} iff executing this microinstruction resulted in an exception
\end{itemize}

The reorder buffer (ROB) behaves like a FIFO queue of ROB lines, each
of which tracks the execution of a single microinstruction. The ROB
keeps these lines in program order, and this ordering is what ensures
that microinstructions are committed in program order even if they
were executed out-of-order.

$\resstations{IC} : \abracket{\rsid, \rsmop, \rsqj, \rsqk, \rsvk, \rsvj, \rscpc, \rsbusy, \rsexec, \rsdst, \rspc}$
\begin{itemize}
  \item $\rsid : \rsids$ is the identifier for this RS
  \item $\rsmop : \uops{IC} $ is the microoperation loaded into this RS
  \item $\rsqj : \robids \cup \{ \texttt{nil} \} $ is the ID of the ROB line to wait on for the J argument, or \texttt{nil} if not needed.
  \item $\rsqk : \robids \cup \{ \texttt{nil} \} $ is the ID of the ROB line to wait on for the K argument, or \texttt{nil} if not needed.
  \item $\rsvj : \natc $ is the value of the J argument
  \item $\rsvk : \natc $ is the value of the K argument
  \item $\rscpc : \natc $ is the cycle at which this RS will finish execution
  \item $\rsbusy : \bools $ is \boolt{} iff the RS is in use
  \item $\rsexec : \bools $ is \boolt{} iff the RS is currently executing a microinstruction
  \item $\rsdst : \robids $ is the ID of the ROB line that the result of this RS should be stored in
  \item $\rspc : \natc $ is the PC value corresponding to this RS' loaded instruction
\end{itemize}

Reservation stations (RSes) are the part of the microarchitecture that
execute ALU operations, memory reads, and memory checks. When a
microinstruction is loaded into a RS and its operands become ready, it
begins to execute the microinstruction. After a number of cycles
(depending on the microinstruction), the RS completes execution, and
the result of the execution is written back to the appropriate ROB
line. Any subsequent microinstructions that may have been waiting on
the value of that ROB line are updated appropriately.

$\regstatentries : \abracket{\regstbusy, \regstreorder}$
\begin{itemize}
  \item $\regstbusy : \bools $ is \boolt{} iff the register mapped to this entry will be written to by an issued and uncompleted microinstruction
  \item $\regstreorder : \robids $ is the ID of the ROB line that contains the instruction that will write to this entry's register
\end{itemize}

The register status file keeps track of the registers that will be
written to by any in-flight microinstructions. This information is
used to handle read-after-write (RAW) hazards: in this context,
situations where an instruction reads a register after a prior
instruction writes to it. If the register-writing instruction is not
committed by the time the register-reading instruction needs to access
the read register's value, the MA must stall on the execution of the
register-reading instruction until the register's value is available.
%In the case of \MAM, keeping a register status file is
%somewhat redundant as enough information is maintained in \rob\ to
%detect these hazards. We include it in \MAM\ as it is often included
%in out-of-order and pipelined microarchitectures.

When a microinstruction is issued, it will be assigned to a ROB
line. Given the sequence of microinstructions to be issued on a
particular cycle and the ROB at the start of the cycle, it is possible
to compute the ID for the ROB line that each microinstruction will be
assigned to.
$\funcfont{rob-ids-ic} : \uinsts{IC}^\ast \times \roblines{IC}^\ast \rightarrow
\robids^\ast$ is a function that will do exactly this.

\subsubsection{Additional Definitions}

\paragraph{\insts{IC}} is the set of instructions that the machine
supports. Each instruction consists of an operation plus zero, one, or
two source operands and zero or one destination operands. The
destination operand (if provided) is always a register specifier, and
the source operands may either be register specifiers or constant
values. The set of operations supported by the machine is \ops{IC}, and
$\instop : \insts{IC} \rightarrow \ops{IC}$ is a function that gets an
instruction's operation.

\paragraph{\uinsts{IC}} is the set of microinstructions that the machine
supports. Each instruction in \insts{IC}\ decodes to a sequence of one
or two microinstructions in \uinsts{IC}. Each microinstruction consists
of a microoperation plus zero, one, or two source operands and zero or
one destination operands, just like an instruction. The set of
microoperations supported by the machine is \uops{IC}, and
$\uinstop : \uinsts{IC} \rightarrow \uops{IC}$ is a function that gets an
microinstruction's microoperation.

\paragraph{\funcfont{rs-needed?}}
Not all microoperations require a reservation station. The predicate
$\funcfont{rs-needed?} : \uinsts{IC} \rightarrow \bools$ holds only for those
microinstructions that require a reservation station.

\begin{flalign*}
  &\funcfont{rs-needed?}(\textit{op}) \iff \\
  &\textit{op} \in \{ \texttt{mnoop}, \texttt{mloadi}, \texttt{maddi}, \texttt{madd}, \texttt{mmul}, \texttt{mand}, \texttt{mcmp}, \texttt{mjg}, \texttt{mjge},\\
  &\texttt{mldri}, \texttt{memi-check}, \texttt{mldr}, \texttt{mem-check}, \texttt{min-cache} \}
\end{flalign*}

\paragraph{\funcfont{reg-write?}}
Not all microoperations will write to a register. The predicate
$\funcfont{reg-write?} : \uinsts{IC} \rightarrow \bools$ holds only for
those microoperations that will write to a register.

\begin{flalign*}
  &\funcfont{reg-write?}(\textit{op}) \iff\\& \textit{op} \in \{ \texttt{mloadi}, \texttt{maddi}, \texttt{madd}, \texttt{mmul}, \texttt{mand}, \texttt{mcmp}, \texttt{mldri}, \texttt{mldr}, \texttt{min-cache} \}
\end{flalign*}

\paragraph{\funcfont{reg-dst}}
For microinstructions with microoperations satisfying
$\funcfont{reg-write?}$, the function
$\funcfont{reg-dst} : \uinsts{IC} \rightharpoonup \mathcal{R}$
determines which register will be written to.

\paragraph{$\funcfont{reg-op}_1$} $: \uinsts{IC} \rightharpoonup \mathcal{R}$ denotes
the register specifier for the first operand of a microinstruction (if
it has at least one operand and the first operand is a register
specifier).

\paragraph{$\funcfont{reg-op}_2$} $: \uinsts{IC} \rightharpoonup \mathcal{R}$ denotes
the register specifier for the second operand of a microinstruction
(if it has two operands and the second operand is a register
specifier).

\paragraph{$\funcfont{const-op}_1$} $:\uinsts{IC} \rightharpoonup \natc$
and $\funcfont{const-op}_2 : \uinsts{IC} \rightharpoonup \natc$ are
similar functions for constant operands instead of register operands.

\paragraph{\funcfont{barrier-op?}}
Some microoperations should behave as though they are \emph{memory
  barriers}. These microoperations are special in that they should not
begin executing while there are uncommitted in-flight memory access
microoperations (those satisfying \funcfont{memory-op?} as described
below). The predicate
$\funcfont{barrier-op?} : \uops{IC} \rightarrow \bools$ holds only
for those microoperations that behave as memory barriers.

$\funcfont{barrier-op?}(\textit{op}) \iff \textit{op} \in \{ \texttt{min-cache} \}$

\paragraph{\funcfont{memory-op?}}
Some microoperations access memory, and should be affected by the
memory barriers described above. In particular, these microoperations
should not begin executing while there are uncommitted in-flight
memory barrier microoperations (those satisfying
\funcfont{barrier-op?} as described above). The predicate
$\funcfont{memory-op?} : \uops{IC} \rightarrow \bools$ holds only
for those microoperations that access memory.

Note that we assume that \funcfont{barrier-op?} and
\funcfont{memory-op?} are mutually exclusive, \eg\ that no
microoperation exists that satisfies both predicates.

$\funcfont{memory-op?}(\textit{op}) \iff \textit{op} \in \{ \texttt{mldri}, \texttt{mldr} \}$

% \paragraph{\funcfont{jump-mop?}} $: \uops{IC} \rightarrow \bools$
% holds for those microoperations that correspond to jumps.

\paragraph{\funcfont{decode-one-ic}} $: \insts{IC} \rightarrow \uinsts{IC}^\ast$ is a
function that decodes an instruction into the appropriate sequence of
microinstructions.
\begin{flalign*}
  &\funcfont{decode-one-ic}(\textit{inst}) =\\ &\begin{cases}
    \sequence{\texttt{memi-check}\ r_1\ c,  \texttt{mldri}\ r_d\ r_1\ c} & \text{if } \textit{inst} = \texttt{ldri}\ r_d\ r_1\ c\\
    \sequence{\texttt{mem-check}\ r_1\ r_2,  \texttt{mldr}\ r_d\ r_1\ r_2} & \text{if } \textit{inst} = \texttt{ldr}\ r_d\ r_1\ r_2\\
    \sequence{\texttt{m}\textit{op}\ \textit{operands}...} & \text{if } \textit{inst} = \textit{op}\ \textit{operands}...
  \end{cases}
\end{flalign*}

\paragraph{\funcfont{comp-val-ic}}
$: \resstations{IC} \times S_{\MAICM} \rightarrow \natc$ is a function that
computes the result of the microoperation inside the given RS,
assuming the RS is ready.
\begin{flalign*}
  &\funcfont{comp-val-ic}(\textit{rs}, s) =\\&\begin{cases}
    \dmem_s(\rsvj \plusc \rsvk) & \text{if } \rsmop_\textit{rs} \in \{ \texttt{mldri}, \texttt{mldr} \}\\
    \rsvj \andc \rsvk & \text{if } \rsmop_\textit{rs} = \texttt{mand}\\
    \rsvj \plusc \rsvk & \text{if } \rsmop_\textit{rs} \in \{ \texttt{maddi}, \texttt{madd} \} \\
    \rsvj \timesc \rsvk & \text{if } \rsmop_\textit{rs} = \texttt{mmul}\\
    \rsvk & \text{if } \rsmop_\textit{rs} = \texttt{mloadi}\\
    1 & \text{if } \rsmop_\textit{rs} = \texttt{min-cache} \wedge \cmem(\rsvj \plusc \rsvk)\downarrow\\
    0 & \text{if } \rsmop_\textit{rs} = \texttt{min-cache} \wedge \cmem(\rsvj \plusc \rsvk)\uparrow\\
    \funcfont{compare}(\rsvk, \rsvj) & \text{if } \rsmop_\textit{rs} = \texttt{mcmp}\\
    \rspc \plusc \rsvk & \text{if } \rsmop_\textit{rs} \in \{ \texttt{mjge}, \texttt{mjg} \} \wedge \rsvj = 2\\
    \rspc \plusc \rsvk & \text{if } \rsmop_\textit{rs} = \texttt{mjge} \wedge \rsvj = 1\\
    \rspc \plusc 1 & \text{if } \rsmop_\textit{rs} = \texttt{mjg} \wedge \rsvj = 1\\
    \rspc \plusc 1 & \text{if } \rsmop_\textit{rs} \in \{ \texttt{mjge}, \texttt{mjg} \} \wedge \rsvj \notin \{ 1, 2 \}\\
    0 & \text{otherwise}
  \end{cases}
\end{flalign*}

\paragraph{\funcfont{to-fetch}} $: S_{\MAICM} \times \natc$ is a
relation that pairs states with a number of instructions to fetch. The
number of instructions to fetch must always be less than or equal than
$\mafetchnum$, and it also must be the case that the sequence of $n$
instructions to be fetched in the associated state is issuable
(described in Section~\ref{sec:issuable}).

$\funcfont{to-fetch} = \{ (s, \funcfont{max-fetch-n}(s)) \vert s \in S_\MAICM \}$

\paragraph{\funcfont{comp-exc}}
$: \resstations{IC} \times S_{\MAICM} \rightarrow \bools$ is a function that
determines whether executing the microoperation inside the given RS
resulted in an exception, assuming the RS is ready.

\begin{flalign*}
  &\funcfont{comp-exc}(\textit{rs}, s) =\\&\begin{cases}
    \boolt & \text{if } \rsmop_\textit{rs} \in \{ \texttt{memi-check}, \texttt{mem-check} \} \wedge \neg\goodaddrfn{\rsvj \plusc \rsvk}\\
    \boolf & \text{otherwise}
  \end{cases}
\end{flalign*}

\paragraph{Issuable}
\label{sec:issuable}

$\funcfont{free-rob} : \roblines{IC}^\ast \rightarrow \natc$, where
$\funcfont{free-rob}(\sigma) = \marobsize - \vert \fndom(\sigma)
\vert$ is the number of free ROB entries.

$\funcfont{idle-rses} : \resstations{IC}^\ast \rightarrow
\resstations{IC}^\ast$ collects the reservation stations that are not
busy.

A sequence of microinstructions $\sigma$ is issuable in a state $s$ if
$\sigma$ is empty or if all of the following hold:
\begin{itemize}
\item $\vert\sequence{u \from u \in \sigma \from \funcfont{rs-needed?}(u)}\vert \leq \vert\funcfont{idle-rses}(\rsf_s)\vert$
\item $\vert\sigma\vert \leq \funcfont{free-rob}(\rob_s)$
\end{itemize}

A sequence of instructions $\sigma$ is issuable in a state $s$ if the
sequence of microinstructions produced by decoding each instruction
and concatenating the resulting sequences of microinstructions
together is issuable in $s$.

\paragraph{$\funcfont{fetch}$} $: (\natc \rightharpoonup \insts{IC}) \times \natc \rightarrow \insts{IC}$ is a function such that:
\[
  \fetchfunc{\textit{imem}}{a} = \begin{cases}
    \textit{imem}(a) &\text{if } \textit{imem}(a)\downarrow\\
    \texttt{noop} &\text{otherwise}
  \end{cases}
\]

\paragraph{$\funcfont{fetch-n}$} $: (\natc \rightharpoonup \insts{IC}) \times \natc
\times \natc \rightarrow \insts{IC}^\ast$ is a function that returns
the first $n$ instructions in the given instruction memory starting
from a particular address.
$\funcfont{fetch-n}(\textit{imem}, \textit{pc}, n) = \sigma$ such that
$\abracket{\forall i \from i \in \{ 0, ..., n-1 \} \from \sigma(i+1) =
  \funcfont{fetch}(\textit{imem}, \textit{pc} \plusc n)}$

\paragraph{\funcfont{max-fetch-n}}
\MAICM\ is multi-issue, but it may not be able to fetch and issue all
\mafetchnum\ instructions on a particular cycle if there are not
sufficient resources available. For example, the ROB may not have
enough capacity to store the ROB entries that issuing all of the
instructions would give rise to, or it could be that all of the RSes
are busy and one of the instructions that would be fetched requires a
RS. $\funcfont{max-fetch-n} : S_{\MAICM} \rightarrow \natc$ is a
function that returns the maximum number of instructions $n$ such that
$\sequence{\funcfont{fetch}(\pc, \imem), ..., \funcfont{fetch}(\pc
  \plusc (n - 1), \imem)}$ is issuable in the given state.

\paragraph{\funcfont{decode-ic}}
$: \mathcal{I}_{\mathit{ic}}^\ast \times \roblines{IC}^\ast \rightarrow (\uinsts{IC} \times
\robids)^\ast$ is a function that applies $\funcfont{decode-one-ic}$ to all of the
given instructions and appends the resulting sequences of
microinstructions together in the same order, then runs
$\funcfont{rob-ids-ic}$ on the resulting sequence of microinstructions
and pairs each microinstruction with the ROB line it will be issued
to.

\paragraph{$\funcfont{rob-get}$} $: \robids \times \roblines{IC}^\ast
\rightharpoonup \roblines{IC}$ is a function that finds the first ROB
line that has a particular ID in a sequence of ROB lines. That is,
\begin{flalign*}
  &\funcfont{rob-get}(x, \sigma) =\\ &\begin{cases}
    \sigma(i)&\parbox{6cm}{$\text{if } \abracket{\exists j \from j \in \mathbb{N} \from \robid_{\sigma(j)} = x} \wedge i = \min_{j \in \mathbb{N} \wedge \robid_{\sigma(j)} = x} j$}\\
    \uparrow &\text{otherwise}
  \end{cases}
\end{flalign*}

\paragraph{$\funcfont{rob-before}$}
$: \robids \times \roblines{IC}^\ast \rightarrow \roblines{IC}^\ast$ is a
function that returns all of the ROB lines prior to the ROB line with
the given ID, retaining order. If no such ROB line exists, the given
ROB lines are returned, retaining order.

\subsubsection{Semantics}
\label{sec:maicm-semantics}

Most of the components of the \MAICM\ state are updated in
parallel. Where one component depends on the value of another
component, it depends on the value of that component in the
``current'' state, before any updates are applied to it. The only
exception is that instruction fetching and decoding occurs before any
components are updated, so that component updates have access to the
sequence of microinstructions to issue. We describe how each component
is updated with its own auxiliary transition relation, and the
\textsc{stepall} transition rule below combines all of those steps of
individual components together.

Note that none of the components of $S_{\MAICM}$ change when
$S_{\MAICM}$ is stepped and \halted\ is set. For brevity, none of the
auxiliary transition systems have rules describing their behavior when
\halted\ holds. The behavior of the auxiliary transition systems in
such a situation is to transition to an identical state. At the top
level, the transition rule \textsc{halted} describes the behavior of
all of the components when the system is halted.

{
  \setlength{\parindent}{0pt}

\[
  \inferrule[halted]{\halted}{S \xtrans{\MAICM} S}
\]

\[
  \inferrule[stepall]{\neg\halted \\ \text{Let } n = \funcfont{max-fetch-n}(S) \\
    \abracket{S, n} \xtrans{\MAICM-\regstat} \abracket{\abracket{..., \regstat^\prime, ...}, n} \\
    S \xtrans{\MAICM-\pc} \abracket{..., \pc^\prime, ...} \\
    S \xtrans{\MAICM-\tsx} \abracket{..., \tsx^\prime, ...} \\
    S \xtrans{\MAICM-\rfv} \abracket{..., \rfv^\prime, ...} \\
    S \xtrans{\MAICM-\rob} \abracket{..., \rob^\prime, ...} \\
    \abracket{S, n} \xtrans{\MAICM-\rsf} \abracket{\abracket{..., \rsf^\prime, ...}, n}\\
    S \xtrans{\MAICM-\cmem} \abracket{..., \cmem^\prime, ...}
  }{S \xtrans{\MAICM} [\regstat \mapsto \regstat^\prime, \fetchpc \mapsto \fetchpc \plusc n, \pc \mapsto \pc^\prime,\\ \tsx \mapsto \tsx^\prime, \rfv \mapsto \rfv^\prime, \rob \mapsto \rob^\prime, \rsf \mapsto \rsf^\prime, \cmem \mapsto \cmem^\prime]S}
\]

}

\paragraph{\regstat}
\[
  \trs{\mrsi{\MAICM}}
\] is a transition system, where $S_{\mrsi{\MAICM}} : S_{\MAICM} \times (\uinsts{IC} \times
\robids)^\ast$.

\[
  \inferrule[regstat-issue-wr]{Q = \abracket{u, rb} \bullet Q^\prime \\ \funcfont{reg-write?}(\funcfont{minst-op}(u)) \\ r = \funcfont{reg-dst}(u) \\ \neg\halted}{\abracket{S, Q} \xtrans{\mrsi{\MAICM}} \abracket{[\regstat \mapsto [r \mapsto \abracket{\boolt, rb}]\regstat]S, Q^\prime}}
\]

\[
  \inferrule[regstat-issue-nowr]{Q = \abracket{u, rb} \bullet Q^\prime \\ \neg\funcfont{reg-write?}(\funcfont{minst-op}(u)) \\ \neg\halted}{\abracket{S, Q} \xtrans{\mrsi{\MAICM}} \abracket{S, Q^\prime}}
\]

\[
  \trs{\mrsc{\MAICM}}
\] is a transition system, where
\[
  S_{\mrsc{\MAICM}} : S_{\MAICM} \times \roblines{IC}^\ast
\]

\[
  \inferrule[regstat-commit-ready-rm]{Q = \textit{rl} \bullet Q^\prime \\ \robrdy_\textit{rl} \\ \regstat(\robdst_\textit{rl}) \downarrow \\ \abracket{\textit{bsy}, \textit{reord}} = \regstat(\robdst_\textit{rl}) \\ \robid_\textit{rl} = \textit{reord} \\ \neg\halted}
  {\abracket{S, Q} \xtrans{\mrsc{\MAICM}}\\ \abracket{[\regstat \mapsto [\robdst_\textit{rl} \mapsto \uparrow]\regstat]S, Q^\prime}}
\]

\[
  \inferrule[regstat-commit-ready-in-nomatch]{Q = \textit{rl} \bullet Q^\prime \\ \robrdy_\textit{rl} \\ \regstat(\robdst_\textit{rl}) \downarrow \\ \abracket{\textit{bsy}, \textit{reord}} = \regstat(\robdst_\textit{rl}) \\ \robid_\textit{rl} \neq \textit{reord} \\ \neg\halted}
  {\abracket{S, Q} \xtrans{\mrsc{\MAICM}} \abracket{S, Q^\prime}}
\]

\[
  \inferrule[regstat-commit-ready-notin]{Q = \textit{rl} \bullet Q^\prime \\ \robrdy_\textit{rl} \\ \regstat(\robdst_\textit{rl}) \uparrow \\ \neg\halted}
  {\abracket{S, Q} \xtrans{\mrsc{\MAICM}} \abracket{S, Q^\prime}}
\]

\[
  \inferrule[regstat-commit-notready]{Q = \textit{rl} \bullet Q^\prime \\ \neg\robrdy_\textit{rl} \\ \neg\halted}
  {\abracket{S, Q} \xtrans{\mrsc{\MAICM}} \abracket{S, \emptyseq}}
\]

Let \trs{\MAICM-\regstat} be a transition system, where $S_{\MAICM-\regstat} = S_{\MAICM} \times \natc$.

\[
  \inferrule[regstat]{\abracket{S, \funcfont{decode-ic}(\funcfont{fetch-n}(\imem, \pc, n), \rob)} \xtrans{\mrsi{\MAICM}}^\ast \abracket{S^\prime, \emptyseq} \\
    \abracket{S^\prime, \rob} \xtrans{\mrsc{\MAICM}}^\ast \abracket{S^{\prime\prime}, \emptyseq} \\ \neg\halted}
  {\abracket{S, n} \xtrans{\MAICM-\regstat}\\ \abracket{[\regstat \mapsto \regstat_{S^{\prime \prime}}]S, n}}
\]

\paragraph{\pc}

Let \trs{\mpcc{\MAICM}} be a transition system, where
$S_{\mpcc{\MAICM}} : S_{\MAICM} \times \roblines{IC}^\ast$.

\[
  \inferrule[pc-commit-excp-tsx]{Q = \textit{rl} \bullet Q^\prime \\ \robrdy_\textit{rl} \\ \robexcp_\textit{rl} \\ \tsxactive \\ \neg\halted}
  {\abracket{S, Q} \xtrans{\mpcc{\MAICM}}\\ \abracket{[\pc \mapsto \tsxfallback]S, \emptyseq}}
\]

\[
  \inferrule[pc-commit-excp-notsx]{Q = \textit{rl} \bullet Q^\prime \\ \robrdy_\textit{rl} \\ \robexcp_\textit{rl} \\ \neg\tsxactive \\ \neg\halted}
  {\abracket{S, Q} \xtrans{\mpcc{\MAICM}} \abracket{S, \emptyseq}}
\]

\[
  \inferrule[pc-commit-mem]{Q = \textit{rl} \bullet Q^\prime \\ \robrdy_\textit{rl} \\ \neg\robexcp_\textit{rl} \\ \robmop_\textit{rl} = \texttt{mem-check} \vee \robmop_\textit{rl} = \texttt{memi-check} \\ \neg\halted}
  {\abracket{S, Q} \xtrans{\mpcc{\MAICM}} \abracket{S, Q^\prime}}
\]

\[
  \inferrule[pc-commit-jmp]{Q = \textit{rl} \bullet Q^\prime \\ \robrdy_\textit{rl} \\ \neg\robexcp_\textit{rl} \\ \robmop_\textit{rl} = \texttt{mjg} \vee \robmop_\textit{rl} = \texttt{mjge} \\ \neg\halted}
  {\abracket{S, Q} \xtrans{\mpcc{\MAICM}} \abracket{[\pc \mapsto \robval_\textit{rl}]S, \emptyseq}}
\]

\[
  \inferrule[pc-commit-halt]{Q = \textit{rl} \bullet Q^\prime \\ \robrdy_\textit{rl} \\ \neg\robexcp_\textit{rl} \\ \robmop_\textit{rl} = \texttt{mhalt} \\ \neg\halted}
  {\abracket{S, Q} \xtrans{\mpcc{\MAICM}} \abracket{[\pc \mapsto \pc \plusc 1]S, \emptyseq}}
\]

\[
  \inferrule[pc-commit-other]{Q = \textit{rl} \bullet Q^\prime \\ \robrdy_\textit{rl} \\ \neg\robexcp_\textit{rl} \\ \robmop_\textit{rl} \notin \{\texttt{mem-check}, \texttt{memi-check}, \texttt{mjge}, \texttt{mjg}, \texttt{mhalt} \} \\ \neg\halted}
  {\abracket{S, Q} \xtrans{\mpcc{\MAICM}} \abracket{[\pc \mapsto \pc \plusc 1]S, Q^\prime}}
\]

\[
  \inferrule[pc-commit-notrdy]{Q = \textit{rl} \bullet Q^\prime \\ \neg\robrdy_\textit{rl} \\ \neg\halted}
  {\abracket{S, Q} \xtrans{\mpcc{\MAICM}} \abracket{S, \emptyseq}}
\]

Let \trs{\MAICM-\pc} be a transition system, where
$S_{\MAICM-\pc} = S_{\MAICM}$.

\[
  \inferrule[pc]{\abracket{S, \rob} \xtrans{\mpcc{\MAICM}}^\ast \abracket{S^\prime, \emptyseq} \\ \neg\halted}
  {S \xtrans{\MAICM-\pc} [\pc \mapsto \pc_{S^\prime}]S}
\]

\paragraph{\tsx}

Let \trs{\mtsxc{\MAICM}} be a transition system, where
$S_{\mtsxc{\MAICM}} : S_{\MAICM} \times \roblines{IC}^\ast$.

\[
  \inferrule[tsx-commit-excp]{Q = \textit{rl} \bullet Q^\prime \\ \robrdy_\textit{rl} \\ \robexcp_\textit{rl} \\ \neg\halted}
  {\abracket{S, Q} \xtrans{\mtsxc{\MAICM}}\\ \abracket{[\tsxactive \mapsto \boolf]S, \emptyseq}}
\]

\[
  \inferrule[tsx-commit-start]{Q = \textit{rl} \bullet Q^\prime \\ \robrdy_\textit{rl} \\ \neg\robexcp_\textit{rl} \\ \robmop_\textit{rl} = \texttt{mtsx-start} \\ \neg\halted}
  {\abracket{S, Q} \xtrans{\mtsxc{\MAICM}}\\ \abracket{[\tsxactive \mapsto \boolt, \tsxrf \mapsto \rfv, \tsxfallback \mapsto \robval_\textit{rl}]S, Q^\prime}}
\]

\[
  \inferrule[tsx-commit-end]{Q = \textit{rl} \bullet Q^\prime \\ \robrdy_\textit{rl} \\ \neg\robexcp_\textit{rl} \\ \robmop_\textit{rl} = \texttt{mtsx-end} \\ \neg\halted}
  {\abracket{S, Q} \xtrans{\mtsxc{\MAICM}}\\ \abracket{[\tsxactive \mapsto \boolf]S, Q^\prime}}
\]

\[
  \inferrule[tsx-commit-halt]{Q = \textit{rl} \bullet Q^\prime \\ \robrdy_\textit{rl} \\ \neg\robexcp_\textit{rl} \\ \robmop_\textit{rl} = \texttt{mhalt} \\ \neg\halted}
  {\abracket{S, Q} \xtrans{\mtsxc{\MAICM}} \abracket{S, \emptyseq}}
\]

\[
  \inferrule[tsx-commit-other]{Q = \textit{rl} \bullet Q^\prime \\ \robrdy_\textit{rl} \\ \neg\robexcp_\textit{rl} \\ \robmop_\textit{rl} \notin \{\texttt{mtsx-start}, \texttt{mtsx-end}, \texttt{mhalt}\} \\ \neg\halted}
  {\abracket{S, Q} \xtrans{\mtsxc{\MAICM}} \abracket{S, Q^\prime}}
\]

\[
  \inferrule[tsx-commit-notrdy]{Q = \textit{rl} \bullet Q^\prime \\ \neg\robrdy_\textit{rl} \\ \neg\halted}
  {\abracket{S, Q} \xtrans{\mtsxc{\MAICM}} \abracket{S, \emptyseq}}
\]

Let \trs{\MAICM-\tsx} be a transition system, where
$S_{\MAICM-\tsx} = S_{\MAICM}$.

\[
  \inferrule[tsx]{\abracket{S, \rob} \xtrans{\mtsxc{\MAICM}}^\ast \abracket{S^\prime, \emptyseq} \\ \neg\halted}
  {S \xtrans{\MAICM-\tsx} [\tsx \mapsto \tsx_{S^\prime}]S}
\]

\paragraph{\rfv}

Let \trs{\mrfvc{\MAICM}} be a transition system, where
$S_{\mrfvc{\MAICM}} : S_{\MAICM} \times \roblines{IC}^\ast$.

\[
  \inferrule[rf-commit-excp-tsx]{Q = \textit{rl} \bullet Q^\prime \\ \robrdy_\textit{rl} \\ \robexcp_\textit{rl} \\ \tsxactive \\ \neg\halted}
  {\abracket{S, Q} \xtrans{\mrfvc{\MAICM}} \abracket{[\rfv \mapsto \tsxrf]S, \emptyseq}}
\]

\[
  \inferrule[rf-commit-excp-notsx]{Q = \textit{rl} \bullet Q^\prime \\ \robrdy_\textit{rl} \\ \robexcp_\textit{rl} \\ \neg\tsxactive \\ \neg\halted}
  {\abracket{S, Q} \xtrans{\mrfvc{\MAICM}} \abracket{S, \emptyseq}}
\]

\[
  \inferrule[rf-commit-halt-jmp]{Q = \textit{rl} \bullet Q^\prime \\ \robrdy_\textit{rl} \\ \neg\robexcp_\textit{rl} \\ \robmop_\textit{rl} \in \{ \texttt{mjge}, \texttt{mjg}, \texttt{mhalt} \} \\ \neg\halted}
  {\abracket{S, Q} \xtrans{\mrfvc{\MAICM}} \abracket{S, \emptyseq}}
\]

\[
  \inferrule[rf-commit-nowrite]{Q = \textit{rl} \bullet Q^\prime \\ \robrdy_\textit{rl} \\ \neg\robexcp_\textit{rl} \\ \neg\funcfont{reg-write?}(\robmop_\textit{rl}) \\ \robmop_\textit{rl} \notin \{ \texttt{mjge}, \texttt{mjg}, \texttt{mhalt} \} \\ \neg\halted}
  {\abracket{S, Q} \xtrans{\mrfvc{\MAICM}} \abracket{S, Q^\prime}}
\]

\[
  \inferrule[rf-commit-other]{Q = \textit{rl} \bullet Q^\prime \\ \robrdy_\textit{rl} \\ \neg\robexcp_\textit{rl} \\ \funcfont{reg-write?}(\robmop_\textit{rl}) \\ \neg\halted}
  {\abracket{S, Q} \xtrans{\mrfvc{\MAICM}}\\ \abracket{[\rfv \mapsto [\robdst_\textit{rl} \mapsto \robval_\textit{rl}]\rfv]S, Q^\prime}}
\]

\[
  \inferrule[rf-commit-notrdy]{Q = \textit{rl} \bullet Q^\prime \\ \neg\robrdy_\textit{rl} \\ \neg\halted}
  {\abracket{S, Q} \xtrans{\mrfvc{\MAICM}} \abracket{S, \emptyseq}}
\]

Let \trs{\MAICM-\rfv} be a transition system, where
$S_{\MAICM-\rfv} = S_{\MAICM}$.

\[
  \inferrule[rf]{\abracket{S, \rob} \xtrans{\mrfvc{\MAICM}}^\ast \abracket{S^\prime, \emptyseq} \\ \neg\halted}
  {S \xtrans{\MAICM-\rfv} [\rfv \mapsto \rfv_{S^\prime}]S}
\]

\paragraph{\rob}

Since \MAICM\ is pipelined, it needs to deal with situations where the
contents of the pipeline must be invalidated as they correspond to
microinstructions that should not be brought to retirement. An
invalidation is necessary on a cycle if there exists a
microinstruction that will be committed on that cycle that satisfies
one of the following conditions: the microinstruction resulted in an
exception, the microinstruction is a jump, or the microinstruction is
a halt. Let $\funcfont{invalidate?} : \roblines{IC}^\ast$ be a predicate
over ROBs that holds iff the ROB indicates that an invalidation will
be required.

Let \trs{\mrbi{\MAICM}} be a transition system, where
$S_{\mrbi{\MAICM}} : S_{\MAICM} \times (\uinsts{IC} \times
\robids)^\ast$.

\[
  \inferrule[rob-issue-jmp]{Q = \abracket{u, \textit{rb}} \bullet Q^\prime \\ \funcfont{minst-op}(u) \in \{ \mathtt{mjg}, \mathtt{mjge} \} \\ \neg\halted}{\abracket{S, Q} \xtrans{\mrbi{\MAICM}}\\ \abracket{[\rob \mapsto \rob \mapp \sequence{\abracket{\textit{rb}, \funcfont{minst-op}(u), \texttt{nil}, \boolf, 0, \boolf}}]S, Q^\prime}}
\]

\noindent\texttt{mtsx-start}, \texttt{mtsx-end}, and \texttt{mhalt} also don't
write to a register, but they are marked as ready immediately upon issue.
\[
  \inferrule[rob-issue-halt-tsx-end]{Q = \abracket{u, \textit{rb}} \bullet Q^\prime \\ \funcfont{minst-op}(u) \in \{ \texttt{mtsx-end}, \texttt{mhalt} \} \\ \neg\halted}{\abracket{S, Q} \xtrans{\mrbi{\MAICM}}\\ \abracket{[\rob \mapsto \rob \mapp \sequence{\abracket{\textit{rb}, \funcfont{minst-op}(u), \texttt{nil}, \boolt, 0, \boolf}}]S, Q^\prime}}
\]

\[
  \inferrule[rob-issue-tsx-start]{Q = \abracket{u, \textit{rb}} \bullet Q^\prime \\ u = \instr{\texttt{mtsx-start}\ c} \\ \neg\halted}{\abracket{S, Q} \xtrans{\mrbi{\MAICM}}\\ \abracket{[\rob \mapsto \rob \mapp \sequence{\abracket{\textit{rb}, \funcfont{minst-op}(u), \texttt{nil}, \boolt, c, \boolf}}]S, Q^\prime}}
\]

\[
  \inferrule[rob-issue-other]{Q = \abracket{u, \textit{rb}} \bullet Q^\prime \\ \funcfont{minst-op}(u) \notin \{ \mathtt{mjg}, \mathtt{mjge}, \texttt{mtsx-start}, \texttt{mtsx-end}, \texttt{mhalt} \} \\ \neg\halted}{\abracket{S, Q} \xtrans{\mrbi{\MAICM}}\\ \abracket{[\rob \mapsto \rob \mapp \sequence{\abracket{\textit{rb}, \funcfont{minst-op}(u), \funcfont{reg-dst}(u), \boolf, 0, \boolf}}]S, Q^\prime}}
\]

Let \trs{\mrbw{\MAICM}} be a transition system, where
$S_{\mrbw{\MAICM}} : S_{\MAICM} \times \resstations{IC}^\ast$.

\textsc{rob-wrb-rdy} describes how the ROB is updated when a RS becomes
ready. Its definition hinges on two functions: $\funcfont{comp-val}$,
which uses the source operand values in the RS to compute the result
of the RS's microoperation, and $\funcfont{comp-exc}$ which
determines if the microoperation should result in an exception
instead. The appropriate ROB entry is updated with the result of these
two functions. In a MA state reachable from a clean start state, it
should always be the case that if a RS becomes ready, there exists
exactly one ROB line with that RS's destination ID in the ROB. 

\[
  \inferrule[rob-wrb-rdy]{Q = \textit{rs} \bullet Q^\prime \\ \cyclec = \rscpc_\textit{rs} \\ \rsbusy_\textit{rs} \\ \rsexec_\textit{rs} \\ \abracket{\exists i \from i \in \mathbb{N} \from \robid_{\rob(i)} = \rsdst_\textit{rs}} \\ \text{Let } i = \min_{j \in \mathbb{N} \wedge \robid_{\rob(j)} = \rsdst_\textit{rs}} j \\ \neg\halted}{\abracket{S, Q} \xtrans{\mrbw{\MAICM}} \La[\rob \mapsto [i \mapsto \\[\robval \mapsto \funcfont{comp-val}(rs, S), \robexcp \mapsto \funcfont{comp-exc}(\textit{rs})]\rob(i)]\rob]S\\, Q^\prime\Ra}
\]

\[
  \inferrule[rob-wrb-notrdy]{Q = \textit{rs} \bullet Q^\prime \\ \cyclec \neq \rscpc_\textit{rs} \vee \neg\rsbusy_\textit{rs} \vee \neg \rsexec_\textit{rs} \vee \neg \abracket{\exists i \from i \in \mathbb{N} \from \robid_{\rob(i)} = \rsdst_\textit{rs}} \\ \neg\halted}{\abracket{S, Q} \xtrans{\mrbw{\MAICM}} \abracket{S, Q^\prime}}
\]

%\drew{Note that we are assuming that if an RS is ready, its ROB destination ID corresponds to an entry that is in the ROB}

Let \trs{\mrbc{\MAICM}} be a transition system, where
$S_{\mrbc{\MAICM}} : S_{\MAICM} \times \roblines{IC}^\ast$.

\[
  \inferrule[rob-commit-invl]{Q = \textit{rl} \bullet Q^\prime \\ \robrdy_\textit{rl} \\ \robexcp_\textit{rl} \vee \robmop_\textit{rl} \in \{ \texttt{mhalt}, \texttt{mjg}, \texttt{mjge} \} \\ \neg\halted}
  {\abracket{S, Q} \xtrans{\mrbc{\MAICM}} \abracket{[\rob \mapsto \emptyseq]S, \emptyseq}}
\]

% \drew{The below could be modified to indicate that we're just popping one element off of the ROB in the state if it is unclear}
\[
  \inferrule[rob-commit-ok]{Q = \textit{rl} \bullet Q^\prime \\ \robrdy_\textit{rl} \\ \neg\robexcp_\textit{rl} \\ \robmop_\textit{rl} \notin \{ \texttt{mhalt}, \texttt{mjg}, \texttt{mjge} \} \\ \neg\halted}
  {\abracket{S, Q} \xtrans{\mrbc{\MAICM}} \abracket{[\rob \mapsto Q^\prime]S, Q^\prime}}
\]

\[
  \inferrule[rob-commit-notrdy]{Q = \textit{rl} \bullet Q^\prime \\ \neg\robrdy_\textit{rl} \\ \neg\halted}
  {\abracket{S, Q} \xtrans{\mrbc{\MAICM}} \abracket{S, \emptyseq}}
\]

Let \trs{\MAICM-\rob} be a transition system, where $S_{\MAICM-\rob} = S_{\MAICM} \times \natc$.

\[
  \inferrule[rob]{\abracket{S, \funcfont{decode-ic}(\funcfont{fetch-n}(\imem, \pc, n), \rob)} \xtrans{\mrbi{\MAICM}}^\ast \abracket{S^\prime, \emptyseq} \\ \abracket{S^\prime, \rsf} \xtrans{\mrbw{\MAICM}}^\ast \abracket{S^{\prime\prime}, \emptyseq} \\  \abracket{S^{\prime\prime}, \rob_{S^{\prime\prime}}} \xtrans{\mrbc{\MAICM}}^\ast \abracket{S^{\prime\prime\prime}, \emptyseq} \\ \neg\halted}
  {\abracket{S, n} \xtrans{\MAICM-\rob} \abracket{[\rob \mapsto \rob_{S^{\prime\prime\prime}}]S, n}}
\]

\paragraph{\rsf}

Let $\robids? = \robids \cup \{ \texttt{nil} \}$.

Let
$\funcfont{detect-raw-ic} : (\uinsts{IC} \times \robids)^\ast \rightarrow
(\robids? \times \robids?)^\ast$ be a function that given a sequence
of microinstructions each paired with the ROB entry they will be
assigned to, will determine for each microinstruction in the list that
uses a register whether that register is being written by a prior
microinstruction in the list. That is, if
$\funcfont{detect-raw-ic}(\sigma) = \pi$ and $\pi(i) = \abracket{a, b}$
then:
\[
  a = \begin{cases}
    \sigma(j)(2) &\parbox{6cm}{$\text{if } \funcfont{reg-op}_1(\sigma(i)(1))\downarrow \wedge \text{ for } S = \{ k \in \mathbb{N} \from k < i \wedge \funcfont{reg-dst}(\sigma(k)(1)) = \funcfont{reg-op}_1(\sigma(i)(1)) \}, \vert S \vert > 0 \wedge j = \max(S)$}\\
    \texttt{nil} &\text{otherwise}
  \end{cases}
\]
\[
  b = \begin{cases}
    \sigma(j)(2) &\parbox{6cm}{$\text{if } \funcfont{reg-op}_2(\sigma(i)(1))\downarrow \wedge  \text{ for } S = \{ k \in \mathbb{N} \from k < i \wedge \funcfont{reg-dst}(\sigma(k)(1)) = \funcfont{reg-op}_2(\sigma(i)(1)) \}, \vert S \vert > 0 \wedge j = \max(S)$}\\
    \texttt{nil} &\text{otherwise}
  \end{cases}
\]

Let
$\funcfont{next-idle-ic} : \resstations{IC}^\ast \rightharpoonup \mathbb{N}$
be a function that finds the index of an idle RS in the given sequence
of reservation stations.
\[
  \funcfont{next-idle-ic}(\sigma) = \begin{cases}
    \min(S) &\text{if } S = \{ i \in \fndom(\sigma) \from \neg\rsbusy_{\sigma(i)} \} \wedge \vert S \vert > 0\\
    \uparrow &\text{otherwise}
  \end{cases}
\]

Let
$\funcfont{rs-get-ic} : \rsids \times \resstations{IC}^\ast
\rightharpoonup \resstations{IC}$ be a function that finds the first RS
that has a particular ID in a sequence of RSes. That is,
\[
  \funcfont{rs-get-ic}(x, \sigma) = \begin{cases}
    \sigma(\min(S))&\text{if } \text{let } S = \{ j \in \mathbb{N} \from \rsid_{\sigma(j)} = x\}, \vert S \vert > 0\\
    \uparrow &\text{otherwise}
  \end{cases}
\]

Let \trs{\mrsfi{\MAICM}} be a transition system, where
$S_{\mrsfi{\MAICM}} : S_{\MAICM} \times (\uinsts{IC} \times
\robids \times \robids? \times \robids? \times \natc)^\ast$.

When issuing a microinstruction to a reservation station, \MAICM\ needs
to determine for each source operand of the microinstruction whether
that operand is going to come from another microinstruction, the
register file or a constant.
$\funcfont{setup-op-ic}_1 : \uinsts{IC} \times \robids? \times \resstations{IC} \times S_{\MAICM}
\rightarrow \resstations{IC}$ is a function that will perform this setup
for the first source operand, and $\funcfont{setup-op-ic}_2$ does the same
for the second source operand.

\begin{flalign*}
  &\funcfont{setup-op-ic}_1(u, \textit{dep}, \textit{rs}, S) =\\ &\begin{cases}
    [\rsqj \mapsto \texttt{nil}, \rsvj \mapsto 0]\textit{rs} &\text{if } \funcfont{minst-op}(u) = \texttt{mnoop}\\
    [\rsqj \mapsto \textit{dep}]\textit{rs} &\text{if } \textit{dep} \neq \texttt{nil}\\
    [\rsqj \mapsto \texttt{nil}, \rsvj \mapsto \robval_d]\textit{rs} &\text{if } \text{let } r_1 = \funcfont{reg-op}_1(u),\\&d = \funcfont{rob-get}(\regstreorder_{\regstat(r_1)}, \rob_S); \\&\regstat(r_1) \downarrow \wedge \regstbusy_{\regstat(r_1)} \wedge d \downarrow \wedge \robrdy_d\\
    [\rsqj \mapsto \regstreorder_{\regstat(r_1)}]\textit{rs} &\text{if } \text{let } r_1 = \funcfont{reg-op}_1(u),\\& d = \funcfont{rob-get}(\regstreorder_{\regstat(r_1)}, \rob_S); \\&\regstat(r_1) \downarrow \wedge \regstbusy_{\regstat(r_1)} \wedge (d \uparrow \vee \neg\robrdy_d)\\
    \parbox{3cm}{$[\rsqj \mapsto \texttt{nil}, \rsvj \mapsto \rf{\funcfont{reg-op}_1(u)}]\textit{rs}$} &\text{otherwise}
  \end{cases}
\end{flalign*}

\begin{flalign*}
  &\funcfont{setup-op-ic}_2(u, \textit{dep}, \textit{rs}, S) =\\ &\begin{cases}
    [\rsqk \mapsto \texttt{nil}, \rsvk \mapsto 0]\textit{rs} &\text{if } \funcfont{minst-op}(u) = \texttt{mnoop}\\
    [\rsqk \mapsto \textit{dep}]\textit{rs} &\text{if } \textit{dep} \neq \texttt{nil}\\
    [\rsqk \mapsto \texttt{nil}, \rsvk \mapsto \robval_d]\textit{rs} &\text{if } \text{let } r_2 = \funcfont{reg-op}_2(u),\\& d = \funcfont{rob-get}(\regstreorder_{\regstat(r_2)}, \rob_S); \\&\regstat(r_2) \downarrow \wedge \regstbusy_{\regstat(r_2)} \wedge d \downarrow \wedge \robrdy_d\\
    [\rsqk \mapsto \regstreorder_{\regstat(r_2)}]\textit{rs} &\text{if } \text{let } r_2 = \funcfont{reg-op}_2(u),\\& d = \funcfont{rob-get}(\regstreorder_{\regstat(r_2)}, \rob_S); \\&\regstat(r_2) \downarrow \wedge \regstbusy_{\regstat(r_2)} \wedge (d \uparrow \vee \neg\robrdy_d)\\
    \parbox{3cm}{$[\rsqk \mapsto \texttt{nil}, \rsvk \mapsto \rf{\funcfont{reg-op}_2(u)}]\textit{rs}$} &\text{otherwise}
  \end{cases}
\end{flalign*}

\[
  \inferrule[rsf-issue]{Q = \abracket{u, \textit{rb}, \textit{dep1}, \textit{dep2}, \textit{ipc}} \bullet Q^\prime \\ \funcfont{next-idle-ic}(\rsf)\downarrow \\ \funcfont{rs-needed?}(\funcfont{minst-op}(u)) \\
    \text{let } i = \funcfont{next-idle-ic}(\rsf) \\ \text{let } \textit{rs} = \funcfont{setup-op-ic}_2(u, \textit{dep2}, \funcfont{setup-op-ic}_1(u, \textit{dep1}, \rsf(i), S), S) \\ \neg\halted}
  {\abracket{S, Q} \xtrans{\mrsfi{\MAICM}} \La[\rsf \mapsto [i \mapsto [\rsmop \mapsto \funcfont{minst-op}(u),\\ \rsdst \mapsto \textit{rb}, \rsbusy \mapsto \boolt, \rspc \mapsto \textit{ipc}]\textit{rs}]\rsf]S, Q^\prime\Ra}
\]

\[
  \inferrule[rsf-issue-nors]{Q = \abracket{u, \textit{rb}, \textit{dep1}, \textit{dep2}, \textit{ipc}} \bullet Q^\prime \\ \neg\funcfont{rs-needed?}(\funcfont{minst-op}(u)) \\ \neg\halted}
  {\abracket{S, Q} \xtrans{\mrsfi{\MAICM}} \abracket{S, Q^\prime}}
\]

Let \trs{\mrsfe{\MAICM}} be a transition system, where
$S_{\mrsfe{\MAICM}} : S_{\MAICM} \times \resstations{IC}^\ast$.

Let $\funcfont{mop-time} : \uops{IC} \rightarrow \natc$ be a
function that determines how many cycles it takes to execute the given
microoperation.

Note that this transition system does not handle setting the \rsexec\
field to \boolf. $\mathcal{M}_{\MAICM-\rsf-\text{wr-b}}$ takes care of
this with the rule rsf-wb-ready.

Let
$\funcfont{check-barrier-start} : \robids \times \roblines{IC}^\ast
\rightarrow \bools$ be a predicate that determines whether it's OK for
the barrier microinstruction associated with the given ROB entry to
begin execution. This is true iff there are no uncommitted in-flight
memory access microinstructions prior to the barrier microinstruction
in question in program
order. $\funcfont{check-barrier-start}(\textit{id}, \textit{lines})$
can be computed by determining if any of the ROB lines returned by
$\funcfont{rob-before}(\textit{id}, \textit{lines})$ have
microoperations satisfying \funcfont{memory-op?}.

Let
$\funcfont{check-memory-start} : \robids \times \roblines{IC}^\ast
\rightarrow \bools$ be a predicate that determines whether it's OK for
the memory access microinstruction associated with the given ROB entry
to begin execution. This is true iff there are no uncommitted
in-flight memory barrier microinstructions prior to the barrier
microinstruction in question in program
order. $\funcfont{check-memory-start}(\textit{id}, \textit{lines})$
can be computed by determining if any of the ROB lines returned by
$\funcfont{rob-before}(\textit{id}, \textit{lines})$ have
microoperations satisfying \funcfont{barrier-op?}.

Note that here for simplicity we assume that \funcfont{barrier-op?}
and \funcfont{memory-op?} are mutually exclusive, \eg\ that no
microoperation exists that satisfies both predicates.

\[
  \inferrule[rsf-exec-wait-ready]{Q = \textit{rs} \bullet Q^\prime \\ \rsbusy_\textit{rs} \\ \rsqj_\textit{rs} = \texttt{nil} \\ \rsqk_\textit{rs} = \texttt{nil} \\ \rsexec_\textit{rs} \\ \cyclec \leq \rscpc_\textit{rs} \\ \neg\halted}
  {\abracket{S, Q} \xtrans{\mrsfe{\MAICM}} \abracket{S, Q^\prime}}
\]

% \[
%   \inferrule[rsf-exec-ready]{Q = \textit{rs} \bullet Q^\prime \\ \rsbusy_\textit{rs} \\ \rsqj_\textit{rs} = \texttt{nil} \\ \rsqk_\textit{rs} = \texttt{nil} \\ \rsexec_\textit{rs} \\ \cyclec = \rscpc_\textit{rs} \\
%     \text{let } \textit{rsi} = \text{rs-get-ic}(\rsid_\textit{rs}, \rsf) \\ \neg\halted}
%   {\abracket{S, Q} \xtrans{\mrsfe{\MAICM}} \abracket{[\rsf \mapsto [\textit{rsi} \mapsto [\rsexec \mapsto \boolf]\textit{rs}]\rsf]S, Q^\prime}}
% \]

\[
  \inferrule[rsf-exec-start-bar]{Q = \textit{rs} \bullet Q^\prime \\ \rsbusy_\textit{rs} \\ \rsqj_\textit{rs} = \texttt{nil} \\ \rsqk_\textit{rs} = \texttt{nil} \\ \neg\rsexec_\textit{rs} \\ \neg\halted \\ \funcfont{barrier-op?}(\rsmop_\textit{rs}) \\ \funcfont{check-barrier-start}(\rsid_\textit{rs}, \rob) }
  {\abracket{S, Q} \xtrans{\mrsfe{\MAICM}} \La[\rsf \mapsto [\rsid_{\textit{rs}} \mapsto\\ [\rsexec \mapsto \boolt, \rscpc \mapsto \cyclec \plusc \funcfont{mop-time}(\rsmop_\textit{rs})]\textit{rs}]\rsf]S,\\ Q^\prime\Ra}
\]

\[
  \inferrule[rsf-exec-start-mem]{Q = \textit{rs} \bullet Q^\prime \\ \rsbusy_\textit{rs} \\ \rsqj_\textit{rs} = \texttt{nil} \\ \rsqk_\textit{rs} = \texttt{nil} \\ \neg\rsexec_\textit{rs} \\ \neg\halted \\ \funcfont{memory-op?}(\rsmop_\textit{rs}) \\ \funcfont{check-memory-start}(\rsid_\textit{rs}, \rob)}
  {\abracket{S, Q} \xtrans{\mrsfe{\MAICM}} \La[\rsf \mapsto [\rsid_{\textit{rs}} \mapsto\\ [\rsexec \mapsto \boolt, \rscpc \mapsto \cyclec \plusc \funcfont{mop-time}(\rsmop_\textit{rs})]\textit{rs}]\rsf]S,\\ Q^\prime\Ra}
\]

\[
  \inferrule[rsf-exec-start]{Q = \textit{rs} \bullet Q^\prime \\ \rsbusy_\textit{rs} \\ \rsqj_\textit{rs} = \texttt{nil} \\ \rsqk_\textit{rs} = \texttt{nil} \\ \neg\rsexec_\textit{rs} \\ \neg\halted \\ \neg\funcfont{barrier-op?}(\rsmop_\textit{rs}) \\ \neg\funcfont{memory-op?}(\rsmop_\textit{rs})}
  {\abracket{S, Q} \xtrans{\mrsfe{\MAICM}} \La[\rsf \mapsto [\rsid_{\textit{rs}} \mapsto\\ [\rsexec \mapsto \boolt, \rscpc \mapsto \cyclec \plusc \funcfont{mop-time}(\rsmop_\textit{rs})]\textit{rs}]\rsf]S,\\ Q^\prime\Ra}
\]

\[
  \inferrule[rsf-exec-notready]{Q = \textit{rs} \bullet Q^\prime \\ \neg\rsbusy_\textit{rs} \vee \rsqj_\textit{rs} \neq \texttt{nil} \vee \rsqk_\textit{rs} \neq \texttt{nil} \\ \neg\halted}
  {\abracket{S, Q} \xtrans{\mrsfe{\MAICM}} \abracket{S, Q^\prime}}
\]

Let \trs{\mrsfw{\MAICM}} be a transition system, where
$S_{\mrsfw{\MAICM}} : S_{\MAICM} \times \resstations{IC}^\ast$.

$\funcfont{prop-single-ic} : \robids \times \natc \times
\resstations{IC} \rightarrow \resstations{IC}$ is a function that propagates a
completed RS execution (with the result being written to the ROB entry
with the given ID) to another RS.

\begin{flalign*}
  &\funcfont{prop-single-ic}(\textit{dst}, \textit{val}, \textit{rs}) =\\&\begin{cases}
    [\rsqj \mapsto \texttt{nil}, \rsvj \mapsto \textit{val}, \rsqk \mapsto \texttt{nil}, \rsvk \mapsto \textit{val}]\textit{rs} & \text{if } \rsqj = \textit{dst} \wedge \rsqk = \textit{dst}\\
    [\rsqj \mapsto \texttt{nil}, \rsvj \mapsto \textit{val}]\textit{rs} & \text{if } \rsqj = \textit{dst} \wedge \rsqk \neq \textit{dst}\\
    [\rsqk \mapsto \texttt{nil}, \rsvk \mapsto \textit{val}]\textit{rs} & \text{if } \rsqj \neq \textit{dst} \wedge \rsqk = \textit{dst}\\
    \textit{rs} & \text{otherwise}
  \end{cases}
\end{flalign*}

$\funcfont{prop-val-ic} : \robids \times \natc \times
\resstations{IC}^\ast \rightarrow \resstations{IC}^\ast$ applies
$\funcfont{prop-single-ic}$ with the given arguments to each element
of the given sequence of RSes to produce a new sequence of RSes.

\[
  \inferrule[rsf-wb-ready]{Q = \textit{rs} \bullet Q^\prime \\ \rsbusy_\textit{rs} \\ \rsexec_\textit{rs} \\ \cyclec = \rscpc_\textit{rs} \\
    \text{let } \textit{rsi} = \funcfont{rs-get-ic}(\rsid_\textit{rs}, \rsf) \\ \text{let } \textit{val} = \funcfont{comp-val}(\textit{rs}, S) \\ \text{let } \textit{excp} = \funcfont{comp-exc}(\textit{rs}, S) \\ \textit{rs-f}^\prime = \funcfont{prop-val-ic}(\rsdst_\textit{rs}, \textit{val}, \rsf) \\ \neg\halted}
  {\abracket{S, Q} \xtrans{\mrsfw{\MAICM}} \La[\rsf \mapsto [\textit{rsi} \mapsto\\ [\rsbusy \mapsto \boolf, \rsexec \mapsto \boolf]\textit{rs}]\textit{rs-f}^\prime]S, Q^\prime\Ra}
\]

\[
  \inferrule[rsf-wb-notready]{Q = \textit{rs} \bullet Q^\prime \\ \neg\rsbusy_\textit{rs} \vee \neg\rsexec_\textit{rs} \vee \cyclec \neq \rscpc_\textit{rs} \\ \neg\halted}
  {\abracket{S, Q} \xtrans{\mrsfw{\MAICM}} \abracket{S, Q^\prime}}
\]

Let \trs{\MAICM-\rsf} be a transition system, where
$S_{\MAICM-\rsf} = S_{\MAICM} \times \natc$.

Let
$\funcfont{decode-detect-raw-ic} : \insts{IC}^\ast \rightarrow (\uinsts{IC}
\times \robids \times \robids ? \times \robids ?))^\ast$
be a function that decodes the given sequence of instructions into a
sequence of microinstructions (using $\funcfont{decode-ic}$), then identifies
any read-after-write hazards for the (up to) two source operands for
each microinstruction (using $\funcfont{detect-raw-ic}$).

\begin{align*}
  &\text{Let } \pi = \funcfont{decode-detect-raw-ic}(\sigma).\\
  &\text{Let } \rho = \funcfont{decode-ic}(\sigma), \text{Let } \tau = \funcfont{detect-raw-ic}(\rho).\\
  &\text{Note that } \fndom(\rho) = \fndom(\tau).\\
  &\abracket{\forall i \from i \in \fndom(\rho) \from \pi(i) = \abracket{\rho(i)(1), \rho(i)(2), \tau(i)(1), \tau(i)(2)}}.
\end{align*}

\[
  \inferrule[rsf]{\abracket{S, \funcfont{decode-detect-raw-ic}(\funcfont{fetch-n}(\imem, \pc, n))} \xtrans{\mrsfi{\MAICM}}^\ast \abracket{S^\prime, \emptyseq} \\
    \abracket{S^\prime, \rsf_{S^\prime}} \xtrans{\mrsfe{\MAICM}}^\ast \abracket{S^{\prime\prime}, \emptyseq} \\
    \abracket{S^{\prime\prime}, \rsf_{S^{\prime\prime}}} \xtrans{\mrsfw{\MAICM}}^\ast \abracket{S^{\prime\prime\prime}, \emptyseq} \\ \neg\halted}
  {\abracket{S, n} \xtrans{\MAICM-\rsf} \abracket{[\rsf \mapsto \rsf_{S^{\prime\prime\prime}}]S, n}}
\]

\paragraph{\cmem}

$\funcfont{do-cache} : \powerset{\natc} \times (\natc
\rightharpoonup \natc) \times (\natc \rightharpoonup \natc)
\rightarrow (\natc \rightharpoonup \natc)$ is a function that takes in
a set of addresses to prefetch, a data memory and a cache and returns
the cache after caching all of the given addresses into it.

\[
  \trs{\mcc{\MAICM}}
\] is a transition system, where
$S_{\mcc{\MAICM}} = S_{\MAICM} \times \resstations{IC}^\ast$.

\[
  \inferrule[ma-cmem-commit-ldr]{Q = \textit{rs} \bullet Q^\prime \\ \cyclec = \rscpc_\textit{rs} \\ \rsbusy_\textit{rs} \\ \rsexec_\textit{rs} \\ \rsmop_\textit{rs} \in \{ \texttt{mldri}, \texttt{mldr} \} \\ a = \rsvj_\textit{rs} \plusc \rsvk_\textit{rs} \\ \neg\funcfont{comp-exc}(\textit{rs}) \\ \neg\halted \\ \cmem^\prime = [a \mapsto \partialget{\dmem}{a, 0}]\cmem}{\abracket{S, Q} \xtrans{\mcc{\MAICM}} \\\abracket{[\cmem \mapsto \funcfont{do-cache}(\prefetch(a), \dmem, \cmem^\prime)]S, Q^\prime}}
\]

\[
  \inferrule[ma-cmem-commit-other]{Q = \textit{rs} \bullet Q^\prime \\ \cyclec \neq \rscpc_\textit{rs} \vee \neg\rsbusy_\textit{rs} \vee \neg\rsexec_\textit{rs} \vee \rsmop_\textit{rs} \notin \{ \texttt{mldri}, \texttt{mldr} \} \\ \vee \funcfont{comp-exc}(\textit{rs}) \\ \neg\halted}{\abracket{S, Q} \xtrans{\mcc{\MAICM}} \abracket{[\cmem \mapsto S, Q^\prime}}
\]

Let \trs{\MAICM-\cmem} be a transition system, where
$S_{\MAICM-\cmem} = S_{\MAICM}$.

\[
  \inferrule[ma-cmem]{\abracket{S, \rsf} \xtrans{\mcc{\MAICM}}^\ast \abracket{S^\prime, \emptyseq} \\ \neg\halted}
  {S \xtrans{\MAICM-\cmem} [\cmem \mapsto \cmem_{S^\prime}]S}
\]

\subsection{Formal Semantics of \MAICNM}
\label{sec:maicnm-semantics}
\subsubsection{Transition System}
\trs{\MAICNM} is a nondeterministic transition system.
$S_{\MAICNM} = S_{\MAICM}$.

\subsubsection{Semantics}

The semantics of \M{\MAICNM} are broadly similar to that of
\M{\MAICM}, except where nondeterministic choices are made. We define
the top-level transition rules for \M{\MAICNM} below, and then discuss
the specific transition rules that differ. In general, the
nondeterminism is dealt with in the following way: the
\textsc{stepall} transition rule involves the nondeterministic
selection of a number of instructions to fetch and issue, the
%reservation stations that the issued instructions that need
%reservation stations will be assigned to $\rsmap$,
unavailable reservation station IDs \busyrs, the set of ROB lines
which are allowed to commit during this cycle (if they were eligible
to be committed otherwise) $\allowcommit$ and the set of reservation
station IDs that are allowed to begin execution this cycle (if they
were eligible to begin execution otherwise) $\allowstart$. These
selections are referred to by the subsidiary transition systems that
make up \M{\MAICNM}. Notice that some of the subsidiary transition
systems used to define \M{\MAICM} will be reused here, as their
behavior need not be modified.

\[
  \inferrule[halted]{\halted}{S \xtrans{\MAICNM} S}
\]

\[
  \inferrule[stepall]{\neg\halted \\ \text{Let } n \in \nats, n \leq \funcfont{max-fetch-n}(S) \\ \allowcommit \subseteq \robids  \\ \allowstart \subseteq \rsids \\ \busyrs \subseteq \rsids \\
    \abracket{S, n, \allowcommit} \xtrans{\MAICNM-\regstat} \abracket{\abracket{..., \regstat^\prime, ...}, n, \allowcommit} \\
    \abracket{S, \allowcommit} \xtrans{\MAICNM-\pc} \abracket{\abracket{..., \pc^\prime, ...}, \allowcommit} \\
    \abracket{S, \allowcommit} \xtrans{\MAICNM-\tsx} \abracket{\abracket{..., \tsx^\prime, ...}, \allowcommit} \\
    \abracket{S, \allowcommit} \xtrans{\MAICNM-\rfv} \abracket{\abracket{..., \rfv^\prime, ...}, \allowcommit} \\
    \abracket{S, \allowcommit} \xtrans{\MAICNM-\rob} \abracket{\abracket{..., \rob^\prime, ...}, \allowcommit} \\
    \abracket{S, n, \allowstart, \busyrs} \xtrans{\MAICNM-\rsf} \abracket{\abracket{..., \rsf^\prime, ...}, n, \allowstart, \busyrs}\\
    S \xtrans{\MAICM-\cmem} \abracket{..., \cmem^\prime, ...}
  }{S \xtrans{\MAICNM} [\regstat \mapsto \regstat^\prime, \fetchpc \mapsto \fetchpc \plusc n, \pc \mapsto \pc^\prime,\\ \tsx \mapsto \tsx^\prime, \rfv \mapsto \rfv^\prime, \rob \mapsto \rob^\prime, \rsf \mapsto \rsf^\prime, \cmem \mapsto \cmem^\prime]S}
\]

\paragraph{\regstat}
\[
  \trs{\mrsc{\MAICNM}}
\] is a transition system, where
\[
  S_{\mrsc{\MAICNM}} : S_{\MAICNM} \times \roblines{IC}^\ast \times \powerset{\robids}
\]

\[
  \inferrule[regstat-commit-ready-rm]{Q = \textit{rl} \bullet Q^\prime \\ \robrdy_{\textit{rl}} \\ \robid_{\textit{rl}} \in \allowcommit \\ \regstat(\robdst_\textit{rl}) \downarrow \\ \abracket{\textit{bsy}, \textit{reord}} = \regstat(\robdst_\textit{rl}) \\ \robid_\textit{rl} = \textit{reord} \\ \neg\halted}
  {\abracket{S, Q, \allowcommit} \xtrans{\mrsc{\MAICNM}}\\ \abracket{[\regstat \mapsto [\robdst_\textit{rl} \mapsto \uparrow]\regstat]S, Q^\prime, \allowcommit}}
\]

\[
  \inferrule[regstat-commit-ready-in-nomatch]{Q = \textit{rl} \bullet Q^\prime \\ \robrdy_{\textit{rl}} \\ \robid_{\textit{rl}} \in \allowcommit \\ \regstat(\robdst_\textit{rl}) \downarrow \\ \abracket{\textit{bsy}, \textit{reord}} = \regstat(\robdst_\textit{rl}) \\ \robid_\textit{rl} \neq \textit{reord} \\ \neg\halted}
  {\abracket{S, Q, \allowcommit} \xtrans{\mrsc{\MAICNM}} \abracket{S, Q^\prime, \allowcommit}}
\]

\[
  \inferrule[regstat-commit-ready-notin]{Q = \textit{rl} \bullet Q^\prime \\ \robrdy_{\textit{rl}} \\ \robid_{\textit{rl}} \in \allowcommit \\ \regstat(\robdst_\textit{rl}) \uparrow \\ \neg\halted}
  {\abracket{S, Q, \allowcommit} \xtrans{\mrsc{\MAICNM}} \abracket{S, Q^\prime, \allowcommit}}
\]

\[
  \inferrule[regstat-commit-notready]{Q = \textit{rl} \bullet Q^\prime \\ \neg\robrdy_\textit{rl} \vee \robid_{\textit{rl}} \notin \allowcommit \\ \neg\halted}
  {\abracket{S, Q, \allowcommit} \xtrans{\mrsc{\MAICNM}} \abracket{S, \emptyseq, \allowcommit}}
\]

\[
  \trs{\MAICNM-\regstat}
\] is a transition system, where
\[
  S_{\MAICNM-\regstat} = S_{\MAICNM} \times \natc \times \powerset{\robids}
\]

\[
  \inferrule[regstat]{\abracket{S, \funcfont{decode-ic}(\funcfont{fetch-n}(\imem, \pc, n), \rob)} \xtrans{\mrsi{\MAICM}}^\ast \abracket{S^\prime, \emptyseq} \\
    \abracket{S^\prime, \rob} \xtrans{\mrsc{\MAICNM}}^\ast \abracket{S^{\prime\prime}, \emptyseq} \\ \neg\halted}
  {\abracket{S, n} \xtrans{\MAICNM-\regstat}\\ \abracket{[\regstat \mapsto \regstat_{S^{\prime \prime}}]S, n}}
\]

\paragraph{\pc}

\[
  \trs{\mpcc{\MAICNM}}
\] is a transition system, where
\[
  S_{\mpcc{\MAICNM}} : S_{\MAICNM} \times \roblines{IC}^\ast \times \powerset{\robids}
\]

\[
  \inferrule[pc-commit-excp-tsx]{Q = \textit{rl} \bullet Q^\prime \\ \robrdy_{\textit{rl}} \\ \robid_{\textit{rl}} \in \allowcommit \\ \robexcp_\textit{rl} \\ \tsxactive \\ \neg\halted}
  {\abracket{S, Q, \allowcommit} \xtrans{\mpcc{\MAICNM}}\\ \abracket{[\pc \mapsto \tsxfallback]S, \emptyseq, \allowcommit}}
\]

\[
  \inferrule[pc-commit-excp-notsx]{Q = \textit{rl} \bullet Q^\prime \\ \robrdy_{\textit{rl}} \\ \robid_{\textit{rl}} \in \allowcommit \\ \robexcp_\textit{rl} \\ \neg\tsxactive \\ \neg\halted}
  {\abracket{S, Q, \allowcommit} \xtrans{\mpcc{\MAICNM}} \abracket{S, \emptyseq, \allowcommit}}
\]

\[
  \inferrule[pc-commit-mem]{Q = \textit{rl} \bullet Q^\prime \\ \robrdy_{\textit{rl}} \\ \robid_{\textit{rl}} \in \allowcommit \\ \neg\robexcp_\textit{rl} \\ \robmop_\textit{rl} = \texttt{mem-check} \vee \robmop_\textit{rl} = \texttt{memi-check} \\ \neg\halted}
  {\abracket{S, Q, \allowcommit} \xtrans{\mpcc{\MAICNM}} \abracket{S, Q^\prime, \allowcommit}}
\]

\[
  \inferrule[pc-commit-jmp]{Q = \textit{rl} \bullet Q^\prime \\ \robrdy_{\textit{rl}} \\ \robid_{\textit{rl}} \in \allowcommit \\ \neg\robexcp_\textit{rl} \\ \robmop_\textit{rl} = \texttt{mjg} \vee \robmop_\textit{rl} = \texttt{mjge} \\ \neg\halted}
  {\abracket{S, Q, \allowcommit} \xtrans{\mpcc{\MAICNM}} \abracket{[\pc \mapsto \robval_\textit{rl}]S, \emptyseq, \allowcommit}}
\]

\[
  \inferrule[pc-commit-halt]{Q = \textit{rl} \bullet Q^\prime \\ \robrdy_{\textit{rl}} \\ \robid_{\textit{rl}} \in \allowcommit \\ \neg\robexcp_\textit{rl} \\ \robmop_\textit{rl} = \texttt{mhalt} \\ \neg\halted}
  {\abracket{S, Q, \allowcommit} \xtrans{\mpcc{\MAICNM}} \abracket{[\pc \mapsto \pc \plusc 1]S, \emptyseq, \allowcommit}}
\]

\[
  \inferrule[pc-commit-other]{Q = \textit{rl} \bullet Q^\prime \\ \robrdy_{\textit{rl}} \\ \robid_{\textit{rl}} \in \allowcommit \\ \neg\robexcp_\textit{rl} \\ \robmop_\textit{rl} \notin \{\texttt{mem-check}, \texttt{memi-check}, \texttt{mjge}, \texttt{mjg}, \texttt{mhalt} \} \\ \neg\halted}
  {\abracket{S, Q, \allowcommit} \xtrans{\mpcc{\MAICNM}} \abracket{[\pc \mapsto \pc \plusc 1]S, Q^\prime, \allowcommit}}
\]

\[
  \inferrule[pc-commit-notrdy]{Q = \textit{rl} \bullet Q^\prime \\ \neg\robrdy_\textit{rl} \vee \robid_{\textit{rl}} \notin \allowcommit \\ \neg\halted}
  {\abracket{S, Q, \allowcommit} \xtrans{\mpcc{\MAICNM}} \abracket{S, \emptyseq, \allowcommit}}
\]

\[
  \trs{\MAICNM-\pc}
\] is a transition system, where
\[
  S_{\MAICNM-\pc} = S_{\MAICNM} \times \powerset{\robids}
\]

\[
  \inferrule[pc]{\abracket{S, \rob, \allowcommit} \xtrans{\mpcc{\MAICNM}}^\ast \abracket{S^\prime, \emptyseq, \allowcommit} \\ \neg\halted}
  {\abracket{S, \allowcommit} \xtrans{\MAICNM-\pc} \abracket{[\pc \mapsto \pc_{S^\prime}]S, \allowcommit}}
\]

\paragraph{\tsx}

\[
  \trs{\mtsxc{\MAICNM}}
\] is a transition system, where
\[
  S_{\mtsxc{\MAICNM}} : S_{\MAICNM} \times \roblines{IC}^\ast \times \powerset{\robids}
\]

\[
  \inferrule[tsx-commit-excp]{Q = \textit{rl} \bullet Q^\prime \\ \robrdy_{\textit{rl}} \\ \robid_{\textit{rl}} \in \allowcommit \\ \robexcp_\textit{rl} \\ \neg\halted}
  {\abracket{S, Q, \allowcommit} \xtrans{\mtsxc{\MAICNM}}\\ \abracket{[\tsxactive \mapsto \boolf]S, \emptyseq, \allowcommit}}
\]

\[
  \inferrule[tsx-commit-start]{Q = \textit{rl} \bullet Q^\prime \\ \robrdy_{\textit{rl}} \\ \robid_{\textit{rl}} \in \allowcommit \\ \neg\robexcp_\textit{rl} \\ \robmop_\textit{rl} = \texttt{mtsx-start} \\ \neg\halted}
  {\abracket{S, Q, \allowcommit} \xtrans{\mtsxc{\MAICNM}}\\ \abracket{[\tsxactive \mapsto \boolt, \tsxrf \mapsto \rfv, \tsxfallback \mapsto \robval_\textit{rl}]S, Q^\prime, \allowcommit}}
\]

\[
  \inferrule[tsx-commit-end]{Q = \textit{rl} \bullet Q^\prime \\ \robrdy_{\textit{rl}} \\ \robid_{\textit{rl}} \in \allowcommit \\ \neg\robexcp_\textit{rl} \\ \robmop_\textit{rl} = \texttt{mtsx-end} \\ \neg\halted}
  {\abracket{S, Q, \allowcommit} \xtrans{\mtsxc{\MAICNM}}\\ \abracket{[\tsxactive \mapsto \boolf]S, Q^\prime, \allowcommit}}
\]

\[
  \inferrule[tsx-commit-halt]{Q = \textit{rl} \bullet Q^\prime \\ \robrdy_{\textit{rl}} \\ \robid_{\textit{rl}} \in \allowcommit \\ \neg\robexcp_\textit{rl} \\ \robmop_\textit{rl} = \texttt{mhalt} \\ \neg\halted}
  {\abracket{S, Q, \allowcommit} \xtrans{\mtsxc{\MAICNM}} \abracket{S, \emptyseq, \allowcommit}}
\]

\[
  \inferrule[tsx-commit-other]{Q = \textit{rl} \bullet Q^\prime \\ \robrdy_{\textit{rl}} \\ \robid_{\textit{rl}} \in \allowcommit \\ \neg\robexcp_\textit{rl} \\ \robmop_\textit{rl} \notin \{\texttt{mtsx-start}, \texttt{mtsx-end}, \texttt{mhalt}\} \\ \neg\halted}
  {\abracket{S, Q, \allowcommit} \xtrans{\mtsxc{\MAICNM}} \abracket{S, Q^\prime, \allowcommit}}
\]

\[
  \inferrule[tsx-commit-notrdy]{Q = \textit{rl} \bullet Q^\prime \\ \neg\robrdy_\textit{rl} \vee \robid_{\textit{rl}} \notin \allowcommit \\ \neg\halted}
  {\abracket{S, Q, \allowcommit} \xtrans{\mtsxc{\MAICNM}} \abracket{S, \emptyseq, \allowcommit}}
\]

Let \trs{\MAICNM-\tsx} be a transition system, where
$S_{\MAICNM-\tsx} = S_{\MAICNM} \times \powerset{\robids}$.

\[
  \inferrule[tsx]{\abracket{S, \rob} \xtrans{\mtsxc{\MAICNM}}^\ast \abracket{S^\prime, \emptyseq} \\ \neg\halted}
  {\abracket{S, \allowcommit} \xtrans{\MAICNM-\tsx} \abracket{[\tsx \mapsto \tsx_{S^\prime}]S, \allowcommit}}
\]

\paragraph{\rfv}

Let \trs{\mrfvc{\MAICNM}} be a transition system, where
$S_{\mrfvc{\MAICNM}} : S_{\MAICNM} \times \roblines{IC}^\ast \times \powerset{\robids}$.

\[
  \inferrule[rf-commit-excp-tsx]{Q = \textit{rl} \bullet Q^\prime \\ \robrdy_{\textit{rl}} \\ \robid_{\textit{rl}} \in \allowcommit \\ \robexcp_\textit{rl} \\ \tsxactive \\ \neg\halted}
  {\abracket{S, Q, \allowcommit} \xtrans{\mrfvc{\MAICNM}} \abracket{[\rfv \mapsto \tsxrf]S, \emptyseq, \allowcommit}}
\]

\[
  \inferrule[rf-commit-excp-notsx]{Q = \textit{rl} \bullet Q^\prime \\ \robrdy_{\textit{rl}} \\ \robid_{\textit{rl}} \in \allowcommit \\ \robexcp_\textit{rl} \\ \neg\tsxactive \\ \neg\halted}
  {\abracket{S, Q, \allowcommit} \xtrans{\mrfvc{\MAICNM}} \abracket{S, \emptyseq, \allowcommit}}
\]

\[
  \inferrule[rf-commit-halt-jmp]{Q = \textit{rl} \bullet Q^\prime \\ \robrdy_{\textit{rl}} \\ \robid_{\textit{rl}} \in \allowcommit \\ \neg\robexcp_\textit{rl} \\ \robmop_\textit{rl} \in \{ \texttt{mjge}, \texttt{mjg}, \texttt{mhalt} \} \\ \neg\halted}
  {\abracket{S, Q, \allowcommit} \xtrans{\mrfvc{\MAICNM}} \abracket{S, \emptyseq, \allowcommit}}
\]

\[
  \inferrule[rf-commit-nowrite]{Q = \textit{rl} \bullet Q^\prime \\ \robrdy_{\textit{rl}} \\ \robid_{\textit{rl}} \in \allowcommit \\ \neg\robexcp_\textit{rl} \\ \neg\funcfont{reg-write?}(\robmop_\textit{rl}) \\ \robmop_\textit{rl} \notin \{ \texttt{mjge}, \texttt{mjg}, \texttt{mhalt} \} \\ \neg\halted}
  {\abracket{S, Q, \allowcommit} \xtrans{\mrfvc{\MAICNM}} \abracket{S, Q^\prime, \allowcommit}}
\]

\[
  \inferrule[rf-commit-other]{Q = \textit{rl} \bullet Q^\prime \\ \robrdy_{\textit{rl}} \\ \robid_{\textit{rl}} \in \allowcommit \\ \neg\robexcp_\textit{rl} \\ \funcfont{reg-write?}(\robmop_\textit{rl}) \\ \neg\halted}
  {\abracket{S, Q, \allowcommit} \xtrans{\mrfvc{\MAICNM}}\\ \abracket{[\rfv \mapsto [\robdst_\textit{rl} \mapsto \robval_\textit{rl}]\rfv]S, Q^\prime, \allowcommit}}
\]

\[
  \inferrule[rf-commit-notrdy]{Q = \textit{rl} \bullet Q^\prime \\ \neg\robrdy_\textit{rl} \vee \robid_{\textit{rl}} \notin \allowcommit \\ \neg\halted}
  {\abracket{S, Q, \allowcommit} \xtrans{\mrfvc{\MAICNM}} \abracket{S, \emptyseq, \allowcommit}}
\]

\[
\trs{\MAICNM-\rfv}
\] is a transition system, where
$S_{\MAICNM-\rfv} = S_{\MAICNM} \times \powerset{\robids}$.

\[
  \inferrule[rf]{\abracket{S, \rob} \xtrans{\mrfvc{\MAICNM}}^\ast \abracket{S^\prime, \emptyseq} \\ \neg\halted}
  {\abracket{S, \allowcommit} \xtrans{\MAICNM-\rfv} \abracket{[\rfv \mapsto \rfv_{S^\prime}]S, \allowcommit}}
\]

\paragraph{\rob}

\[
  \trs{\mrbc{\MAICNM}}
\] is a transition system, where
\[
  S_{\mrbc{\MAICNM}} : S_{\MAICNM} \times \roblines{IC}^\ast \times \powerset{\robids}
\]

\[
  \inferrule[rob-commit-invl]{Q = \textit{rl} \bullet Q^\prime \\ \robrdy_{\textit{rl}} \\ \robid_{\textit{rl}} \in \allowcommit \\ \robexcp_\textit{rl} \vee \robmop_\textit{rl} \in \{ \texttt{mhalt}, \texttt{mjg}, \texttt{mjge} \} \\ \neg\halted}
  {\abracket{S, Q, \allowcommit} \xtrans{\mrbc{\MAICNM}} \abracket{[\rob \mapsto \emptyseq]S, \emptyseq, \allowcommit}}
\]

% \drew{The below could be modified to indicate that we're just popping one element off of the ROB in the state if it is unclear}
\[
  \inferrule[rob-commit-ok]{Q = \textit{rl} \bullet Q^\prime \\ \robrdy_{\textit{rl}} \\ \robid_{\textit{rl}} \in \allowcommit \\ \neg\robexcp_\textit{rl} \\ \robmop_\textit{rl} \notin \{ \texttt{mhalt}, \texttt{mjg}, \texttt{mjge} \} \\ \neg\halted}
  {\abracket{S, Q, \allowcommit} \xtrans{\mrbc{\MAICNM}} \abracket{[\rob \mapsto Q^\prime]S, Q^\prime, \allowcommit}}
\]

\[
  \inferrule[rob-commit-notrdy]{Q = \textit{rl} \bullet Q^\prime \\ \neg\robrdy_\textit{rl} \vee \robid_{\textit{rl}} \notin \allowcommit \\ \neg\halted}
  {\abracket{S, Q, \allowcommit} \xtrans{\mrbc{\MAICNM}} \abracket{S, \emptyseq, \allowcommit}}
\]

\[
  \trs{\MAICNM-\rob}
\] is a transition system, where
\[
  S_{\MAICNM-\rob} = S_{\MAICNM} \times \natc \times \powerset{\robids}
\]

\[
  \inferrule[rob]{\abracket{S, \funcfont{decode-ic}(\funcfont{fetch-n}(\imem, \pc, n), \rob)} \xtrans{\mrbi{\MAICM}}^\ast \abracket{S^\prime, \emptyseq} \\ \abracket{S^\prime, \rsf} \xtrans{\mrbw{\MAICM}}^\ast \abracket{S^{\prime\prime}, \emptyseq} \\  \abracket{S^{\prime\prime}, \rob_{S^{\prime\prime}}, \allowcommit} \xtrans{\mrbc{\MAICNM}}^\ast \abracket{S^{\prime\prime\prime}, \emptyseq, \allowcommit} \\ \neg\halted}
  {\abracket{S, n, \allowcommit} \xtrans{\MAICNM-\rob} \abracket{[\rob \mapsto \rob_{S^{\prime\prime\prime}}]S, n, \allowcommit}}
\]

\paragraph{\rsf}

\[
  \trs{\mrsfi{\MAICNM}}
\] is a transition system, where
\[
  S_{\mrsfi{\MAICNM}} : S_{\MAICNM} \times (\uinsts{IC} \times
  \robids \times \robids? \times \robids? \times \natc)^\ast \times \powerset{\rsids}
\]

\M{\mrsfi{\MAICNM}} uses $\funcfont{setup-op-ic}_1$ and
$\funcfont{setup-op-ic}_2$ as defined in
Appendix~\ref{sec:maicm-semantics}. These functions are used to
determine the reference to use for each source operand of a
microinstruction.

% \[
%   \inferrule[rsf-issue]{Q = \abracket{u, \textit{rb}, \textit{dep1}, \textit{dep2}, \textit{ipc}} \bullet Q^\prime \\ \rsmap(\textit{rb})\downarrow \\ \funcfont{rs-needed?}(\funcfont{minst-op}(u)) \\
%     \text{let } i = \rsmap(\textit{rb}) \\ \text{let } \textit{rs} = \funcfont{setup-op-ic}_2(u, \textit{dep2}, \funcfont{setup-op-ic}_1(u, \textit{dep1}, \rsf(i), S), S) \\ \neg\halted}
%   {\abracket{S, Q, \rsmap} \xtrans{\mrsfi{\MAICNM}} \La[\rsf \mapsto [i \mapsto [\rsmop \mapsto \funcfont{minst-op}(u),\\ \rsdst \mapsto \textit{rb}, \rsbusy \mapsto \boolt, \rspc \mapsto \textit{ipc}]\textit{rs}]\rsf]S, Q^\prime, \rsmap\Ra}
% \]

% \[
%   \inferrule[rsf-issue-nors]{Q = \abracket{u, \textit{rb}, \textit{dep1}, \textit{dep2}, \textit{ipc}} \bullet Q^\prime \\ \neg\funcfont{rs-needed?}(\funcfont{minst-op}(u)) \\ \neg\halted}
%   {\abracket{S, Q, \rsmap} \xtrans{\mrsfi{\MAICNM}} \abracket{S, Q^\prime, \rsmap}}
% \]

Let
$\funcfont{rm-rs-ic} : \resstations{IC}^\ast \times \powerset{\rsids}
\rightarrow \resstations{IC}^\ast$ be a function that returns a modified
version of the given sequence, where all RSes with $\rsid$s in the
given set are removed.

\[
  \inferrule[rsf-issue]{Q = \abracket{u, \textit{rb}, \textit{dep1}, \textit{dep2}, \textit{ipc}} \bullet Q^\prime \\ \funcfont{next-idle-ic}(\funcfont{rm-rs-ic}(\rsf, \busyrs))\downarrow \\ \funcfont{rs-needed?}(\funcfont{minst-op}(u)) \\
    \text{let } i = \funcfont{next-idle-ic}(\funcfont{rm-rs-ic}(\rsf, \busyrs)) \\ \text{let } \textit{rs} = \funcfont{setup-op-ic}_2(u, \textit{dep2}, \funcfont{setup-op-ic}_1(u, \textit{dep1}, \rsf(i), S), S) \\ \neg\halted}
  {\abracket{S, Q, \busyrs} \xtrans{\mrsfi{\MAICNM}} \La[\rsf \mapsto [i \mapsto [\rsmop \mapsto \funcfont{minst-op}(u),\\ \rsdst \mapsto \textit{rb}, \rsbusy \mapsto \boolt, \rspc \mapsto \textit{ipc}]\textit{rs}]\rsf]S, Q^\prime, \busyrs\Ra}
\]

\[
  \inferrule[rsf-issue-nors]{Q = \abracket{u, \textit{rb}, \textit{dep1}, \textit{dep2}, \textit{ipc}} \bullet Q^\prime \\ \neg\funcfont{rs-needed?}(\funcfont{minst-op}(u)) \\ \neg\halted}
  {\abracket{S, Q, \busyrs} \xtrans{\mrsfi{\MAICNM}} \abracket{S, Q^\prime, \busyrs}}
\]

\[
  \trs{\mrsfe{\MAICNM}}
\] is a transition system, where
\[
  S_{\mrsfe{\MAICNM}} : S_{\MAICNM} \times \resstations{IC}^\ast \times \powerset{\rsids}
\]

$\funcfont{mop-time}$, $\funcfont{check-barrier-start}$ and
$\funcfont{check-memory-start}$ as defined in
Appendix~\ref{sec:maicm-semantics} are used here.

\[
  \inferrule[rsf-exec-wait-ready]{Q = \textit{rs} \bullet Q^\prime \\ \rsbusy_\textit{rs} \\ \rsqj_\textit{rs} = \texttt{nil} \\ \rsqk_\textit{rs} = \texttt{nil} \\ \rsexec_\textit{rs} \\ \cyclec \leq \rscpc_\textit{rs} \\ \neg\halted}
  {\abracket{S, Q, \allowstart} \xtrans{\mrsfe{\MAICNM}} \abracket{S, Q^\prime, \allowstart}}
\]

% \[
%   \inferrule[rsf-exec-ready]{Q = \textit{rs} \bullet Q^\prime \\ \rsbusy_\textit{rs} \\ \rsqj_\textit{rs} = \texttt{nil} \\ \rsqk_\textit{rs} = \texttt{nil} \\ \rsexec_\textit{rs} \\ \cyclec = \rscpc_\textit{rs} \\
%     \text{let } \textit{rsi} = \text{rs-get-ic}(\rsid_\textit{rs}, \rsf) \\ \neg\halted}
%   {\abracket{S, Q} \xtrans{\mrsfe{\MAICNM}} \abracket{[\rsf \mapsto [\textit{rsi} \mapsto [\rsexec \mapsto \boolf]\textit{rs}]\rsf]S, Q^\prime}}
% \]

\[
  \inferrule[rsf-exec-start-bar]{Q = \textit{rs} \bullet Q^\prime \\ \rsbusy_\textit{rs} \\ \rsqj_\textit{rs} = \texttt{nil} \\ \rsqk_\textit{rs} = \texttt{nil} \\ \neg\rsexec_\textit{rs} \\ \neg\halted \\ \funcfont{barrier-op?}(\rsmop_\textit{rs}) \\ \funcfont{check-barrier-start}(\rsid_\textit{rs}, \rob) \\ \rsid_\textit{rs} \in \allowstart }
  {\abracket{S, Q, \allowstart} \xtrans{\mrsfe{\MAICNM}} \La[\rsf \mapsto [\rsid_{\textit{rs}} \mapsto\\ [\rsexec \mapsto \boolt, \rscpc \mapsto \cyclec \plusc \funcfont{mop-time}(\rsmop_\textit{rs})]\textit{rs}]\rsf]S,\\ Q^\prime, \allowstart\Ra}
\]

\[
  \inferrule[rsf-exec-start-mem]{Q = \textit{rs} \bullet Q^\prime \\ \rsbusy_\textit{rs} \\ \rsqj_\textit{rs} = \texttt{nil} \\ \rsqk_\textit{rs} = \texttt{nil} \\ \neg\rsexec_\textit{rs} \\ \neg\halted \\ \funcfont{memory-op?}(\rsmop_\textit{rs}) \\ \funcfont{check-memory-start}(\rsid_\textit{rs}, \rob) \\ \rsid_\textit{rs} \in \allowstart}
  {\abracket{S, Q, \allowstart} \xtrans{\mrsfe{\MAICNM}} \La[\rsf \mapsto [\rsid_{\textit{rs}} \mapsto\\ [\rsexec \mapsto \boolt, \rscpc \mapsto \cyclec \plusc \funcfont{mop-time}(\rsmop_\textit{rs})]\textit{rs}]\rsf]S,\\ Q^\prime, \allowstart\Ra}
\]

\[
  \inferrule[rsf-exec-start]{Q = \textit{rs} \bullet Q^\prime \\ \rsbusy_\textit{rs} \\ \rsqj_\textit{rs} = \texttt{nil} \\ \rsqk_\textit{rs} = \texttt{nil} \\ \neg\rsexec_\textit{rs} \\ \neg\halted \\ \neg\funcfont{barrier-op?}(\rsmop_\textit{rs}) \\ \neg\funcfont{memory-op?}(\rsmop_\textit{rs}) \\ \rsid_\textit{rs} \in \allowstart}
  {\abracket{S, Q, \allowstart} \xtrans{\mrsfe{\MAICNM}} \La[\rsf \mapsto [\rsid_{\textit{rs}} \mapsto\\ [\rsexec \mapsto \boolt, \rscpc \mapsto \cyclec \plusc \funcfont{mop-time}(\rsmop_\textit{rs})]\textit{rs}]\rsf]S,\\ Q^\prime, \allowstart\Ra}
\]

\[
  \inferrule[rsf-exec-notready]{Q = \textit{rs} \bullet Q^\prime \\ \neg\rsbusy_\textit{rs} \vee \rsqj_\textit{rs} \neq \texttt{nil} \vee \rsqk_\textit{rs} \neq \texttt{nil} \vee \rsid_\textit{rs} \notin \allowstart \\ \neg\halted}
  {\abracket{S, Q, \allowstart} \xtrans{\mrsfe{\MAICNM}} \abracket{S, Q^\prime, \allowstart}}
\]

\[
  \trs{\MAICNM-\rsf}
\] is a transition system, where
\[
  S_{\MAICNM-\rsf} = S_{\MAICNM} \times \natc \times \powerset{\rsids} \times \powerset{\rsids}
\]

$\funcfont{decode-detect-raw-ic}$ is defined as in
Appendix~\ref{sec:maicm-semantics}.

\[
  \inferrule[rsf]{\text{Let } \textit{decoded} =  \funcfont{decode-detect-raw-ic}(\funcfont{fetch-n}(\imem, \pc, n))\\ \abracket{S, \textit{decoded}, \busyrs} \xtrans{\mrsfi{\MAICNM}}^\ast \abracket{S^\prime, \emptyseq, \busyrs} \\
    \abracket{S^\prime, \rsf_{S^\prime}, \allowstart} \xtrans{\mrsfe{\MAICNM}}^\ast \abracket{S^{\prime\prime}, \emptyseq, \allowstart} \\
    \abracket{S^{\prime\prime}, \rsf_{S^{\prime\prime}}} \xtrans{\mrsfw{\MAICM}}^\ast \abracket{S^{\prime\prime\prime}, \emptyseq} \\ \neg\halted}
  {\abracket{S, n, \allowstart, \busyrs} \xtrans{\MAICNM-\rsf} \abracket{[\rsf \mapsto \rsf_{S^{\prime\prime\prime}}]S, n, \allowstart, \busyrs}}
\]

\subsection{Formal Semantics of \MAICHM}
\label{sec:maichm-semantics}
\subsubsection{Transition System}
\[
  \trs{\MAICHM}
\] is a deterministic transition system.  States of \M{\MAICHM} are
\M{\MAICM} states augmented with history information:
$S_\MAICHM = S_\MAICM \times H_\MAICM$.

\[
  H_\MAICM : \abracket{\mahcommitcyc, \mahstartcyc, \mahcommitcache, \mahcacheeffects, \mahlines}
\]

\begin{itemize}
  \item $\mahcommitcyc : \natc$ is the cycle during which the most recent commit occurred
  \item $\mahstartcyc : \natc$ is the first cycle for which this history state has data
  \item $\mahcommitcache : \natc \rightharpoonup \natc$ is the cache state, without any updates that may have occurred since the last instruction commit
  \item $\mahcacheeffects : \robids \rightharpoonup (\natc \rightharpoonup \natc)$ maps a ROB identifier to the cache entries that should be added to the cache after committing that ROB line's microinstruction
  \item $\mahlines : \mahstatuslines^\ast$ contains information about the progress of all in-flight microinstructions
\end{itemize}

$\mahstatuslines : \abracket{\mahslrob, \mahslpc, \mahsllines}$

\begin{itemize}
\item $\mahslrob : \robids$ is the ID of the ROB line that this status
  information is for
\item $\mahslpc : \natc$ is the PC corresponding to the instruction
  loaded into the ROB line with an ID equal to $\mahslrob$
\item $\mahsllines : \mahstatuses^\ast$ is the sequence of statuses
  corresponding to this ROB's progress
\end{itemize}

$\mahstatuses ::= \texttt{fetch}\ \textit{pc}\ \textit{rsi} \ \vert\ \texttt{exec}\ \vert\ \texttt{wr-b}\ \textit{cache}\ \vert\ \texttt{delay} \vert\ \texttt{post-comm}$

\begin{itemize}
\item $\texttt{fetch} \ \textit{pc}\ \textit{rsi}$\ where
  $\textit{pc} \in \natc$ and
  $\textit{rsi} \in \rsids \cup \{ \texttt{nil} \}$ indicates that the
  microinstruction was fetched and issued. If
  $\textit{rsi} \neq \texttt{nil}$ then it indicates the RS to which
  this instruction was issued. If $\textit{rsi} = \texttt{nil}$, the
  microinstruction does not require a RS. $\textit{pc}$ indicates the
  PC of the instruction corresponding to the microinstruction that was
  fetched and issued.
\item \texttt{exec} indicates that the microinstruction was executing
  in an RS.
\item $\texttt{wr-b}\ \textit{cache}$ where
  $\textit{cache} \in \natc \rightharpoonup \natc$ indicates that the
  instruction wrote back, and $\textit{cache}$ indicates the value of
  the cache at the time of the write back.
\item \texttt{delay} indicates that the microinstruction either had
  not started execution because it was waiting on a dependency, or it
  had completed execution and written back to the ROB, but the ROB
  line had not yet been committed as another in-flight instruction
  that comes earlier in program order had not yet been committed.
\item \texttt{post-comm} indicates that the microinstruction is a
  \texttt{mem-check} or \texttt{memi-check} microinstruction that has
  been committed before its corresponding \texttt{mldr} or
  \texttt{mldri} instruction. This is the only case where a
  microinstruction's status line is retained after it is retired.
\end{itemize}

%It is expected that entries in $\mahlines$ will not have duplicate
%$\mahslrob$ values.

The history information gathered by \MAICHM{} will be used to
determine whether an arbitrary \MAICHM{} state is ``\good'', where
``\good'' means essentially that when the state is invalidated back to
the point at which the earliest in-flight instruction was issued and
run forward, it is possible to reach that state. All invalidated
states are considered \good, so the only states that we need to worry
about here are those that have in-flight instructions (a nonempty
pipeline).

% \[
%   \inferrule[halted]{\halted}{\abracket{S, H} \xtrans{\MAICHM} \abracket{S, H}}
% \]

% \[
%   \inferrule[stepall]{\neg\halted \\ \text{Let } n = \funcfont{max-fetch-n}(S) \\
%     \abracket{S, n} \xtrans{\MAICM-\regstat} \abracket{\abracket{..., \regstat^\prime, ...}, n} \\
%     S \xtrans{\MAICM-\pc} \abracket{..., \pc^\prime, ...} \\
%     S \xtrans{\MAICM-\tsx} \abracket{..., \tsx^\prime, ...} \\
%     S \xtrans{\MAICM-\rfv} \abracket{..., \rfv^\prime, ...} \\
%     S \xtrans{\MAICM-\rob} \abracket{..., \rob^\prime, ...} \\
%     \abracket{S, n} \xtrans{\MAICM-\rsf} \abracket{\abracket{..., \rsf^\prime, ...}, n}\\
%     S \xtrans{\MAICM-\cmem} \abracket{..., \cmem^\prime, ...}\\
%     \abracket{S, H} \xtrans{\MAICHM-\textit{hist}} \abracket{S^\prime, H^\prime}
%   }{\abracket{S, H} \xtrans{\MAICHM} \La[\regstat \mapsto \regstat^\prime, \fetchpc \mapsto \fetchpc \plusc n, \pc \mapsto \pc^\prime,\\ \tsx \mapsto \tsx^\prime, \rfv \mapsto \rfv^\prime, \rob \mapsto \rob^\prime, \rsf \mapsto \rsf^\prime, \cmem \mapsto \cmem^\prime]S, H^\prime\Ra}
% \]

$\xtrans{\MAICHM}$ treats the first component of the state in the
same way that \MAICM\ does. That is:
\[
  \abracket{s, h} \xtrans{\MAICHM} \abracket{s^\prime, h^\prime}
  \implies s \xtrans{\MAICM} s^\prime
\]

Let $\funcfont{will-commit?-ic} : S_\MAICM \rightarrow \bools$ be a function
that returns \boolt\ iff \rob\ is nonempty and given $q$ is the first
ROB line, $\robrdy_q$. This indicates that at least one
microinstruction will be committed on the next cycle.

Let
$\funcfont{to-commit-ic} : \roblines{IC}^\ast \rightarrow \roblines{IC}^\ast$
be a function that returns the sequence of ROB lines that will be
committed in the next step.

$\funcfont{to-commit-ic}(\sigma) = \sequence{\sigma_k}_{k < \max(S)}$, where:
\begin{flalign*}
  S = \{&i \in \fndom(\sigma) \from\\
        &\abracket{\forall x \from x \in \fndom(\sigma) \wedge x < i-1 \from \robrdy_{\sigma(x)} \wedge \neg\robexcp_{\sigma(x)} \wedge\\
        &\quad\robmop_{\sigma(x)} \notin \{ \texttt{mhalt}, \texttt{mjge}, \texttt{mjg} \} } \wedge\\
        &(\neg\robrdy_{\sigma(i)} \vee \robexcp_{\sigma(i)} \vee \robmop_{\sigma(i)} \in \{ \texttt{mhalt}, \texttt{mjge}, \texttt{mjg} \} \vee\\
        &i = \max(\fndom(\sigma))) \}
\end{flalign*}

Let $\funcfont{will-invld?-ic} : \roblines{IC}^\ast \rightarrow \bools$ be a
function that returns \boolt\ iff a microinstruction will be committed
that will result in an invalidation. Note that it is defined to only
check the last element of $\funcfont{to-commit-ic}(\sigma)$, as the
definition of $\funcfont{to-commit-ic}$ is such that if the returned
sequence contains an invalidation, it will always be the final element
of the sequence.

\begin{flalign*}
  &\funcfont{will-invld?-ic}(\sigma) \iff\\
  &\pi \neq \emptyseq \wedge (\robexcp_{\pi(i)} \vee \robmop_{\pi(i)} \in \{ \texttt{mhalt}, \texttt{mjge}, \texttt{mjg} \})\\
  &\text{ where } \pi = \funcfont{to-commit-ic}(\sigma),  i = \max(\fndom(\pi))
\end{flalign*}

\[
  \inferrule[ic-h-halted]{\halted}{\abracket{S, H} \xtrans{\MAICHM} \abracket{S, H}}
\]

\[
  \inferrule[ic-h-stepall]{\neg\halted \\ S \xtrans{\MAICM} S^\prime \\
    \abracket{S, H} \xtrans{\MAICHM-\mahcommitcyc} \abracket{S, \abracket{..., \mahcommitcyc^\prime, ...}} \\
    \abracket{S, H} \xtrans{\mhccc{\MAICHM}} \abracket{S, \abracket{..., \mahcommitcache^\prime, ...}} \\
    \abracket{S, H} \xtrans{\MAICHM-\mahcacheeffects} \abracket{S, \abracket{..., \mahcacheeffects^\prime, ...}} \\
    \abracket{S, H} \xtrans{\mhl{\MAICHM}} \abracket{S, \abracket{..., \mahlines^\prime, ...}} \\
    \abracket{S, H} \xtrans{\MAICHM-\mahstartcyc} \abracket{S, \abracket{..., \mahstartcyc^\prime, ...}}
  }
  {S \xtrans{\MAICHM} \La S^\prime, [\mahcommitcyc \mapsto \mahcommitcyc^\prime, \\\mahcommitcache \mapsto \mahcommitcache^\prime, \mahcacheeffects \mapsto \mahcacheeffects^\prime,\\ \mahlines \mapsto \mahlines^\prime, \mahstartcyc \mapsto \mahstartcyc^\prime]H\Ra}
\]

Note that in the below transition systems, it's not necessary to
handle ROB entries corresponding to ready jumps, ready halts, or ready
entries where the exception flag is set since the history is going to
be invalidated in those cases anyways. So, the below rules are not
going to handle those cases.

\paragraph{\mahcommitcyc}
\[
  \trs{\MAICHM-\mahcommitcyc}
\] is a transition
system. $S_{\MAICHM-\mahcommitcyc} = S_\MAICHM$.

\[
  \inferrule[ccyc-will-commit]{\neg\halted \\ \funcfont{will-commit?-ic}(S)}
  {\abracket{S, H} \xtrans{\MAICHM-\mahcommitcyc} \abracket{S, [\mahcommitcyc \mapsto \cyclec]H}}
\]

\[
  \inferrule[ccyc-will-not-commit]{\halted \vee \neg\funcfont{will-commit?-ic}(S)}
  {\abracket{S, H} \xtrans{\MAICHM-\mahcommitcyc} \abracket{S, H}}
\]

\paragraph{\mahcommitcache}

\[
  \trs{\mhccc{\MAICHM}}
\] is a transition
system.
$S_{\mhccc{\MAICHM}} = S_\MAICHM \times \roblines{IC}^\ast$.

\[
  \inferrule[commit-cache-commit]{Q = \textit{rl} \bullet Q^\prime \\ \robrdy_\textit{rl} \\ \robmop_\textit{rl} \in \{ \texttt{mldri}, \texttt{mldr} \} \\ \neg\funcfont{will-invld?-ic}(\rob) \\ \neg\halted \\
    \text{let } \textit{eff} = \partialget{\mahcacheeffects}{\robid_\textit{rl}, \emptyset} }
  {\abracket{\abracket{S, H}, Q} \xtrans{\mhccc{\MAICHM}}\\ \abracket{\abracket{S, [\mahcommitcache \mapsto \mahcommitcache \cup \textit{eff}]H}, Q^\prime}}
\]

\[
  \inferrule[commit-cache-rdy]{Q = \textit{rl} \bullet Q^\prime \\ \robrdy_\textit{rl} \\ \robmop_\textit{rl} \notin \{ \texttt{mldri}, \texttt{mldr} \} \\ \neg\funcfont{will-invld?-ic}(\rob) \\ \neg\halted}
  {\abracket{\abracket{S, H}, Q} \xtrans{\mhccc{\MAICHM}} \abracket{\abracket{S, H}, Q^\prime}}
\]

\[
  \inferrule[commit-cache-notrdy]{Q = \textit{rl} \bullet Q^\prime \\ \neg\robrdy_\textit{rl} \\ \neg\funcfont{will-invld?-ic}(\rob) \\ \neg\halted}
  {\abracket{\abracket{S, H}, Q} \xtrans{\mhccc{\MAICHM}} \abracket{\abracket{S, H}, \emptyseq}}
\]

% \[
%   \inferrule[commit-cache-commit]{Q = \textit{rl} \bullet Q^\prime \\ \robrdy_\textit{rl} \\ \robmop_\textit{rl} \notin \{ \texttt{mhalt}, \texttt{mjge}, \texttt{mjg} \} \\ \neg\halted}{\abracket{\abracket{S, H}, Q} \xtrans{\mhccc{\MAICHM}} \abracket{\abracket{S, [\mahcacheeffects \mapsto [\robid_\textit{rl} \mapsto \uparrow]\mahcacheeffects]H}, Q^\prime}}
% \]

\paragraph{\mahcacheeffects}

\[
  \trs{\mhceffc{\MAICHM}}
\] is a transition
system.
$S_{\mhceffc{\MAICHM}} = S_\MAICHM \times \roblines{IC}^\ast$.

\[
  \inferrule[cache-effects-committed-rm]{Q = \textit{rl} \bullet Q^\prime \\ \robrdy_\textit{rl} \\ \robmop_\textit{rl} \neq \texttt{mhalt} \\ \neg\funcfont{will-invld?-ic}(\rob) \\ \neg\halted}{\abracket{\abracket{S, H}, Q} \xtrans{\mhceffc{\MAICHM}}\\ \abracket{\abracket{S, [\mahcacheeffects \mapsto [\robid_\textit{rl} \mapsto \uparrow]\mahcacheeffects]H}, Q^\prime}}
\]

\[
  \inferrule[cache-effects-committed-other]{Q = \textit{rl} \bullet Q^\prime \\ \neg\robrdy_\textit{rl} \vee \robmop_\textit{rl} = \texttt{mhalt} \\ \neg\funcfont{will-invld?-ic}(\rob) \\ \neg\halted}{\abracket{\abracket{S, H}, Q} \xtrans{\mhceffc{\MAICHM}} \abracket{\abracket{S, H}, Q^\prime}}
\]

\trs{\mhceffw{\MAICHM}} is a transition
system.
$S_{\mhceffw{\MAICHM}} = S_\MAICHM \times \resstations{IC}^\ast$.

\[
  \inferrule[cache-effects-wr-b-ldr]{Q = \textit{rs} \bullet Q^\prime \\ \rsbusy_\textit{rs} \\ \rscpc_\textit{rs} = \cyclec \\ \rsmop_\texttt{rs} \in \{ \texttt{mldri}, \texttt{mldr} \} \\ \neg\funcfont{will-invld?-ic}(\rob) \\ \neg\halted \\ \text{let } \textit{ea} = \rsvj_\textit{rs} \plusc \rsvk_\textit{rs} }{\abracket{\abracket{S, H}, Q} \xtrans{\mhceffw{\MAICHM}}\\ \abracket{\abracket{S, [\mahcacheeffects \mapsto [\rsdst_\textit{rs} \mapsto [\textit{ea} \mapsto \dmem(\textit{ea})]]\mahcacheeffects]H}, Q^\prime}}
\]

\[
  \inferrule[cache-effects-wr-b-no-ldr]{Q = \textit{rs} \bullet Q^\prime \\ \neg\rsbusy_\textit{rs} \vee \rscpc_\textit{rs} \neq \cyclec \vee \rsmop_\texttt{rs} \notin \{ \texttt{mldri}, \texttt{mldr} \} \\ \neg\funcfont{will-invld?-ic}(\rob) \\ \neg\halted}{\abracket{\abracket{S, H}, Q} \xtrans{\mhceffw{\MAICHM}} \abracket{\abracket{S, H}, Q^\prime}}
\]

\[
  \trs{\MAICHM-\mahcacheeffects}
\] is a transition
system.
$S_{\MAICHM-\mahcacheeffects} = S_\MAICHM $.

\[
  \inferrule[cache-effects]{\abracket{\abracket{S, H}, \rob} \xtrans{\mhceffc{\MAICHM}}^\ast \abracket{\abracket{x, H^\prime}, \emptyseq} \\ \abracket{\abracket{S, H^\prime}, \rsf} \xtrans{\mhceffw{\MAICHM}}^\ast \abracket{\abracket{z, H^{\prime\prime}}, \emptyseq} \\ \neg\funcfont{will-invld?-ic}(\rob) \\ \neg\halted}
  {\abracket{S, H} \xtrans{\MAICHM-\mahcacheeffects} \abracket{S, [\mahcacheeffects \mapsto \mahcacheeffects_{H^{\prime\prime}}]H}}
\]

\paragraph{\mahlines}
\label{sec:mahlines-rules}

\[
  \trs{\mhlrm{\MAICHM}}
\] is a transition system. $S_{\mhlrm{\MAICHM}} = S_\MAICHM \times \roblines{IC}^\ast$.

$\funcfont{rm-hist-line} : \robids \times \mahstatuslines^\ast \rightarrow
\mahstatuslines^\ast$ is a function that removes any status line in the
given sequence that has the given ROB ID.

\[
  \funcfont{rm-hist-line}(\textit{id}, \sigma) = \pi
\]
where for $A = \{ i \in \fndom(\sigma) \from \mahslrob_{\sigma(i)} \neq \textit{id} \}$ and $\tau$ such that $\tau$ is a sequence consisting of the elements of $A$ in monotonically increasing order,
\[
  \abracket{\forall i \from i \in \fndom(\tau) \from \pi(i) = \sigma(\tau(i))}
\]

% \begin{flalign*}
%   &\funcfont{rm-hist-line}(\textit{id}, \sigma) = \pi \text{ such that:}\\
%   &\text{Let } A = \{ i \in \fndom(\sigma) \from \mahslrob_{\sigma(i)} \neq \textit{id} \}\\
%   &\text{Let } \tau \text{ be a sequence consisting of the elements of } A \text{ in monotonically increasing order.}\\
%   &\abracket{\forall i \from i \in \fndom(\tau) \from \pi(i) = \sigma(\tau(i))}
% \end{flalign*}

% \[
%   \inferrule[lines-rm-commit-invld]{Q = \textit{rb} \bullet Q^\prime \\ \robrdy_\textit{rb} \\ \robmop_\textit{rb} \in \{ \texttt{mhalt}, \texttt{mjge}, \texttt{mjg} \} \\ \neg\halted }
%   {\abracket{\abracket{S, H}, Q} \xtrans{\mhscrm{\MAICHM}} \abracket{\abracket{S, [\mahlines \mapsto \funcfont{rm-hist-line}(\robid_\textit{rb}, \mahlines)]H}, \emptyseq}}
% \]

\[
  \inferrule[lines-rm-commit-partial-ldr]{Q = \textit{rb} \bullet \textit{rb}^\prime \bullet Q^\prime \\ \robrdy_\textit{rb} \\ \neg\robrdy_{\textit{rb}^\prime} \\ \robmop_{\textit{rb}} \in \{ \texttt{mem-check}, \texttt{memi-check} \}\\
    \neg\funcfont{will-invld?-ic}(\rob) \\ \neg\halted }
  {\abracket{\abracket{S, H}, Q} \xtrans{\mhscrm{\MAICHM}} \abracket{\abracket{S, H}, \emptyseq}}
\]

\[
  \inferrule[lines-rm-commit-complete-ldr]{Q = \textit{rb} \bullet Q^\prime \\ \robrdy_\textit{rb} \\ \robmop_{\textit{rb}} \in \{ \texttt{ldr}, \texttt{ldri} \}\\
    \neg\funcfont{will-invld?-ic}(\rob) \\ \neg\halted\\
  \text{Let } H^\prime = [\mahlines \mapsto \funcfont{rm-hist-line}(\funcfont{prev}_{\robids}(\robid_{\textit{rb}}),\\ \funcfont{rm-hist-line}(\robid_\textit{rb}, \mahlines))]H}
  {\abracket{\abracket{S, H}, Q} \xtrans{\mhscrm{\MAICHM}} \abracket{\abracket{S, H^\prime}, Q^\prime}}
\]

\[
  \inferrule[lines-rm-commit-both-ldr]{Q = \textit{rb} \bullet \textit{rb}^\prime \bullet Q^\prime \\ \robrdy_\textit{rb} \\ \robrdy_{\textit{rb}^\prime} \\ \robmop_{\textit{rb}} \in \{  \texttt{mem-check}, \texttt{memi-check} \} \\
    \neg\funcfont{will-invld?-ic}(\rob) \\ \neg\halted\\
    \text{Let } H^\prime = [\mahlines \mapsto \funcfont{rm-hist-line}(\robid_{\textit{rb}^\prime},\\ \funcfont{rm-hist-line}(\robid_\textit{rb}, \mahlines))]H}
  {\abracket{\abracket{S, H}, Q} \xtrans{\mhscrm{\MAICHM}} \abracket{\abracket{S, H^\prime}, Q^\prime}}
\]

\[
  \inferrule[lines-rm-commit-no-invld]{Q = \textit{rb} \bullet Q^\prime \\ \robrdy_\textit{rb} \\ \robmop_{\textit{rb}} \notin \{  \texttt{mem-check}, \texttt{memi-check}, \texttt{ldr}, \texttt{ldri} \} \\
    \neg\funcfont{will-invld?-ic}(\rob) \\ \neg\halted\\
    \text{Let } H^\prime = [\mahlines \mapsto \funcfont{rm-hist-line}(\robid_\textit{rb}, \mahlines)]H}
  {\abracket{\abracket{S, H}, Q} \xtrans{\mhscrm{\MAICHM}} \abracket{\abracket{S, H^\prime}, Q^\prime}}
\]

\[
  \inferrule[lines-rm-commit-notrdy]{Q = \textit{rb} \bullet Q^\prime \\ \neg\robrdy_\textit{rb} \\ \neg\halted }
  {\abracket{\abracket{S, H}, Q} \xtrans{\mhscrm{\MAICHM}} \abracket{\abracket{S, H}, \emptyseq}}
\]

\trs{\mhhlw{\MAICHM}} is a transition system. \\
$S_{\mhhlw{\MAICHM}} = S_\MAICHM \times \roblines{IC}^\ast \times \bools$.

$\funcfont{add-status} : \mahstatuses \times \robids \times
\mahstatuslines^\ast \rightarrow \mahstatuslines^\ast$ is a function
that will add the given status to the statuses of the status line
associated with the given ROB ID. If no such status line exists, it
will be created and associated with the given ROB ID.

\[
  \inferrule[lines-skip-rdy]{Q = \textit{rb} \bullet Q^\prime \\ \robrdy_\textit{rb} \\ \textit{skip?} \\ \neg\funcfont{will-invld?-ic}(\rob) \\ \neg\halted}
  {\abracket{\abracket{S, H}, Q, \textit{skip?}} \xtrans{\mhhlw{\MAICHM}} \abracket{\abracket{S, H}, Q^\prime, \textit{skip?}}}
\]

% \[
%   \inferrule[lines-partial-ldr]{Q = \textit{rb} \bullet \textit{rb}^\prime \bullet Q^\prime \\ \robrdy_\textit{rb} \\ \neg\robrdy_{\textit{rb}^\prime} \\ \textit{skip?} \\ \neg\funcfont{will-invld?-ic}(\rob) \\ \neg\halted\\
%   \text{Let } H^\prime = [\mahlines \mapsto \funcfont{add-status}()]H}
%   {\abracket{\abracket{S, H}, Q, \textit{skip?}} \xtrans{\mhhlw{\MAICHM}} \abracket{\abracket{S, H^\prime}, Q^\prime, \boolf}}
% \]

\[
  \inferrule[lines-wait-first-notrdy]{Q = \textit{rb} \bullet Q^\prime \\ \neg\robrdy_\textit{rb} \\ \textit{skip?} \\ \neg\funcfont{will-invld?-ic}(\rob) \\ \neg\halted}
  {\abracket{\abracket{S, H}, Q, \textit{skip?}} \xtrans{\mhhlw{\MAICHM}} \abracket{\abracket{S, H}, Q^\prime, \boolf}}
\]

\[
  \inferrule[lines-waiting-rdy]{Q = \textit{rb} \bullet Q^\prime \\ \robrdy_\textit{rb} \\ \neg\textit{skip?} \\ \neg\funcfont{will-invld?-ic}(\rob) \\ \neg\halted}
  {\abracket{\abracket{S, H}, Q, \textit{skip?}} \xtrans{\mhhlw{\MAICHM}}\\ \abracket{\abracket{S, [\mahlines \mapsto \funcfont{add-status}(\texttt{delay}, \robid_\textit{rb}, \mahlines)]H}, Q^\prime, \textit{skip?}}}
\]

\[
  \inferrule[lines-waiting-nordy]{Q = \textit{rb} \bullet Q^\prime \\ \neg\robrdy_\textit{rb} \\ \neg\textit{skip?} \\ \neg\funcfont{will-invld?-ic}(\rob) \\ \neg\halted}
  {\abracket{\abracket{S, H}, Q, \textit{skip?}} \xtrans{\mhhlw{\MAICHM}}\\ \abracket{\abracket{S, H}, Q^\prime, \textit{skip?}}}
\]

% \[
%   \inferrule[startcyc-commit-no-invld]{Q = \textit{rb} \bullet Q^\prime \\ \robrdy_\textit{rb} \\ \robmop_\textit{rb} \notin \{ \texttt{mhalt}, \texttt{mjge}, \texttt{mjg} \} \\ \textit{rdy?} \\ \neg\halted\\\textit{new-start-cyc} = \funcfont{startcyc-remove-rb-from-lines}(\robid_\textit{rb}, \mahstartcyc_H, \mahlines_H) }
%   {\abracket{\abracket{S, H}, Q, \textit{rdy?}} \xtrans{\MAICHM-\mahstartcyc-\text{commit}} \abracket{\abracket{S, [\mahstartcyc \mapsto \textit{new-start-cyc}]H}, Q^\prime, \textit{rdy?}}}
% \]

% \[
%   \inferrule[startcyc-commit-not-rdy]{Q = \textit{rb} \bullet Q^\prime \\ \neg\robrdy_\textit{rb} \vee \neg\textit{rdy?} \\ \neg\halted\\\textit{new-start-cyc} = \funcfont{startcyc-remove-rb-from-lines}(\robid_\textit{rb}, \mahstartcyc_H, \mahlines_H) }
%   {\abracket{\abracket{S, H}, Q, \textit{rdy?}} \xtrans{\MAICHM-\mahstartcyc-\text{commit}} \abracket{\abracket{S, [\mahstartcyc \mapsto \textit{new-start-cyc}]H}, Q^\prime, \textit{rdy?}}}
% \]

\[
  \trs{\mhli{\MAICHM}}
\] is a transition system. 
$S_{\mhli{\MAICHM}} = S_\MAICHM \times (\uinsts{IC} \times
\robids \times (\rsids \cup \{ \texttt{nil} \}) \times \natc)^\ast$.

\[
  \inferrule[lines-issue]{Q = \abracket{u, \textit{rb}, \textit{rsi?}, \textit{pc}} \bullet Q^\prime \\ \neg\halted}{\abracket{\abracket{S, H}, Q} \xtrans{\mhli{\MAICHM}}\\ \abracket{\abracket{S, [\mahlines \mapsto \funcfont{add-status}(\texttt{fetch}\ \textit{pc}\ \textit{rsi?}, \textit{rb}, \mahlines)]H}, Q^\prime}}
\]

\[
  \trs{\mhlrs{\MAICHM}}
\] is a transition system.
$S_{\mhlrs{\MAICHM}} = S_\MAICHM \times \resstations{IC}$.

\[
  \inferrule[lines-rs-ready-wrb]{Q = \textit{rs} \bullet Q^\prime \\ \rsbusy_\textit{rs} \\ \rscpc_\textit{rs} = \cyclec \\ \neg\halted}{\abracket{\abracket{S, H}, Q} \xtrans{\mhlrs{\MAICHM}}\\ \abracket{\abracket{S, [\mahlines \mapsto \funcfont{add-status}(\texttt{wr-b}\ \cmem, \rsdst_\textit{rs}, \mahlines)]H}, Q^\prime}}
\]

\[
  \inferrule[lines-rs-exec-start]{Q = \textit{rs} \bullet Q^\prime \\ \rsbusy_\textit{rs} \\ \rsqj_\textit{rs} = \texttt{nil} \\ \rsqk_\textit{rs} = \texttt{nil} \\ \neg\rsexec_\textit{rs} \\ \neg\halted}
  {\abracket{\abracket{S, H}, Q} \xtrans{\mhlrs{\MAICHM}}\\ \abracket{\abracket{S, [\mahlines \mapsto \funcfont{add-status}(\texttt{exec}, \rsdst_\textit{rs}, \mahlines)]H}, Q^\prime}}
\]

\[
  \inferrule[lines-rs-exec-continue]{Q = \textit{rs} \bullet Q^\prime \\ \rsbusy_\textit{rs} \\ \rsexec_\textit{rs} \\ \rscpc_\textit{rs} \neq \cyclec \\ \neg\halted}
  {\abracket{\abracket{S, H}, Q} \xtrans{\mhlrs{\MAICHM}}\\ \abracket{\abracket{S, [\mahlines \mapsto \funcfont{add-status}(\texttt{exec}, \rsdst_\textit{rs}, \mahlines)]H}, Q^\prime}}
\]

\[
  \inferrule[lines-rs-delay]{Q = \textit{rs} \bullet Q^\prime \\ \rsbusy_\textit{rs} \\ \neg\rsexec_\textit{rs} \\ \rsqj_\textit{rs} \neq \texttt{nil} \vee \rsqk_\textit{rs} \neq \texttt{nil} \\ \neg\halted}
  {\abracket{\abracket{S, H}, Q} \xtrans{\mhlrs{\MAICHM}}\\ \abracket{\abracket{S, [\mahlines \mapsto \funcfont{add-status}(\texttt{delay}, \rsdst_\textit{rs}, \mahlines)]H}, Q^\prime}}
\]\

\[
  \inferrule[lines-rs-idle]{Q = \textit{rs} \bullet Q^\prime \\ \neg\rsbusy_\textit{rs} \\ \neg\halted}
  {\abracket{\abracket{S, H}, Q} \xtrans{\mhlrs{\MAICHM}} \abracket{\abracket{S, H}, Q^\prime}}
\]

\trs{\mhlmc{\MAICHM}} is a transition system. \\
$S_{\mhlmc{\MAICHM}} = S_\MAICHM$.

\[
  \inferrule[lines-mem-check-dly]{\rob = \textit{rb} \bullet \rob^\prime \\ \robrdy_\textit{rb} \\ \robmop_{\textit{rb}} \in \{ \texttt{mldr}, \texttt{mldri} \} \\ \neg\funcfont{will-invld?-ic}(\rob) \\ \neg\halted\\
    \text{Let } H^\prime = [\mahlines \mapsto \\\funcfont{add-status}(\texttt{post-comm}, \funcfont{prev}_{\robids}(\robid_{\textit{rb}}), \mahlines)]H}
  {\abracket{S, H} \xtrans{\mhlmc{\MAICHM}} \abracket{S, H^\prime}}
\]

\[
  \inferrule[lines-mem-check-empty]{\rob = \emptyseq \\ \neg\halted}
  {\abracket{S, H} \xtrans{\mhlmc{\MAICHM}} \abracket{S, H}}
\]

\[
  \inferrule[lines-mem-check-not-rdy-or-ldr]{\rob = \textit{rb} \bullet \rob^\prime \\ \neg\robrdy_\textit{rb} \vee \robmop_{\textit{rb}} \notin \{ \texttt{mldr}, \texttt{mldri} \} \vee \funcfont{will-invld?-ic}(\rob) \\ \neg\halted}
  {\abracket{S, H} \xtrans{\mhlmc{\MAICHM}} \abracket{S, H}}
\]

\trs{\mhl{\MAICHM}} is a transition system.
$S_{\mhl{\MAICHM}} = S_\MAICHM$.

$\funcfont{idle-rs-ids} : \resstations{IC}^\ast \rightharpoonup
\mathbb{N}^\ast$ is a function that finds the indices of idle RSes in
the given sequence of reservation stations. The indices are in order
with respect to the given sequence of reservation stations.

$\funcfont{fetch-with-pc} : S_\MAICM \rightarrow
(\insts{IC}, \natc)^\ast$ is a function that fetches the
appropriate number of instructions (based on $\funcfont{max-fetch-n}$) from
\imem, and produces a sequence pairing each instruction with its PC.
\begin{flalign*}
  &\text{Let } \sigma = \funcfont{fetch-with-pc}(\abracket{s, h})\\
  &\La\forall i \from i \in \{ 1, ..., \funcfont{max-fetch-n}(s)\}\ \from\\
  &\text{\quad\quad}\sigma(i) = \funcfont{fetch}_\textit{IC}(\imem, \fetchpc \plusc (i-1))\Ra
\end{flalign*}

$\funcfont{decode-rs-and-pc} : (\insts{IC},
\natc)^\ast \times \roblines{IC}^\ast \rightarrow (\uinsts{IC} \times \robids
\times (\rsids \cup \{ \textit{nil} \})\times \natc))^\ast$ is a
function that given a sequence of instructions to be issued and their
PCs, returns the sequence of microinstructions that will be issued,
the ROB ID they will be assigned to, the ID of the RS (if any) that
the microinstruction will be assigned to, and the associated PC.

%\drew{TODO flesh out the above.}

\[
  \inferrule[lines-no-invld]{
    \abracket{\abracket{S, H}, \rob} \xtrans{\mhlrm{\MAICHM}}^\ast \abracket{\abracket{S^\prime, H^\prime}, \emptyseq} \\
    \abracket{\abracket{S^\prime, H^\prime}, \rob} \xtrans{\mhhlw{\MAICHM}}^\ast \abracket{\abracket{S^{\prime\prime}, H^{\prime\prime}}, \emptyseq} \\
    \text{Let } \textit{dec} = \funcfont{decode-rs-and-pc}(\funcfont{fetch-n}_\textit{IC}(\imem, \pc, \funcfont{max-fetch-n}(S)))\\
    \abracket{\abracket{S^{\prime\prime}, H^{\prime\prime}}, \textit{dec}} \xtrans{\mhli{\MAICHM}}^\ast \abracket{\abracket{S^{\prime\prime\prime}, H^{\prime\prime\prime}}, \emptyseq} \\
    \abracket{\abracket{S^{\prime\prime\prime}, H^{\prime\prime\prime}}, \rsf} \xtrans{\mhlrs{\MAICHM}}^\ast \abracket{\abracket{S^{\prime\prime\prime\prime}, H^{\prime\prime\prime\prime}}, \emptyseq} \\
     \abracket{S^{\prime\prime\prime\prime}, H^{\prime\prime\prime\prime}} \xtrans{\mhlmc{\MAICHM}} \abracket{S^{\prime\prime\prime\prime\prime}, H^{\prime\prime\prime\prime\prime}} \\
    \neg\funcfont{will-invld?-ic}(\rob) \\ \neg\halted}
  {\abracket{S, H} \xtrans{\mhl{\MAICHM}} \abracket{S, [\mahlines \mapsto \mahlines_{H^{\prime\prime\prime\prime\prime}}]H}}
\]

\paragraph{\mahstartcyc}

As can be seen in Section~\ref{sec:mahlines-rules}, there are three
situations in which the cycle at which the earliest microinstruction
in $\mahlines$ was issued may change: (1) if the sequence of lines is
empty and a microinstruction is issued, (2) if a microinstruction is
committed, and (3) if the MA is invalidated.

$\funcfont{sc-rem-rb} : \robids \times \natc
\times \mahstatuses^\ast \rightarrow \natc$ is a function that
determines what the start cycle should be after removing the history
line corresponding to the given ROB ID. If the history line to be
removed is the oldest in the history (the first entry) and there are
at least two history lines, the start cycle must be adjusted by the
difference in cycles between the cycle during which the
microinstruction corresponding to the first entry was issued and the
cycle during which the microinstruction corresponding to the second
entry was issued.

\begin{flalign*}
  &\funcfont{sc-rem-rb}(\textit{id}, \textit{cy}, \sigma) = \\&\begin{cases}
    \textit{cy} &\text{if } \sigma = \emptyseq\\
    \textit{cy} &\text{if } \mahslrob_{\sigma(1)} \neq \textit{id}\\
    0 &\text{if } \mahslrob_{\sigma(1)} = \textit{id} \wedge \vert \sigma \vert = 1\\
    \textit{cy} \plusc (\vert \mahsllines_{\sigma(1)} \vert \minusc \vert \mahsllines_{\sigma(2)} \vert) &\text{if } \mahslrob_{\sigma(1)} = \textit{id} \wedge \vert \sigma \vert > 1\\
  \end{cases}
\end{flalign*}

\[
  \trs{\mhscc{\MAICHM}}
\] is a transition system. $S_{\mhscc{\MAICHM}} = S_\MAICHM \times \roblines{IC}^\ast$.
\[
  \inferrule[startcyc-commit-invld]{Q = \textit{rb} \bullet Q^\prime \\ \robrdy_\textit{rb} \\ \robmop_\textit{rb} \in \{ \texttt{mhalt}, \texttt{mjge}, \texttt{mjg} \} \\ \neg\halted\\\textit{new-start-cyc} = \funcfont{sc-rem-rb}(\robid_\textit{rb}, \mahstartcyc_H, \mahlines_H) }
  {\abracket{\abracket{S, H}, Q} \xtrans{\mhscc{\MAICHM}} \abracket{\abracket{S, [\mahstartcyc \mapsto \textit{new-start-cyc}]H}, \emptyseq}}
\]

\[
  \inferrule[startcyc-commit-no-invld]{Q = \textit{rb} \bullet Q^\prime \\ \robrdy_\textit{rb} \\ \robmop_\textit{rb} \notin \{ \texttt{mhalt}, \texttt{mjge}, \texttt{mjg} \} \\ \neg\halted\\\textit{new-start-cyc} = \funcfont{sc-rem-rb}(\robid_\textit{rb}, \mahstartcyc_H, \mahlines_H) }
  {\abracket{\abracket{S, H}, Q} \xtrans{\mhscc{\MAICHM}} \abracket{\abracket{S, [\mahstartcyc \mapsto \textit{new-start-cyc}]H}, Q^\prime}}
\]

\[
  \inferrule[startcyc-commit-not-rdy]{Q = \textit{rb} \bullet Q^\prime \\ \neg\robrdy_\textit{rb} \\ \neg\halted}
  {\abracket{\abracket{S, H}, Q} \xtrans{\mhscc{\MAICHM}} \abracket{\abracket{S, H}, \emptyseq}}
\]

\[
  \trs{\mhsci{\MAICHM}}
\] is a transition system.
$S_{\mhsci{\MAICHM}} = S_\MAICHM \times (\uinsts{IC} \times
\robids)^\ast$.

\[
  \inferrule[startcyc-issue-empty]{Q = \abracket{u, \textit{rb}} \bullet Q^\prime \\ \mahlines = \emptyseq \\ \neg\halted}{\abracket{\abracket{S, H}, Q} \xtrans{\mhsci{\MAICHM}} \abracket{\abracket{S, [\mahstartcyc \mapsto \cyclec]H}, Q^\prime}}
\]

\[
  \inferrule[startcyc-issue-nonempty]{Q = \abracket{u, \textit{rb}} \bullet Q^\prime \\ \mahlines \neq \emptyseq \\ \neg\halted}{\abracket{\abracket{S, H}, Q} \xtrans{\mhsci{\MAICHM}} \abracket{\abracket{S, H}, Q^\prime}}
\]

\[
  \trs{\MAICHM-\mahstartcyc}
\] is a transition system.
$S_{\MAICHM-\mahstartcyc} = S_\MAICHM$.

\[
  \inferrule[startcyc-invld]{\funcfont{will-invld?-ic}(\rob) \\ \neg\halted}{\abracket{S, H} \xtrans{\MAICHM-\mahstartcyc} \abracket{S, [\mahstartcyc \mapsto \cyclec \plusc 1]H}}
\]
%[\mahcommitcyc \mapsto \cyclec \plusc 1, \mahstartcyc \mapsto \cyclec \plusc 1, \mahcommitcache \mapsto \emptyset, \mahcacheeffects \mapsto \emptyset, \mahlines \mapsto \emptyseq]H

\[
  \inferrule[startcyc-no-invld]{\abracket{\abracket{S, H}, \rob} \xtrans{\mhscc{\MAICHM}}^\ast \abracket{\abracket{S^\prime, H^\prime}, \emptyseq} \\
    \text{Let } \textit{decoded} = \funcfont{decode-ic}(\funcfont{fetch-n}_\textit{IC}(\imem, \pc, \funcfont{max-fetch-n}(S)))\\
    \abracket{\abracket{S^\prime, H^\prime}, \textit{decoded}, \rob)} \xtrans{\mhsci{\MAICHM}}^\ast \abracket{\abracket{S^{\prime\prime}, H^{\prime\prime}}, \emptyseq} \\ \neg\funcfont{will-invld?-ic}(\rob) \\
    \neg\halted}{\abracket{S, H} \xtrans{\MAICHM-\mahstartcyc} \abracket{S, [\mahstartcyc \mapsto \mahstartcyc_{H^{\prime\prime}}]H}}
\]

\subsection{Formal Semantics of \MAICAM}
\label{sec:maicam-semantics}
\subsubsection{Transition System}
\mbox{}\\
\atrs{\MAICAM} is an action labeled transition system.
$S_{\MAICAM} = S_{\MAICM}$.

\subsubsection{Semantics}

\[
  \inferrule[halted]{\halted}{S \atrans{\MAICAM}{\emptyseq} S}
\]

\[
  \inferrule[stepall]{\neg\halted \\ S \xtrans{\MAICM} S^\prime \\
    a = \funcfont{auth-actions}(S, S^\prime)
  }{S \atrans{\MAICAM}{a} S^\prime}
\]

\section{Meltdown Proof Obligations}
\label{sec:appendix-meltdown-obligations}

We will now describe the proof obligations that arise from using our
notion of correctness for Meltdown on \M{\ISAICM} and
\M{\MAICM}. First, we will instantiate the set of entangled states
with $\mathcal{X} = \MAICM$. We use the formal definition from
Section~\ref{sec:entangled-states}, which requires that we provide
$\M{\MAICNM}$, $\M{\MAICHM}$, $\funcfont{step-using-h}_{\MAICNM}$,
$\funcfont{invl}_{\MAICM}$, $\funcfont{init-h}_{\MAICM}$ and
$S^{\textit{init}}_{\MAICM}$. We briefly discussed $\M{\MAICNM}$ and
$\M{\MAICHM}$ above and full definitions can be found in
Appendices~\ref{sec:maicnm-semantics} and \ref{sec:maichm-semantics}
respectively. The rest of the functions are defined below.

\[
  \funcfont{reset-rs}(\textit{rs}) = [\rsbusy \mapsto \boolf, \rsexec \mapsto \boolf]\textit{rs}
\]

\[
  \funcfont{reset-rs-f}(\sigma) = \pi \text{ such that } \abracket{\forall i \from i \in \fndom(\sigma) \from \pi(i) = \funcfont{reset-rs}(\sigma(i))}
\]

\[
  \funcfont{comp-start-cyc}_{\MAICM}(s, h) = \begin{cases}
    \cyclec &\text{if } \vert\mahlines\vert = 0\\
    \mahstartcyc &\text{otherwise}
  \end{cases}
\]

\begin{flalign*}
  \funcfont{invl}_{\MAICM}(s, h) &= [\\
                                                &\fetchpc \mapsto \pc,\\
                                                &\rob \mapsto \emptyseq,\\
                                                &\regstat \mapsto \emptyseq,\\
                                                &\rsf \mapsto \funcfont{reset-rs-f}(\rsf)\\
                                                &\cmem \mapsto \mahcommitcache\\
                                                &\cyclec \mapsto \funcfont{comp-start-cyc}_{\MAICM}(s, h)\\
                           ]s&
\end{flalign*}

$\funcfont{init-h}_{\MAICM}(s) = \abracket{\cyclec_s, \cyclec_s, \emptyset, \emptyset, \emptyset}$

\begin{flalign*}
  S^{\textit{init}}_{\MAICM} =  \{ s \in &S_\MAICM \from \fetchpc_s = \pc_s \wedge \rob_s = \emptyseq \wedge \regstat_s = \emptyseq \wedge\\
                                                        &\abracket{\forall i \from i \in \fndom(\rsf) \from \neg\rsbusy_{\rsf(i)} \wedge \neg\rsexec_{\rsf(i)}} \}
\end{flalign*}

\[
  \funcfont{steps-to-take}_{\MAICHM}(\abracket{s, h}) = \begin{cases}
    0 &\text{if } \mahlines_h = \emptyseq\\
    \cyclec_s \minusc \mahstartcyc_h &\text{otherwise}
  \end{cases}
\]

$\funcfont{step-using-h}_{\MAICNM}$ operates by calculating the
appropriate values for $n$, $\allowcommit$, $\allowstart$ and
$\busyrs$, and then using the \textsc{stepall} transition rule for
\M{\MAICNM} with those values.

Let
$\funcfont{get-h} : H_{\MAICM} \times \natc \rightarrow (\robids
\times \natc \times \mahstatuses)^\ast$ be a function that given
history information and a cycle, gets a sequence of tuples, where each
tuple describes the status of one of the ROB lines during the given
cycle.

$n$ can be calculated for a state $\abracket{s, h}$ by counting the
number of ROB lines with a \texttt{fetch} status in
$\funcfont{get-h}(h, \cyclec_s)$.

$\busyrs$ can be calculated for a state $\abracket{s, h}$ by computing
$\rsids \setminus \mathit{used}$ where $\mathit{used}$ is a set
computed by taking all of the \texttt{fetch} status in
$\funcfont{get-h}(h, \cyclec_s)$, selecting only those statuses that
indicate an assignment to an RS, and collecting the RS IDs from such
statuses.

$\allowcommit$ can be calculated for a state $\abracket{s, h}$ by
computing $\robids \setminus \mathit{used}$ where $\mathit{used}$ is a
set containing all of the ROB IDs in $\funcfont{get-h}(h, \cyclec_s)$.
%that are associated with a status that is not \texttt{post-comm}

$\allowstart$ can be calculated for a state $\abracket{s, h}$ by
finding all of the reservation stations $\mathit{rs}$ in $\rsf_s$ such that
$\rsbusy_{\mathit{rs}}$ and $\rsdst_{\mathit{rs}}$ is one of the ROB
IDs that has a \texttt{exec} status in
$\funcfont{get-h}(h, \cyclec_s)$, and then collecting the $\rsid$ for
all such RSes.

%$\funcfont{step-using-h}_{\MAICNM}(s, h)$
%\drew{TODO finish up step-using-h}

Then, we get that:

\begin{flalign*}
  S^\textit{\goodshort}_{\MAICHM} = \{ &\abracket{s, h} \in S_{\MAICHM}, i = \funcfont{steps-to-take}_{\MAICHM}(\abracket{s, h}) \from\\
                                       &\abracket{\exists h \from h^\prime \in H \from \funcfont{step-using-h}^i_{\MAICNM}(\funcfont{invl}_{\mathcal{X}}(s, h), h) = \abracket{s, h^\prime}}\}
\end{flalign*}

We are claiming that \M{\ISAICM}, \M{\MAICM}, \M{\MAICNM} and
\M{\MAICHM} are all TRSes. This means that they must all be well-typed
and left-total, as is required by the definition of a TRS.

From the use of the definition of the set of entangled states, we now
must discharge the following obligations:
\begin{equation}
  \label{eqn:apdx-maicn-superset-behavior-maic}
  \abracket{\forall s, u \from s, u \in S_{\MAICM} \wedge s
    \xtrans{\MAICM} u \from s \xtrans{\MAICNM} u}
\end{equation}
\begin{equation}
  \label{eqn:apdx-maicn-bisim-maich}
  \begin{aligned}
  &\M{\MAICHM}\sim_{\funcfont{hist}}\M{\MAICNM} \text{ where }
  \funcfont{hist} \text{ is a function such that }\\
  &\abracket{\forall s, h \from \abracket{s, h} \in S_{\MAICHM}
    \from \funcfont{hist}(\abracket{s, h}) = s}
  \end{aligned}
\end{equation}
\begin{equation}
  \label{eqn:apdx-maic-init-states-entangled}
  \abracket{\forall s \from s \in S^{\textit{init}}_{\MAICM} \from \abracket{s, \funcfont{init-h}_{\MAICM}(s)} \in S^\textit{\goodshort}_{\MAICHM}}
\end{equation}
\begin{equation}
  \label{eqn:apdx-maich-closed-under-entangled}
  \abracket{\forall s \from s \in S^{\textit{\goodshort}}_{\MAICHM} \from \abracket{\forall w \from s \xtrans{\MAICHM} w \from w \in S^\textit{\goodshort}_{\MAICHM}}}
\end{equation}
In addition, our notion of correctness for Meltdown requires that
\M{\MAGICM} is a witness skipping refinement of \M{\ISAICM} with
respect to our refinement map $\funcfont{r-ic}$, defined below. This
is proved by showing the existence of a witness skipping relation on
the transition system produced by taking the ``disjoint union'' of
\M{\MAGICM} and \M{\ISAICM}. Let
$\M{\textit{ic}} = \disjtrs{\MAGICM}{\ISAICM}{}$ be this system. Let
$S_{\textit{ic}} = S_{\MAGICM} \uplus S_{\ISAICM}$ and
$\xtrans{\textit{ic}} = \xtrans{\MAGICM} \uplus \xtrans{\ISAICM}$. We
instantiate Definition~\ref{def:witness-sk}, providing
$\funcfont{skip-wit-ic} : S_{\textit{ic}} \times S_{\textit{ic}}
\rightarrow \nats \setminus \{ 0 \}$ for $\funcfont{skip-wit}$,
$\funcfont{stutter-wit-ic} : S_{\textit{ic}} \times S_{\textit{ic}}
\rightarrow \nats$ for $\funcfont{stutter-wit}$,
$\funcfont{run-ic} : S_{\textit{ic}} \times S_{\textit{ic}} \times
S_{\textit{ic}} \rightarrow S_{\textit{ic}} $ for $\funcfont{run}$,
and
$\skiprel{ic} \subseteq S_{\textit{ic}} \times S_{\textit{ic}} $
for $B$. The obligations generated are as follows:
\begin{equation}
  \label{eqn:apdx-wsk1-meltdown}
  \La \forall s \in S_{\MAGICM} :: s \skiprel{ic} \funcfont{r-ic}.s\Ra
\end{equation}
\begin{equation}
  \label{eqn:apdx-wsk2-meltdown}
  \abracket{\forall w, s, u \from s \skiprel{ic} w \wedge s \xtrans{\textit{ic}} u \from w \xtrans{\textit{ic}}^{\funcfont{skip-wit-ic}(s, u)} \funcfont{run-ic}(w, s, u)}
\end{equation}
\begin{equation}
  \label{eqn:apdx-wsk3-meltdown}
  \begin{aligned}
  &\forall s,u,w \in S_{\textit{ic}}: s \skiprel{ic} w \And s \xtrans{\textit{ic}} u: &\\
  &\text{\quad } (u \skiprel{ic} w \And \funcfont{stutter-wit-ic}(u, w) < \funcfont{stutter-wit-ic}(s, w))\ \vee&\\
  &\text{\quad } u \skiprel{ic} (\funcfont{run-ic}(w, s, u))&
  \end{aligned}
\end{equation}
% \begin{align}
%   \La \forall s \in S_{\MAGICM} :: s \skiprel{ic} \funcfont{r-ic}.s\Ra \label{eqn:apdx-wsk1-meltdown}\\
%   \abracket{\forall w, s, u \from s \skiprel{ic} w \wedge s \xtrans{\textit{ic}} u \from w \xtrans{\textit{ic}}^{\funcfont{skip-wit-ic}(s, u)} \funcfont{run-ic}(w, s, u)} \label{eqn:apdx-wsk2-meltdown}\\
%   \begin{aligned}
%   &\forall s,u,w \in S_{\textit{ic}}: s \skiprel{ic} w \And s \xtrans{\textit{ic}} u: &\\
%   &\text{\quad } (u \skiprel{ic} w \And \funcfont{stutter-wit-ic}(u, w) < \funcfont{stutter-wit-ic}(s, w))\ \vee&\\
%   &\text{\quad } u \skiprel{ic} (\funcfont{run-ic}(w, s, u))&
%   \end{aligned}   \label{eqn:apdx-wsk3-meltdown}
% \end{align}
Recall that
\[
   S_{\ISAICM} : \La\pc, \rfv, \tsx, \halted, \imem, \dmem,
   \goodaddr, \cmem\Ra
\]

The refinement map for \M{\MAGICM} and label function for \M{\ISAICM}
are as follows:
\[
  \funcfont{r-ic}(\abracket{s,h}) = \La \pc_s, \rfv_s, \tsx_s, \halted_s, \imem_s, \dmem_s, \goodaddr_s, \emptyseq \Ra
\]
\[
  L_{\ISAICM}(s) = [\cmem \mapsto \emptyseq]s
\]

We then define $\skiprel{ic}$ in the following way:
% \[
%   \funcfont{to-isa-melt}(s) = \begin{cases}
%     s &\text{if } s \in S_{\ISAICM}\\
%     \funcfont{r-ic}(s) &\text{if } s \in S_{\MAGICM}
%   \end{cases}
% \]
\begin{flalign*}
  &\skiprel{ic}(s, w) \iff \\&\begin{cases}
    s = w &\text{if } s, w \in S_{\ISAICM} \vee s, w \in S_{\MAGICM}\\
    L_{\ISAICM}(s) = L_{\ISAICM}(\funcfont{r-ic}(w)) &\text{if } s \in S_{\ISAICM} \wedge w \in S_{\MAGICM}\\
    L_{\ISAICM}(\funcfont{r-ic}(s)) = L_{\ISAICM}(w) &\text{otherwise}
    %L_{\ISAICM}(\funcfont{to-isa-melt}(s)) = L_{\ISAICM}(\funcfont{to-isa-melt}(w)) &\text{otherwise}
  \end{cases}
\end{flalign*}

Note that $\skiprel{ic}$ and the above obligations are stated in a
way that is agnostic of whether the two related states $s$ and $w$ are
both from $S_{\MAGICM}$ or $S_{\ISAICM}$, or whether they are from
different systems. For the sake of brevity, we will only give a short
discussion regarding handling the case where the two states are in the
same system: for all $s, w \in S_{\textit{ic}}$ such that
$s \in S_{\MAGICM} \wedge w \in S_{\MAGICM}$ or
$s \in S_{\ISAICM} \wedge w \in S_{\ISAICM}$, the following hold:
$\funcfont{skip-wit-ic}(s, w) = 1$ and
$\funcfont{stutter-wit-ic}(s, w) = 0$.
%$\funcfont{run-ic}$ is straightforward in both

We now discuss the behavior when the two states are in different
systems. We focus primarily on the case where $s \in S_{\MAGICM}$ and
$w \in S_{\ISAICM}$.

We define $\funcfont{stutter-wit-ic}(s, w)$ to be a function that
returns the number of steps it will take starting at the state $s$
before at least one instruction is retired. By inspecting the
transitions of \M{\MAGICM} it is straightforward to produce a method
for computing this value.
%Intuitively this definition
%makes sense since as soon as t

$\funcfont{skip-wit-ic}(s, u)$ is a function that returns the number
of instructions that are committed in the transition from $s$ to
$u$. This is exactly the number of \M{\ISAICM} steps that should be
required to match the behavior of the \M{\MAGICM} step.

$\funcfont{run-ic}(w, s, u)$ is a function that steps $w$
$\funcfont{skip-wit-ic}(s, u)$ times, using $s$ and $u$ to resolve
nondeterminism when there are multiple successors to the \M{\ISAICM}
state. The goal is, for each instruction, ensure that the
\M{\ISAICM}'s cache prior to executing that instruction is equivalent
to the cache that the \M{\MAGICM} had when that instruction was
executed. This can be gleaned from the history information gathered by
\M{\MAGICM}. Once the desired cache state prior to instruction
execution is known, it is possible to choose the first
\textsc{isa-ic-c} transition that is part of an \textsc{isa-ic}
transition in such a way that the desired cache state is achieved
prior to instruction execution. A similar technique can be used to
determine what the state of the cache should be after each instruction
is executed, so the second \textsc{isa-ic-c} transition can be chosen
to achieve it.

\end{document}